\RequirePackage[mathlines]{lineno} 
\documentclass[a4paper,11pt]{article}

\usepackage{jheppub} 

\usepackage[T1]{fontenc} 

\usepackage{threeparttable}
\usepackage{multirow}
\usepackage{multicol}
\usepackage{subfigure}
\usepackage{amssymb,bm,mathrsfs,bbm,amscd}
\usepackage{float}
\let\oldequation\equation
\let\oldendequation\endequation
\renewenvironment{equation}
  {\linenomathNonumbers\oldequation}
  {\oldendequation\endlinenomath}

\def \ee   {e^{+}e^{-}}
\def \piz  {\pi^{0}}
\def \pip  {\pi^{+}}
\def \pim  {\pi^{-}}

\def \ifb  {\mbox{fb$^{-1}$}}
\def \gev  {\mbox{GeV}}
\def \gevc {\mbox{GeV/$c$}}
\def \gevcc{\mbox{GeV/$c^2$}}
\def \mev  {\mbox{MeV}}

\def \alphaA {\alpha_{\Lambda\rho(770)^+}}
\def \alphaB {\alpha_{\Sigma(1385)^{+}\pi^0}}
\def \alphaC {\alpha_{\Sigma(1385)^{0}\pi^+}}

\def \romanOne   {\uppercase\expandafter{\romannumeral1}}
\def \romanTwo   {\uppercase\expandafter{\romannumeral2}}
\def \romanThree {\uppercase\expandafter{\romannumeral3}}
\def \romanFour  {\uppercase\expandafter{\romannumeral4}}
\def \romanFive  {\uppercase\expandafter{\romannumeral5}}
\def \romanSix   {\uppercase\expandafter{\romannumeral6}}
\def \romanSeven {\uppercase\expandafter{\romannumeral7}}
\def \romanEight {\uppercase\expandafter{\romannumeral8}}

\def \pippiz {\pi^{+}\pi^{0}}

\def \mbc {M_{\rm{BC}}}
\def \dE {\Delta E}
\def \ebeam {E_{\mathrm{beam}}}

\def \lcp {\Lambda_{c}^{+}}
\def \lcm {\bar{\Lambda}_{c}^{-}}

\newcommand{\BR}{\mathcal{B}}

\begin{document}

\title{\boldmath Partial wave analysis of the charmed baryon hadronic decay $\Lambda_c^+\to\Lambda\pi^+\pi^0$}

\collaborationImg{\includegraphics[width=0.15\textwidth, angle=90]{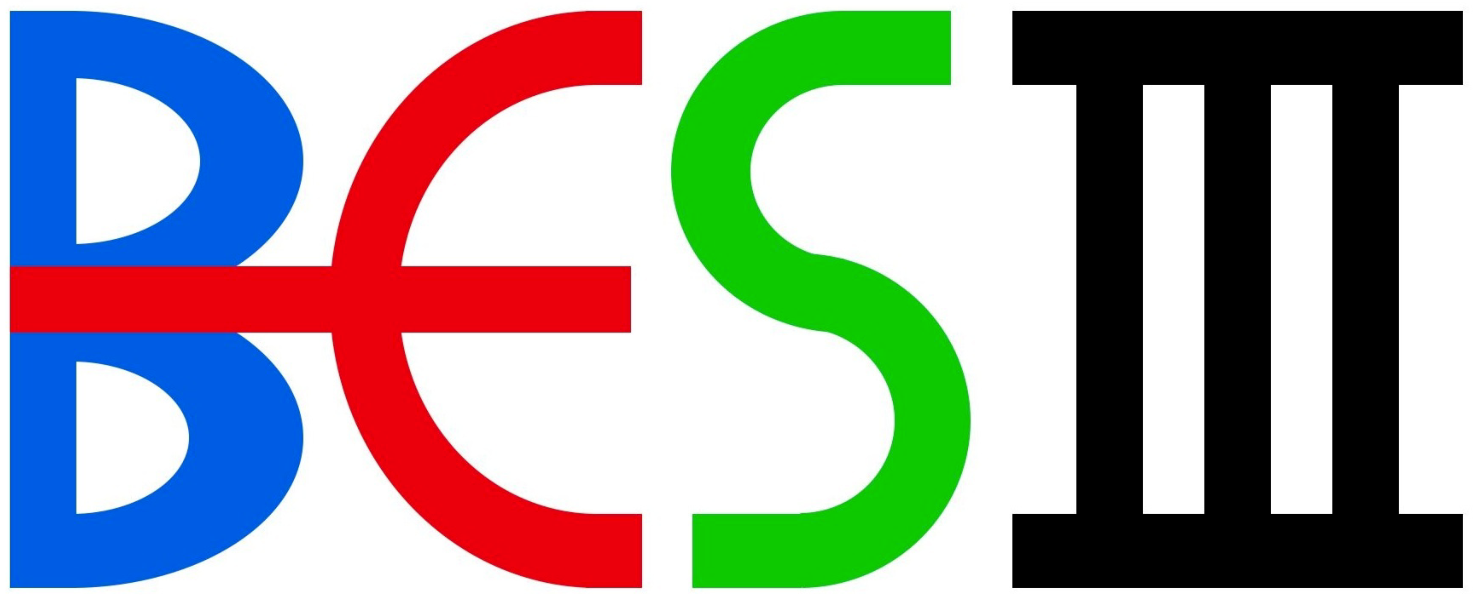}}
\collaboration{The BESIII collaboration}
\emailAdd{besiii-publications@ihep.ac.cn}

\abstract{ Based on $e^+e^-$ collision samples corresponding to an
  integrated luminosity of 4.4 $\mbox{fb$^{-1}$}$ collected with the
  BESIII detector at center-of-mass energies between
  $4.6\,\,\mathrm{GeV}$ and $4.7\,\,\mathrm{GeV}$, a partial wave
  analysis of the charmed baryon hadronic decay
  $\Lambda_c^+\to\Lambda\pi^+\pi^0$ is performed, and the decays
  $\Lambda_c^+\to\Lambda\rho(770)^{+}$ and
  $\Lambda_c^+\to\Sigma(1385)\pi$ are studied for the first time.
  Making use of the world-average branching fraction
  $\mathcal{B}(\Lambda_c^+\to\Lambda\pi^+\pi^0)$, their branching
  fractions are determined to be
\begin{eqnarray*}
\begin{aligned}
\mathcal{B}(\Lambda_c^+\to\Lambda\rho(770)^+)=&(4.06\pm0.30\pm0.35\pm0.23)\times10^{-2},\\
\mathcal{B}(\Lambda_c^+\to\Sigma(1385)^+\pi^0)=&(5.86\pm0.49\pm0.52\pm0.35)\times10^{-3},\\
\mathcal{B}(\Lambda_c^+\to\Sigma(1385)^0\pi^+)=&(6.47\pm0.59\pm0.66\pm0.38)\times10^{-3},\\
\end{aligned}
\end{eqnarray*}
where the first uncertainties are statistical, the second are
systematic, and the third are from the uncertainties of the branching
fractions $\mathcal{B}(\Lambda_c^+\to\Lambda\pi^+\pi^0)$ and
$\mathcal{B}(\Sigma(1385)\to\Lambda\pi)$. 
In addition, 
the decay asymmetry parameters are measured to be
$\alpha_{\Lambda\rho(770)^+}=-0.763\pm0.053\pm0.045$,
$\alpha_{\Sigma(1385)^{+}\pi^0}=-0.917\pm0.069\pm0.056$, and
$\alpha_{\Sigma(1385)^{0}\pi^+}=-0.789\pm0.098\pm0.056$.
}


\arxivnumber{2209.08464}

\maketitle
\flushbottom

\section{Introduction}
\label{sec:introduction}
\hspace{1.5em} The charmed baryon $\lcp$ was discovered 40 years
ago~\cite{LcDiscovery} and has been designated as the ground state of
the charmed baryons with spin-parity
$J^P=\frac{1}{2}^+$~\cite{BESIII:2020kap}.  Recently, remarkable
progress has been achieved in the study of hadronic weak decays of the
charmed baryons, including absolute branching fraction (BF)
measurements~\cite{Cheng:2015,Cheng:2021qpd,Li:2021iwf} and the
re-ordering of the lifetime
hierarchy~\cite{LHCb:2018nfa,LHCb:2019ldj,LHCb:2021vll,Cheng:2021vca}.
The BESIII collaboration is presently one of the most active
contributors to the measurements of the $\lcp$ decay
parameters~\cite{LcTagMode}.  The application of partial wave analysis
(PWA) techniques to multi-body hadronic decays of the $\lcp$ at BESIII
is expected to fill the current lack of information about $\lcp$
decays to an octet baryon and a vector meson, or a decuplet baryon and a pseudo-scalar
meson~\cite{pdg}.
For instance, the decays $\lcp\to\Sigma(1385)\pi$ and
$\lcp\to\Lambda\rho(770)^+$ have not yet been observed~\cite{pdg}, and
only an upper limit has been reported by CLEO ~\cite{CLEO1994}
assuming $\lcp\to\Lambda\rho(770)^+$ contributes 100\% in the
$\lcp\to\Lambda\pip\piz$ process. Therefore, the accurate measurements
of these intermediate states will provide crucial information about
the nature of charmed baryons.

In the field of theoretical calculations, much effort has been made in
recent years to evaluate the non-perturbative contributions to the
charmed baryon hadronic
decays~\cite{Cheng:2015,Cheng:2021qpd,Li:2021iwf}.  Based on different
theoretical approaches~\cite{theo2,theo31,theo32,theo4,theo5}, the BFs
and decay asymmetry parameters of the decays
$\lcp\to\Lambda\rho(770)^+$ and $\lcp\to\Sigma(1385)\pi$ are
predicted.  Table~\ref{tab:theo} summarizes the theoretical
predictions of the BFs of $\lcp\to\Lambda\rho(770)^+$ and
$\lcp\to\Sigma(1385)\pi$, as well as the current world average
published by the Particle Data Group (PDG). Remarkably, no
experimental measurement is available so far.

\vspace{-0.0cm}
\begin{figure}[htbp]
\setlength{\abovecaptionskip}{-1pt}
\setlength{\belowcaptionskip}{10pt}
\centering
\subfigure[\quad\quad\quad]{
\includegraphics[trim = 9mm 0mm 0mm 0mm, width=0.31\textwidth]{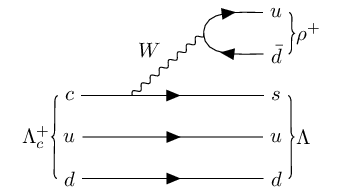}
}
\subfigure[\quad\quad\quad]{
\includegraphics[trim = 9mm 0mm 0mm 0mm, width=0.31\textwidth]{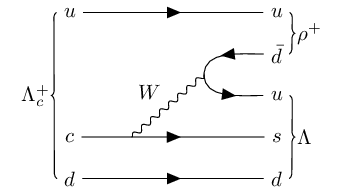}
}
\subfigure[\quad\quad\quad]{
\includegraphics[trim = 9mm 0mm 0mm 0mm, width=0.31\textwidth]{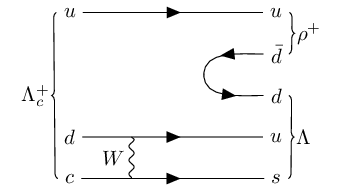}
}\\
\subfigure[\quad\quad\quad]{
\includegraphics[trim = 9mm 0mm 0mm 0mm, width=0.31\textwidth]{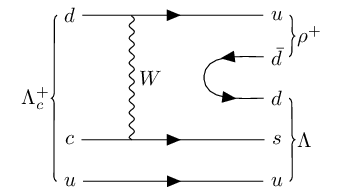}
}
\quad
\subfigure[\quad\quad\quad]{
\includegraphics[trim = 9mm 0mm 0mm 0mm, width=0.31\textwidth]{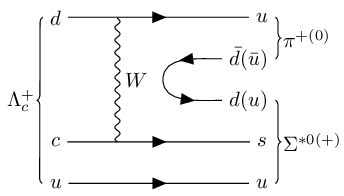}
}
\caption{Topological diagrams for the decays $\lcp\to\Lambda\rho(770)^+$ (a)-(d) and $\lcp\to\Sigma(1385)\pi$ (e). The external $W$-emission diagram (a) corresponds to a factorizable amplitude, while the internal $W$-emission (b) and $W$-exchange diagrams (c)-(e)  correspond to non-factorizable amplitudes. Here, the symbols $\rho^+$ and $\Sigma^*$ denote $\rho(770)^+$ and $\Sigma(1385)$, respectively.}
\label{fig:diagram}
\end{figure}

For the decay $\lcp\to\Lambda\rho(770)^+$, both factorizable and
non-factorizable diagrams will contribute to the
amplitude~\cite{theo2}. The contributing topological diagrams are
shown in Figure~\ref{fig:diagram}(a)-(d), where the external
$W$-emission diagram in Figure~\ref{fig:diagram}(a) contributes as a
factorizable amplitude and the internal $W$-emission in
Figure~\ref{fig:diagram}(b) and the $W$-exchange diagrams in
Figure~\ref{fig:diagram}(c) and (d) contribute as non-factorizable
amplitudes.  The BFs are related to the modulus squared of the sum
of different topological amplitudes, and the decay asymmetry
parameters are relevant to the interference of the internal partial
wave amplitudes. In theoretical calculations, Ref.~\cite{theo2}
adopted an effective Hamiltonian with a factorization approach and
$SU(3)_F$ symmetry, and Refs.~\cite{theo31,theo32} implemented the pole
model, where the baryon decay amplitude is parameterized as baryon and
meson pole contributions.

For the decay $\lcp\to\Sigma(1385)\pi$, only a non-factorizable
contribution~\cite{theo4} is expected in the quark diagram scheme.
The $\lcp$ baryon has spin-parity $J^P=\frac{1}{2}^+$ and its valence
quark structure consists of a charm quark and of a flavor
anti-symmetric $ud$ diquark component. The constituent quarks of the
$\Sigma(1385)$ baryon, instead, have a fully symmetric flavor
structure, due to its spin-parity $J^P=\frac{3}{2}^+$.  Hence, the
external $W$-emission diagram similar to Figure~\ref{fig:diagram}(a)
is forbidden in this decay.  According to the K\"orner-Pati-Woo
theorem~\cite{KPW1,KPW21,KPW22,KPW3}, the quark pair, which is
produced through the weak interaction and then confined into a baryon, is
flavor anti-symmetric and is not allowed in the $\Sigma(1385)$ quark
structure. Therefore, the internal $W$-emission and one of the
$W$-exchange diagrams, similar to Figure~\ref{fig:diagram}(b) and (c),
are also suppressed in this decay.  As a consequence, only the
$W$-exchange diagram in Figure~\ref{fig:diagram}(e) contributes to the
decay $\lcp\to\Sigma(1385)\pi$, which is a pure non-factorizable
contribution. In Ref.~\cite{theo4} the BFs are calculated through the
quark diagram scheme, while in Ref.~\cite{theo5} the BFs along with
decay asymmetry parameters are determined on the basis of an effective
Hamiltonian with $SU(3)_F$ symmetry.

\begin{table*}[!htbp]
\caption{Various theoretical calculations of BFs and decay asymmetry
  parameters of the decays $\lcp\to\Lambda\rho(770)^+$ and
  $\lcp\to\Sigma(1385)\pi$. Reference~\cite{theo2} adopted the
  effective Hamiltonian with factorization approach and $SU(3)_F$
  symmetry, Refs.~\cite{theo31,theo32} implemented the pole model,
  Ref.~\cite{theo4} considered the quark diagram scheme and
  Ref.~\cite{theo5} used the effective Hamiltonian with $SU(3)_F$
  symmetry. The current experimental measurement is
  from the PDG~\cite{pdg}, and ``---'' means unavailable.}
\setlength{\abovecaptionskip}{1.2cm}
\setlength{\belowcaptionskip}{0.2cm}
\label{tab:theo}
\begin{center}
\vspace{-0.0cm}
\resizebox{\textwidth}{23mm}{
\begin{tabular}{c|cccc|c}
\hline\hline
 & \multicolumn{4}{c|}{Theoretical calculation} & \multirow{2}{*}{PDG}\\
 & Ref.~\cite{theo2} & Ref.~\cite{theo31,theo32} & Ref.~\cite{theo4} & Ref.~\cite{theo5} & \\
\hline
$10^2\times\BR(\lcp\to\Lambda\rho(770)^+)$  & $4.81\pm0.58$ & $4.0$ & --- & --- & $<6$\\
$10^3\times\BR(\lcp\to\Sigma(1385)^+\piz)$   & --- & --- & $2.8\pm0.4$ & $2.2\pm0.4$ & ---\\
$10^3\times\BR(\lcp\to\Sigma(1385)^0\pip)$   & --- & --- & $2.8\pm0.4$ & $2.2\pm0.4$ & ---\\
$\alphaA$ & $-0.27\pm0.04$ & $-0.32$ & --- & --- & ---\\
$\alphaB$ & --- & --- &--- & $-0.91^{+0.45}_{-0.10}$ & ---\\
$\alphaC$ & --- & --- &--- & $-0.91^{+0.45}_{-0.10}$ & ---\\
\hline\hline
\end{tabular}
}
\end{center}
\end{table*}
\vspace{-0.0cm}
Since the non-factorizable contribution is more difficult to treat
than the factorizable one in the theoretical calculations,
detailed study of the complicated $\lcp\to\Lambda\rho(770)^+$ process
and the pure non-factorizable $\lcp\to\Sigma(1385)\pi$ process will provide important inputs to improve the theoretical calculation.
Thus, in this paper, the first PWA of the charmed baryon hadronic decay
$\lcp\to\Lambda\pip\piz$ is reported. The analysis is performed on
$\ee$ collision data samples with a total integrated luminosity of 4.4
\ifb~\cite{lumi1,lumi2} collected at center-of-mass (c.m.) energies
$\sqrt{s}$ between 4.6 and 4.7 GeV with the $\uchyph=0$BESIII
detector~\cite{detector} at the $\uchyph=0$BEPCII~\cite{BEPCII}
collider. The luminosities at each c.m. energy are listed in
Table~\ref{tab:lumi}.  From the PWA results, the fit fractions (FFs)
and the partial wave amplitudes of intermediate resonances can be
derived. Combining the FFs with the average value of
$\BR(\lcp\to\Lambda\pip\piz)=(7.1\pm 0.4)\%$ from the PDG~\cite{pdg},
the BFs for the decays $\lcp\to\Lambda\rho(770)^+$ and
$\lcp\to\Sigma(1385)\pi$ are determined for the first time. In
addition, using the partial wave amplitudes obtained in the
PWA, the corresponding decay asymmetry parameters are determined for
the first time. They will provide useful information for testing
theoretical calculations, especially for the descriptions of the
interference effects among the internal different partial
waves~\cite{asy1,asy2}. Charge conjugation is implied throughout this
paper unless mentioned explicitly otherwise.

\begin{table*}[!htbp]
\caption{The c.m. energies and integrated luminosities for the data
  samples~\cite{lumi1,lumi2}.}
\setlength{\abovecaptionskip}{1.2cm}
\setlength{\belowcaptionskip}{0.2cm}
\label{tab:lumi}
\begin{center}
\vspace{-0.0cm}
\begin{tabular}{c|c}
\hline\hline
$\sqrt{s}$ (GeV) & Luminosity ($\mathrm{pb}^{-1}$)\\
\hline
4.600 & $586.9\pm0.1\pm3.9$\\
4.612 & $103.8\pm0.1\pm0.6$\\
4.628 & $521.5\pm0.1\pm2.8$\\
4.641 & $552.4\pm0.1\pm2.9$\\
4.661 & $529.6\pm0.1\pm2.8$\\
4.682 & $1669.3\pm0.2\pm8.8$\\
4.699 & $536.4\pm0.1\pm2.8$\\
\hline\hline
\end{tabular}
\end{center}
\end{table*}
\vspace{-0.0cm}

\section{BESIII experiment and Monte Carlo simulation}
\label{sec:detector}
\hspace{1.5em} The $\uchyph=0$BESIII detector~\cite{detector} records
symmetric $\ee$ collisions provided by the $\uchyph=0$BEPCII storage
ring~\cite{BEPCII}, which operates at c.m energies ranging from 2.0 to
4.95 GeV, with a peak luminosity of $1 \times
10^{33}\;\mathrm{cm}^{-2}\mathrm{s}^{-1}$ achieved at $\sqrt{s} =
3.77\;\mathrm{GeV}$. The $\uchyph=0$BESIII detector has collected
large data samples in this energy region~\cite{Ablikim:2019hff}.  The
cylindrical core of the $\uchyph=0$BESIII detector covers 93\% of the
full solid angle and consists of a helium-based multilayer drift
chamber (MDC), a plastic scintillator time-of-flight system (TOF), and
a CsI(Tl) electromagnetic calorimeter (EMC), which are all enclosed in
a superconducting solenoidal magnet providing a 1.0 T magnetic field.
The solenoid is supported by an octagonal flux-return yoke with
resistive plate counter muon identification modules interleaved with
steel.  The charged-particle momentum resolution at $1\,\gevc$ is
$0.5\%$, and the $\mathrm{d}E/\mathrm{d}x$ resolution is $6\%$ for
electrons from Bhabha scattering. The EMC measures photon energies
with a resolution of $2.5\%$ ($5\%$) at $1\,\gev$ in the barrel (end
cap) region.  The time resolution in the TOF barrel region is 68 ps,
while that in the end cap region was initially 110 ps. The end cap TOF
system was upgraded in 2015 using multi-gap resistive plate chamber
technology, providing a time resolution of 60
ps~\cite{etof1,etof2,etof3}.

Simulated samples produced with {\sc geant4}-based~\cite{geant4} Monte
Carlo (MC) software, which includes the geometric description of the
BESIII detector and the detector response
performance~\cite{GDMLMethod,BesGDML,bes3GeoRecent}, are used to determine detection
efficiencies and to estimate background contributions. The simulation
describes the beam energy spread and initial state radiation (ISR) in
the $e^+e^-$ annihilations with the generator {\sc
  kkmc}~\cite{ref:kkmc1,ref:kkmc2}.  The inclusive MC samples,
corresponding to around 40 times of the number of events of the data
samples, include the production of $\lcp\lcm$ pairs, open charm
processes, the ISR production of vector charmonium(-like) states, and
the continuum processes incorporated in {\sc
  kkmc}~\cite{ref:kkmc1,ref:kkmc2}.  The known decay modes are
modelled with {\sc evtgen}~\cite{ref:evtgen1,ref:evtgen2} using BFs
taken from the PDG~\cite{pdg} and the remaining unknown charmonium
decays are modelled with {\sc
  lundcharm}~\cite{ref:lundcharm1,ref:lundcharm2}. Final state
radiation from charged final state particles is incorporated using
{\sc photos}~\cite{photos}. The MC samples of the signal process
$\lcp\to\Lambda\pip\piz$ are produced with a uniform phase-space
distribution (PHSP).  At each c.m. energy point 800k events are
generated, except at 4.682 GeV, where the sample consists of 1.6M
events to reflect the larger integrated luminosity of the data set.

\section{Event selection}
\label{sec:selection}
\hspace{1.5em} The $\ee$ collision energies of the data sets are just
above the production threshold of the $\lcp\lcm$ pair, providing a
clean environment without additional accompanying hadrons. Taking
advantage of the threshold pair production and of the excellent
performance of the $\uchyph=0$BESIII detector, a single-tag strategy
is used, where only one charmed baryon decay is reconstructed
($\lcp$), improving the detection efficiency and, therefore, providing
a larger data sample.  The signal candidates for
$\lcp\to\Lambda\pip\piz$ are reconstructed from combinations of
charged tracks and photon candidates recorded by the detector that
satisfy the following selection criteria.

Charged particle tracks detected in the MDC are required to have a
polar angle $\theta$ in the range of $|\cos\theta|<0.93$, where
$\theta$ is defined with respect to the $z$-axis, which is the
symmetry axis of the MDC.  For charged tracks not originating from
$\Lambda$ decays, the distance of the closest approach to the
interaction point (IP) is required to be less than 10 cm along the
$z$-axis ($V_z$), and to be less than 1 cm in the perpendicular
plane ($V_r$). Particle identification (PID) for charged tracks is
implemented by combining the information of specific ionization energy
loss in the MDC ($\mathrm{d}E/\mathrm{d}x$) and the time of flight
measured in the TOF into a likelihood value $\mathcal{L}(h)$ for each
hadron $h$ hypothesis, where $h=p$, $K$, or $\pi$. Charged tracks are
identified as protons if the proton hypothesis has the greatest
likelihood ($\mathcal{L}(p)>\mathcal{L}(K)$ and
$\mathcal{L}(p)>\mathcal{L}(\pi)$), or as pions if satisfying
$\mathcal{L}(\pi)>\mathcal{L}(K)$.

Photon candidates from $\piz$ decays are reconstructed using
electromagnetic showers in the EMC. The deposited energy of each
shower is required to be larger than $25\,\mev$ in the barrel region
($|\cos \theta|< 0.80$) and larger than $50\,\mev$ in the end cap
region ($0.86 <|\cos \theta|< 0.92$). To reject fake photons arising
from electronic noise, beam background, and showers unrelated to the
event, the difference between the EMC time and the event start
time~\cite{Guan:2013jua} is required to be within 700 ns. For $\piz$
candidates, the invariant mass of the photon pair is required to be
within $0.115<M_{\gamma\gamma}<0.150\,\gevcc$. To further improve the
momentum resolution, a one-constraint (1C) kinematic fit is performed by
constraining the invariant mass of the photon pair to the nominal
$\piz$ mass~\cite{pdg}. The updated momentum will be used in the
further analysis.

The $\Lambda$ candidates are reconstructed with two oppositely charged
tracks identified as $p$ and $\pi^-$. The tracks are required to
satisfy $V_z < 20$ cm, and no $V_r$ requirement is imposed.  
For proton, the previous mentioned PID requirement is applied, 
while for pion, it is not. The $p\pim$ pairs
are constrained to originate from a common vertex by requiring the
$\chi^2$ of a vertex fit to be less than 100. 
An additional fit is performed by constraining the momentum of $p\pim$ 
pair to be aligned with the direction from the IP to the $\Lambda$ decay vertex, 
and the fitted decay length is required to be larger than twice its uncertainty.
Furthermore, the invariant mass is required to be within
$1.111<M_{p\pim}<1.121\,\gevcc$.

If an event satisfies both the above $\lcp\to\Lambda\pip\piz$
selection criteria and $\lcp\to\Sigma^0\pip$ selection criteria, it
will be rejected to veto $M_{\Lambda\pip}$ peaking background arising
from $\lcp\to\Sigma^0\pip$ decay. Here, the $\lcp\to\Sigma^0\pip$
selection criteria consist of selecting $\pip$, $\Lambda$ and $\gamma$
candidates where the combination $\Lambda\gamma$ is required to be
within $\Sigma^0$ mass window $1.179<M_{\Lambda\gamma}<1.203\;\gevcc$,
as described in detail in Ref.~\cite{LcTagMode}.  
After applying these selections, no peaking background is found in data, 
while only 1.6\% signal detection efficiency is lost, 
as evaluated in the inclusive MC samples.

To further select the signal candidates, the beam-constrained mass
$\mbc$ and the energy difference $\dE$ are used, defined as
\begin{eqnarray}
\mbc \equiv \sqrt{{\ebeam}^2/c^4-\left|\vec{p}\right|^2/c^2}
\label{eq:mbc}
\end{eqnarray}
and
\begin{eqnarray}
\dE \equiv E-\ebeam,
\label{eq:dEtag}
\end{eqnarray}
where $\ebeam$ is the beam energy, $\vec{p}$ and $E$ are the
reconstructed momentum and energy of the signal candidate,
respectively. Signal candidates are expected to have $\mbc$ and $\dE$
consistent with the nominal $\lcp$ mass~\cite{pdg} and zero,
respectively. The combination with the minimum $|\dE|$ is considered
as the $\lcp$ candidate and its $\dE$ is required to satisfy
$-0.03<\dE<0.02\,\gev$. The $\dE$ distributions of data and signal MC
samples are shown in Figure~\ref{fig:deltaE}.
\vspace{-0.0cm}
\begin{figure}[htbp]
\setlength{\abovecaptionskip}{-1pt}
\setlength{\belowcaptionskip}{10pt}
\centering
\includegraphics[trim = 9mm 0mm 0mm 0mm, width=0.31\textwidth]{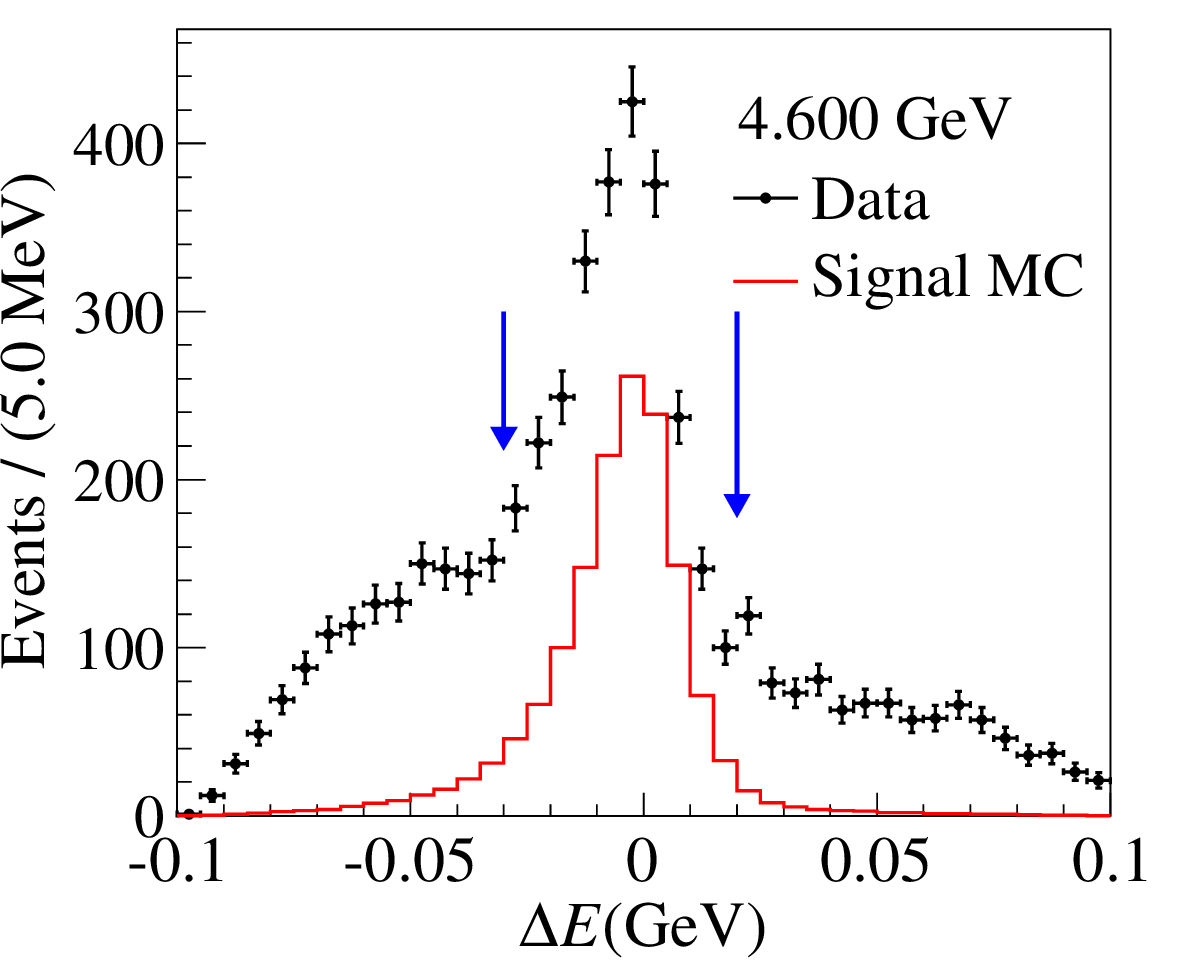}
\includegraphics[trim = 9mm 0mm 0mm 0mm, width=0.31\textwidth]{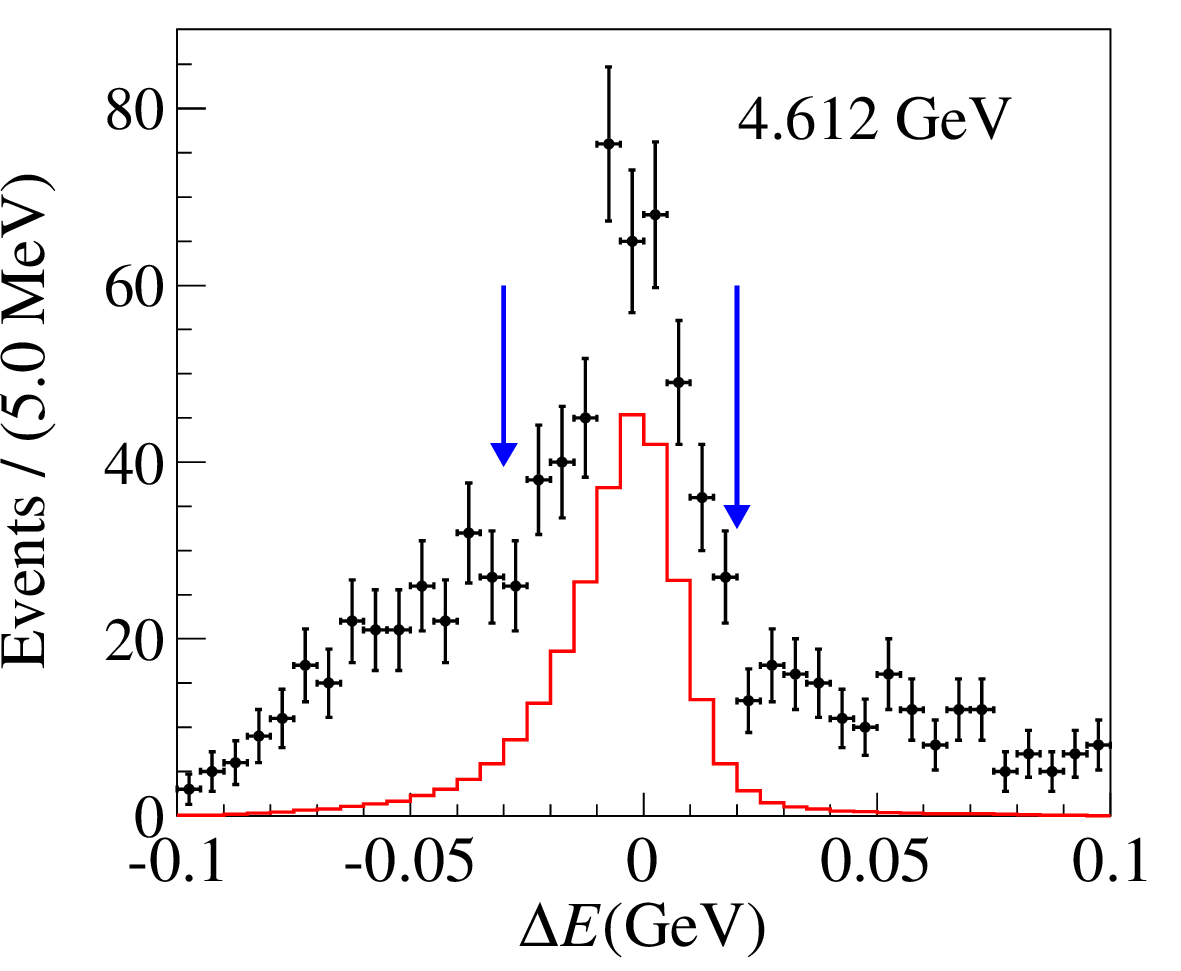}
\includegraphics[trim = 9mm 0mm 0mm 0mm, width=0.31\textwidth]{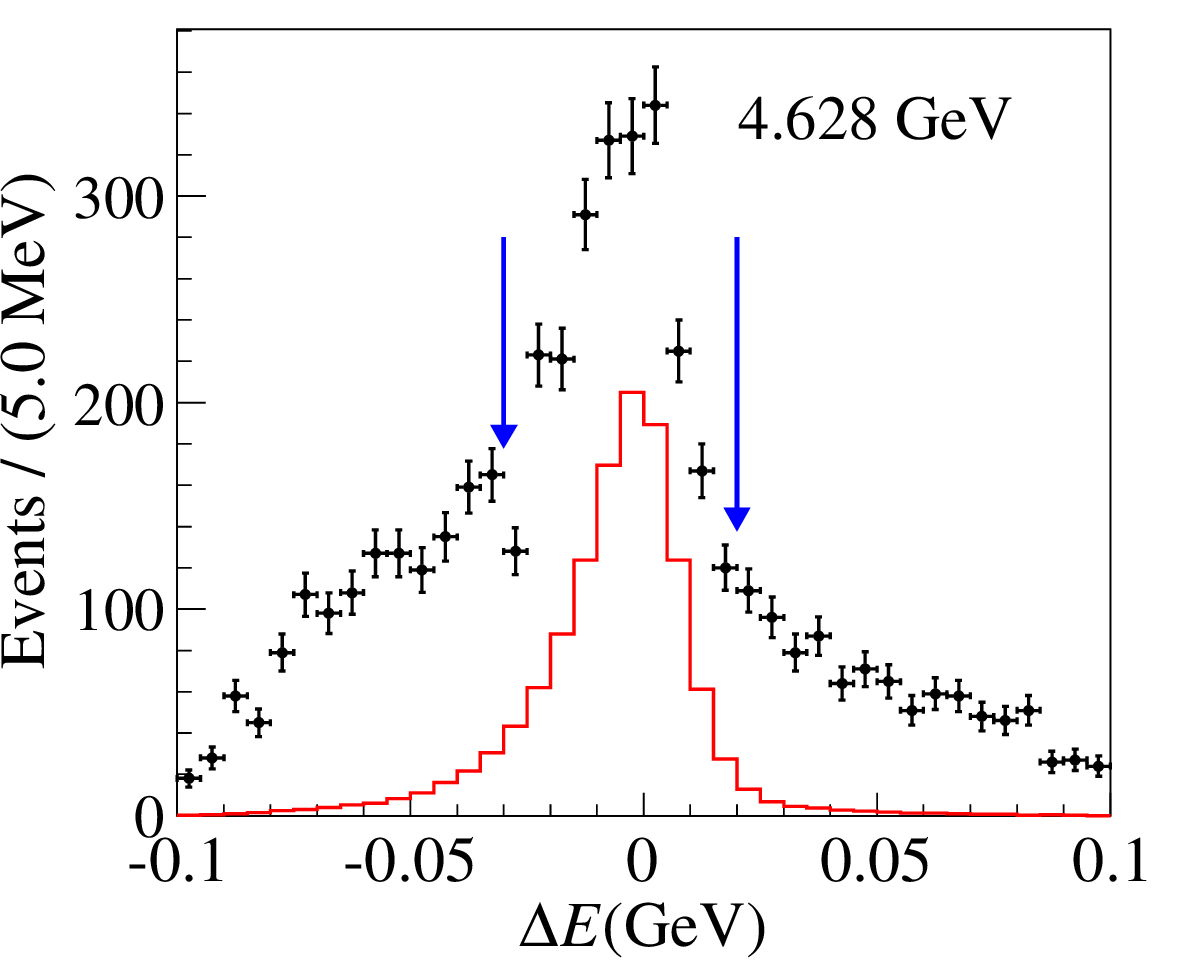}\\
\includegraphics[trim = 9mm 0mm 0mm 0mm, width=0.31\textwidth]{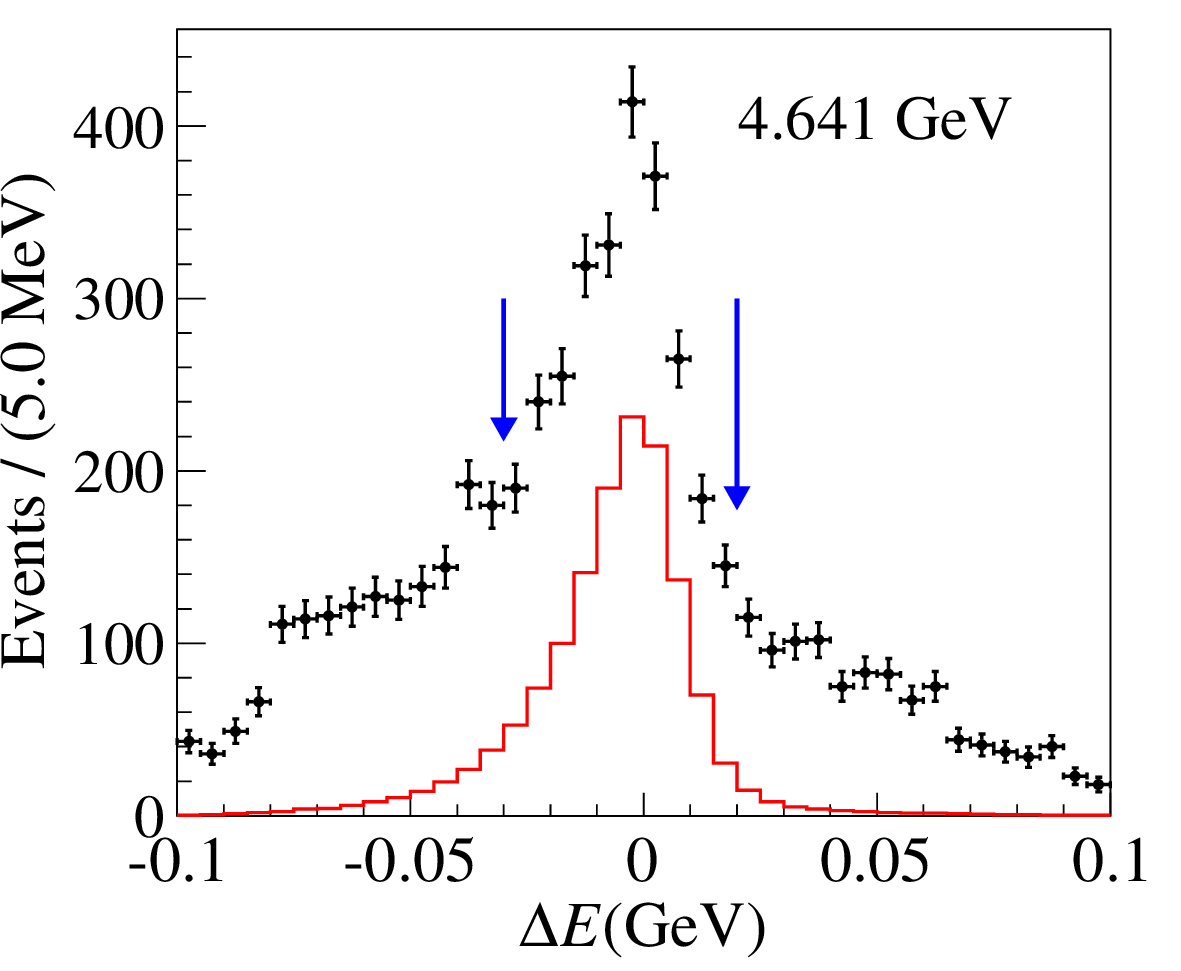}
\includegraphics[trim = 9mm 0mm 0mm 0mm, width=0.31\textwidth]{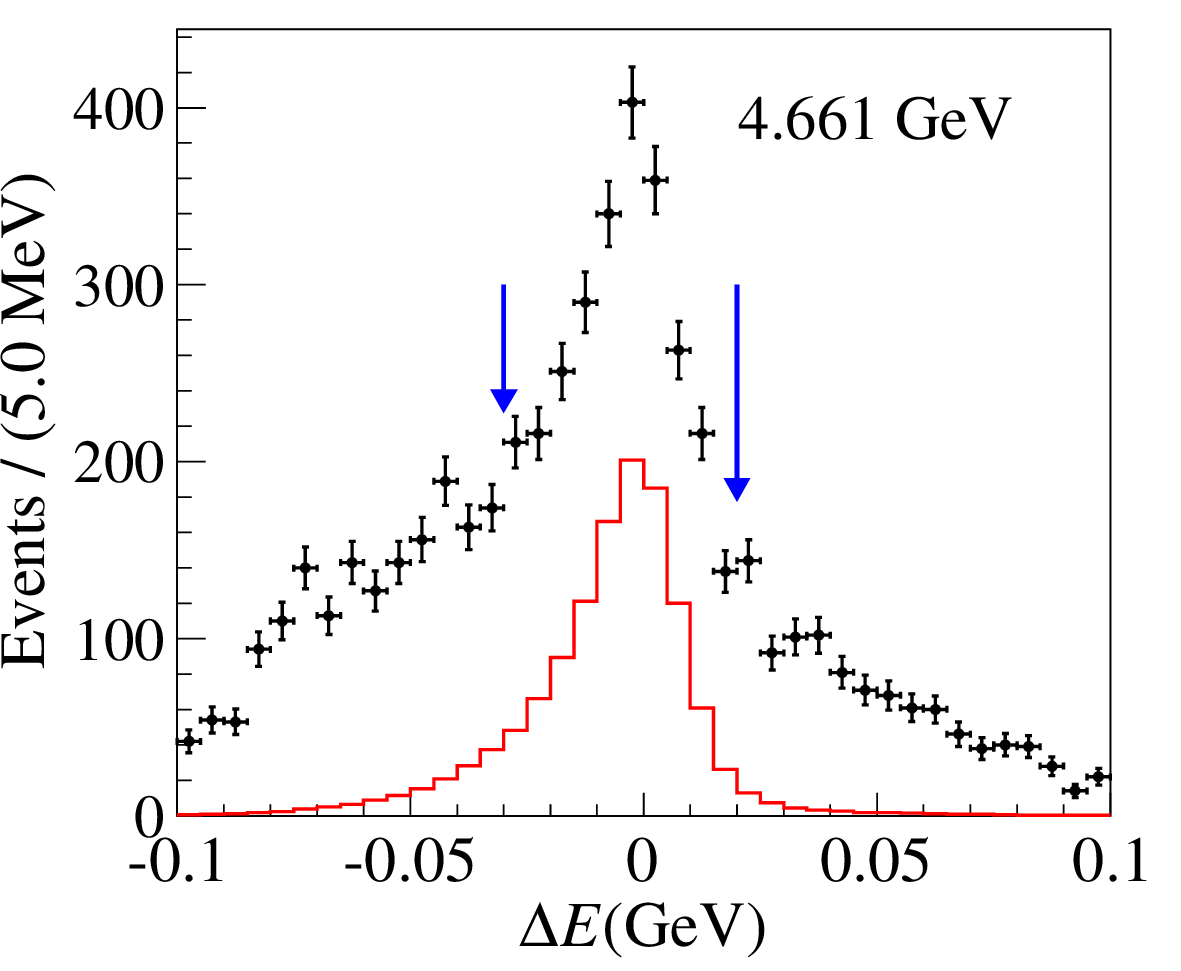}
\includegraphics[trim = 9mm 0mm 0mm 0mm, width=0.31\textwidth]{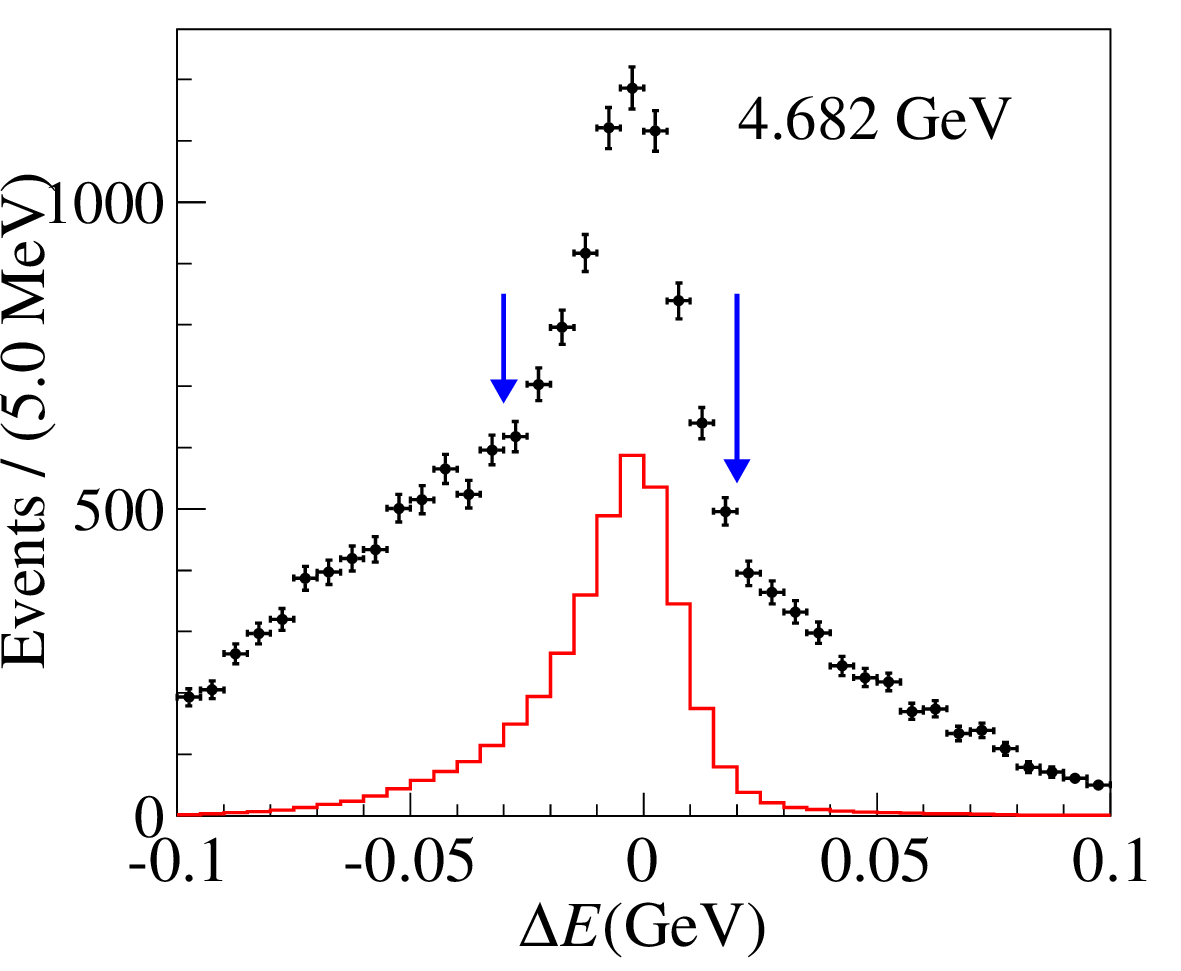}\\
\includegraphics[trim = 9mm 0mm 0mm 0mm, width=0.31\textwidth]{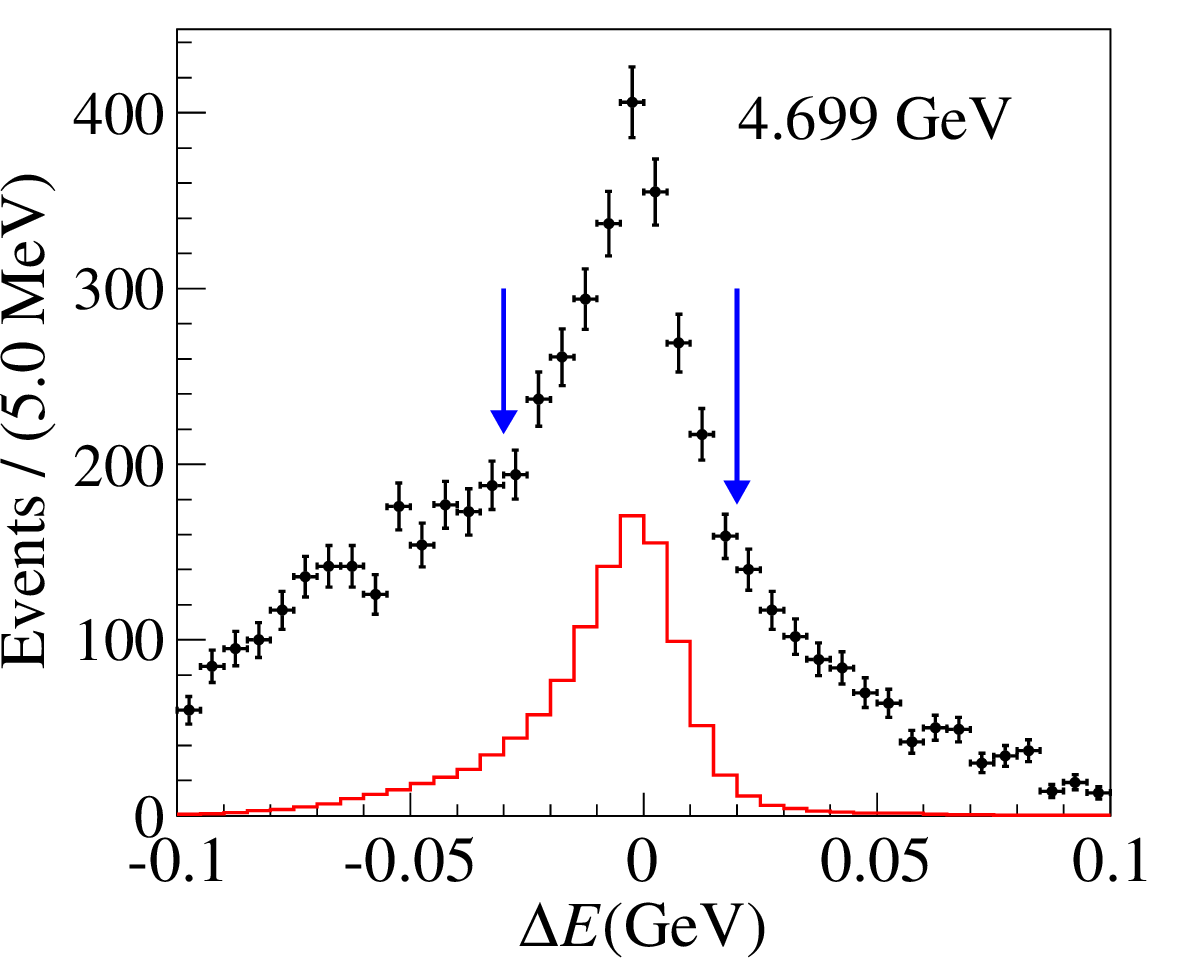}
\caption{The $\dE$ distributions of data (dots with error bars) and
  signal MC samples (thick lines) at seven energy points for
  $\lcp\to\Lambda\pip\piz$ decay. The blue arrows indicate the $\dE$ selection region. The signal MC samples are presented
  in arbitrary scale for illustration.}
\label{fig:deltaE}
\end{figure}

An extended un-binned maximum likelihood fit is performed on the
$\mbc$ distribution of each energy point, as shown in
Figure~\ref{fig:fitmbc}, in order to determine the signal and
background yields. In the fit, the signal shape is derived from the
kernel-estimated non-parametric shape~\cite{keyspdf} based on PHSP MC
samples convolved with a Gaussian function, to account for the
difference between data and MC simulation caused by imperfect
modelling of the detector resolution and beam-energy spread. The
parameters of the Gaussian function are left free in the fit. The
background shape is modelled by an $\uchyph=0$ARGUS
function~\cite{1990argus} defined as
\begin{eqnarray}
f(\mbc,E_0,c,p)=\mbc\left(1-\left(\frac{\mbc}{E_0}\right)^2\right)^p\times\mathrm{e}^{c\cdot\left(1-\frac{\mbc}{E_0}\right)^2} ,
\label{eq:argus}
\end{eqnarray}
where $E_0$ is the endpoint of $\mbc$ and is fixed to the beam energy,
$p$ is the power parameter and is equal to 0.5, and $c$ is a free
parameter in the fit.  The global probability density function
consists of a linear combination of signal and background
contributions.  The fit results, the $\mbc$ signal regions, as well as
the signal purities and background fractions are listed in
Table~\ref{tab:fitmbc}.  Events within the $\mbc$ signal region are
considered as signal candidates in the further PWA fit, and the events
underlying the $\mbc$ sideband region $2.25<\mbc<2.27\,\gevcc$ are
considered as background events.  At each energy point, the signal
purity within the $\mbc$ signal region is larger than 80\%, which is
sufficient to perform a reliable partial wave analysis.
  
\vspace{-0.0cm}
\begin{figure}[htbp]
\setlength{\abovecaptionskip}{-1pt}
\setlength{\belowcaptionskip}{10pt}
\centering
\includegraphics[trim = 9mm 0mm 0mm 0mm, width=0.31\textwidth]{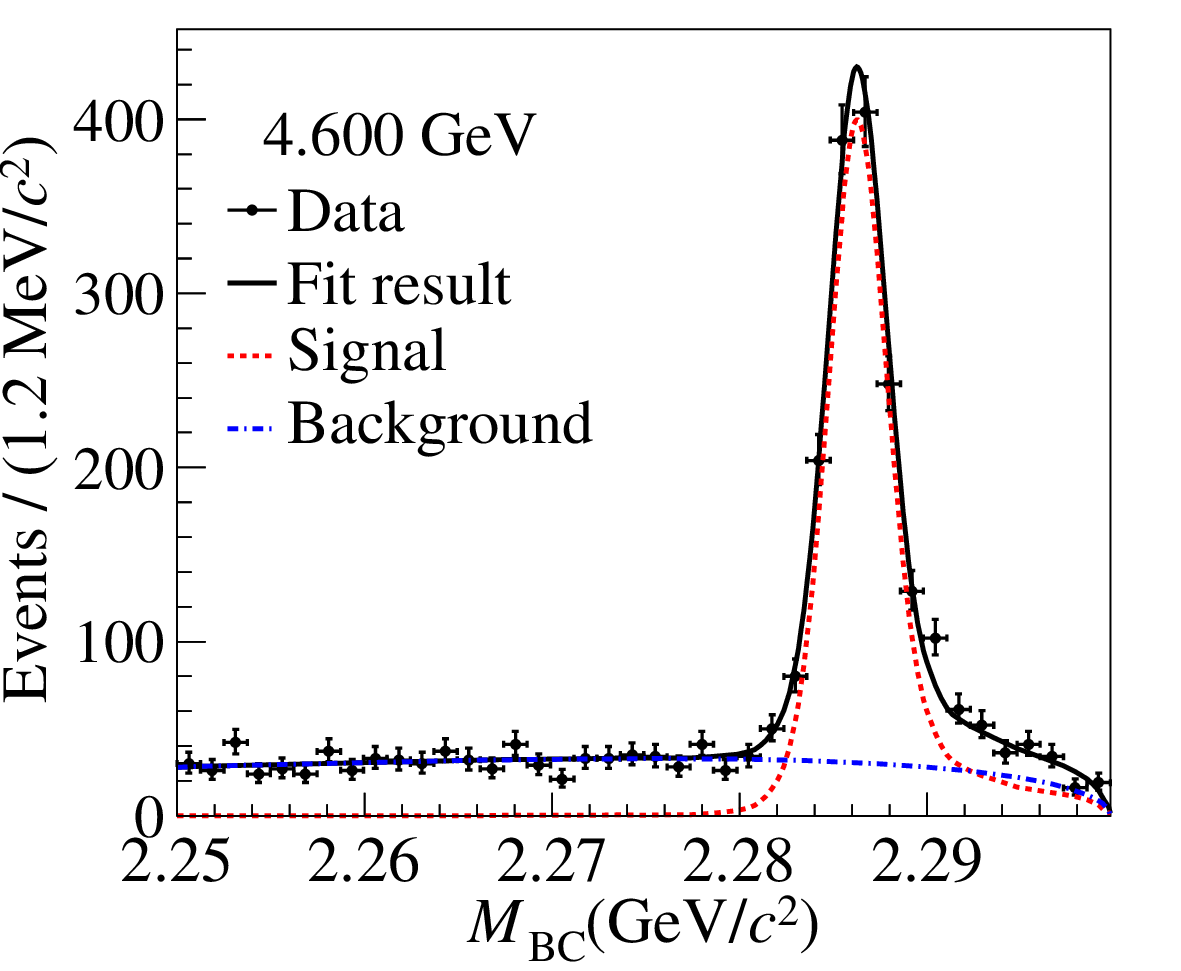}
\includegraphics[trim = 9mm 0mm 0mm 0mm, width=0.31\textwidth]{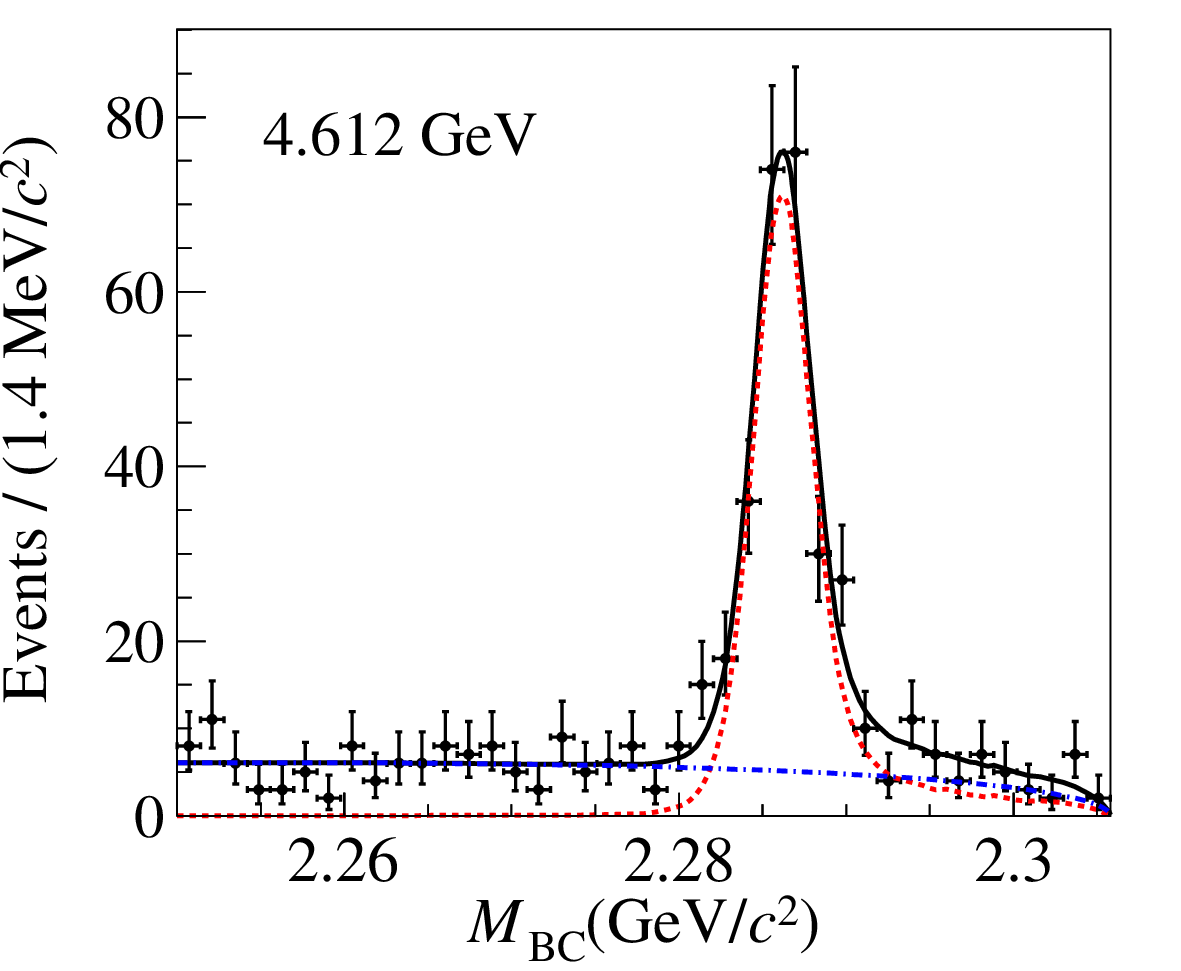}
\includegraphics[trim = 9mm 0mm 0mm 0mm, width=0.31\textwidth]{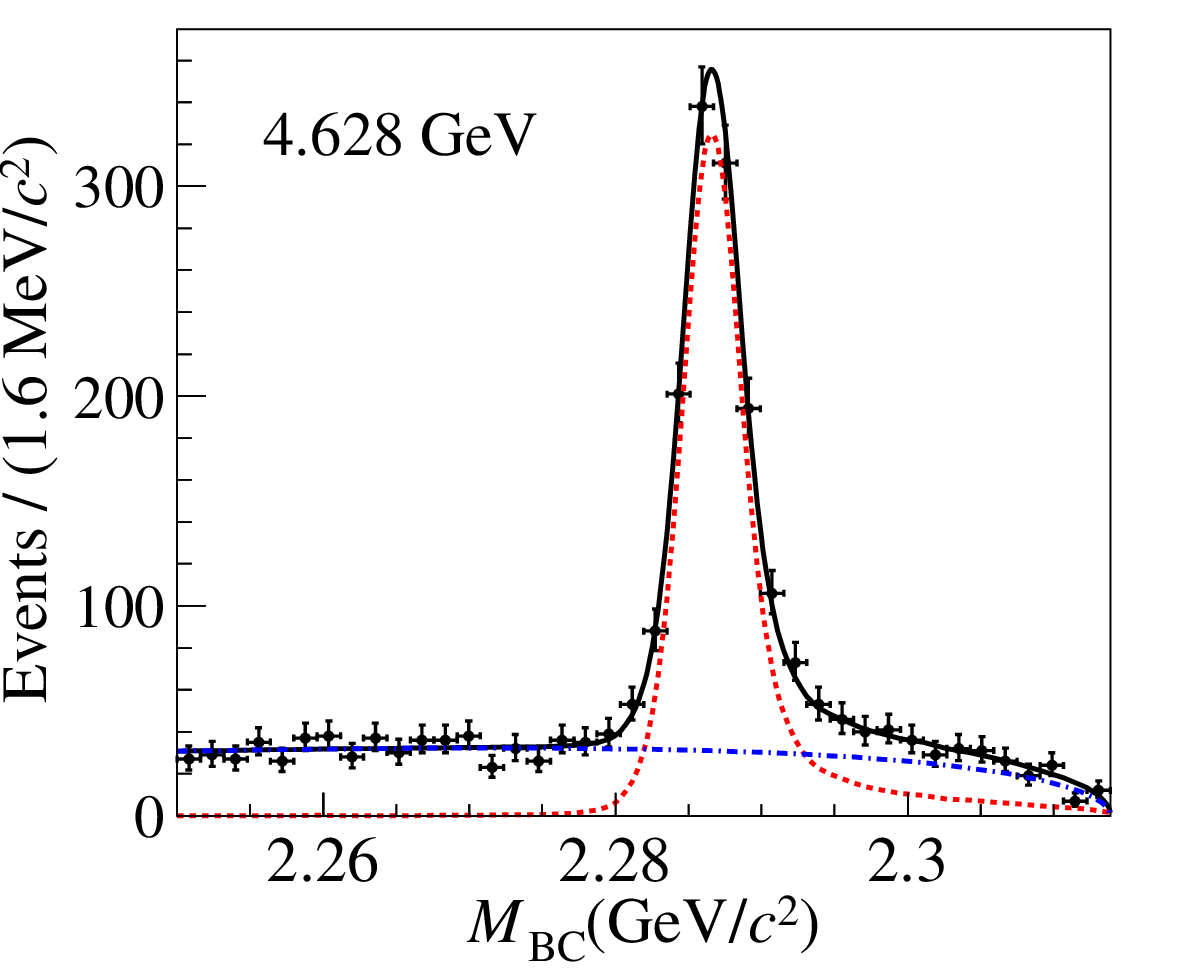}\\
\includegraphics[trim = 9mm 0mm 0mm 0mm, width=0.31\textwidth]{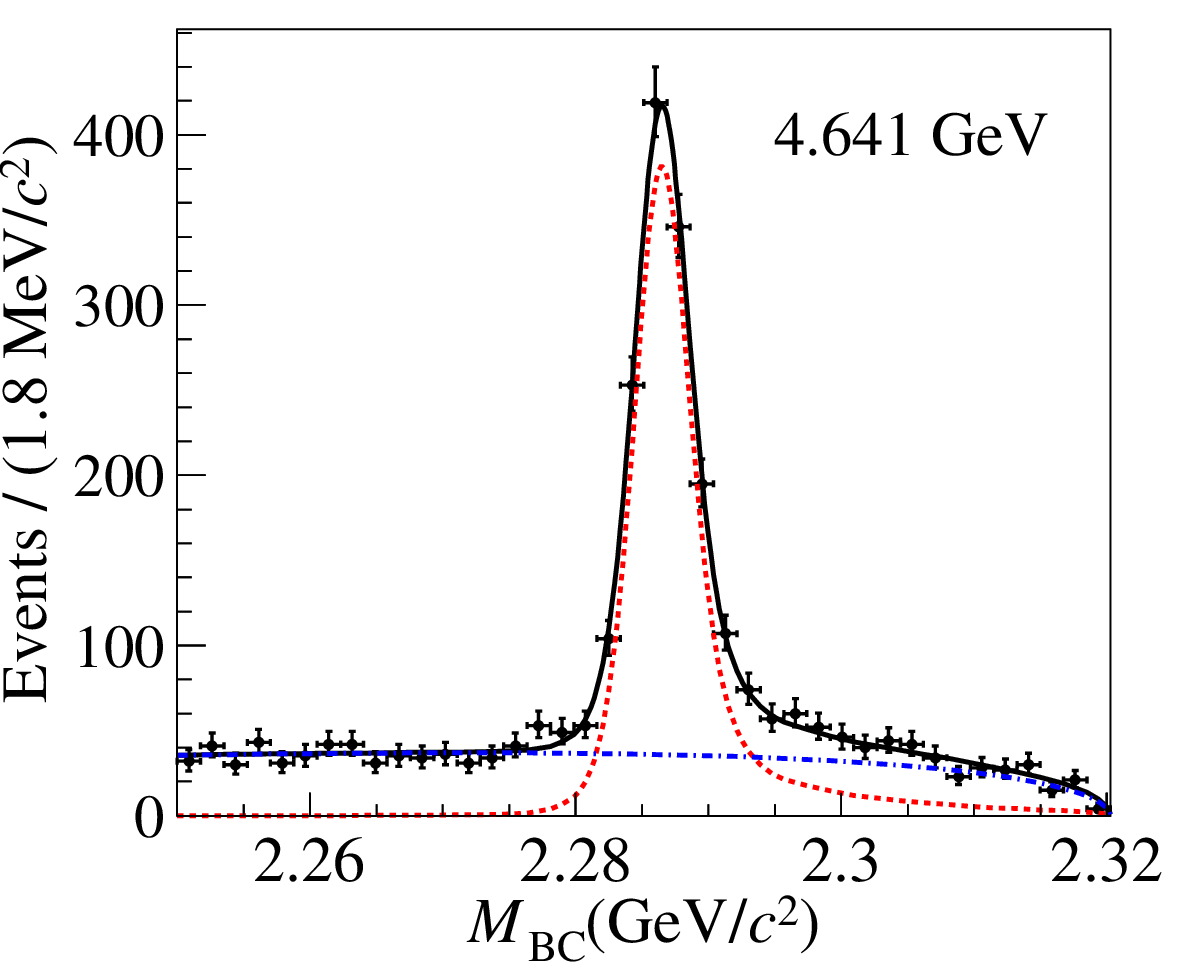}
\includegraphics[trim = 9mm 0mm 0mm 0mm, width=0.31\textwidth]{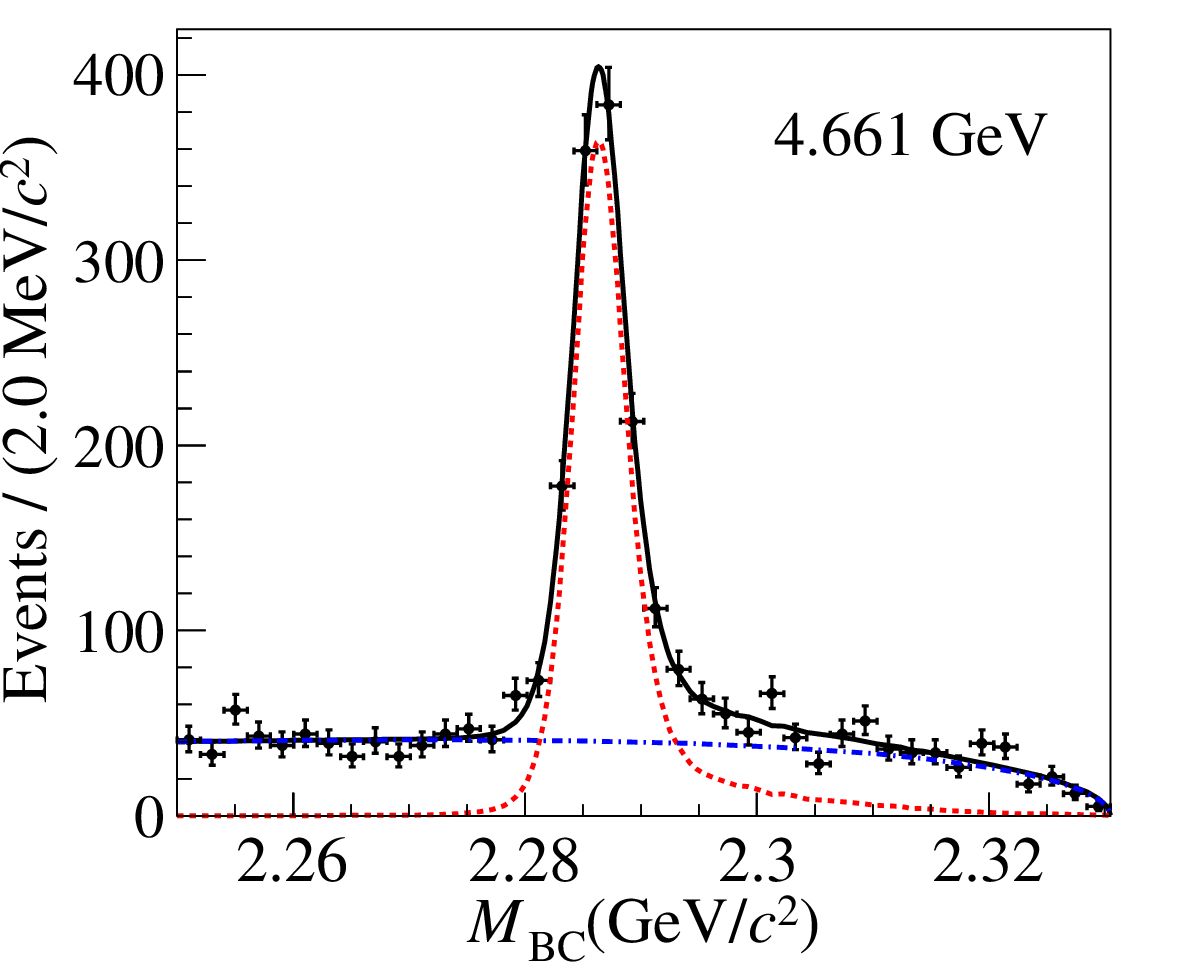}
\includegraphics[trim = 9mm 0mm 0mm 0mm, width=0.31\textwidth]{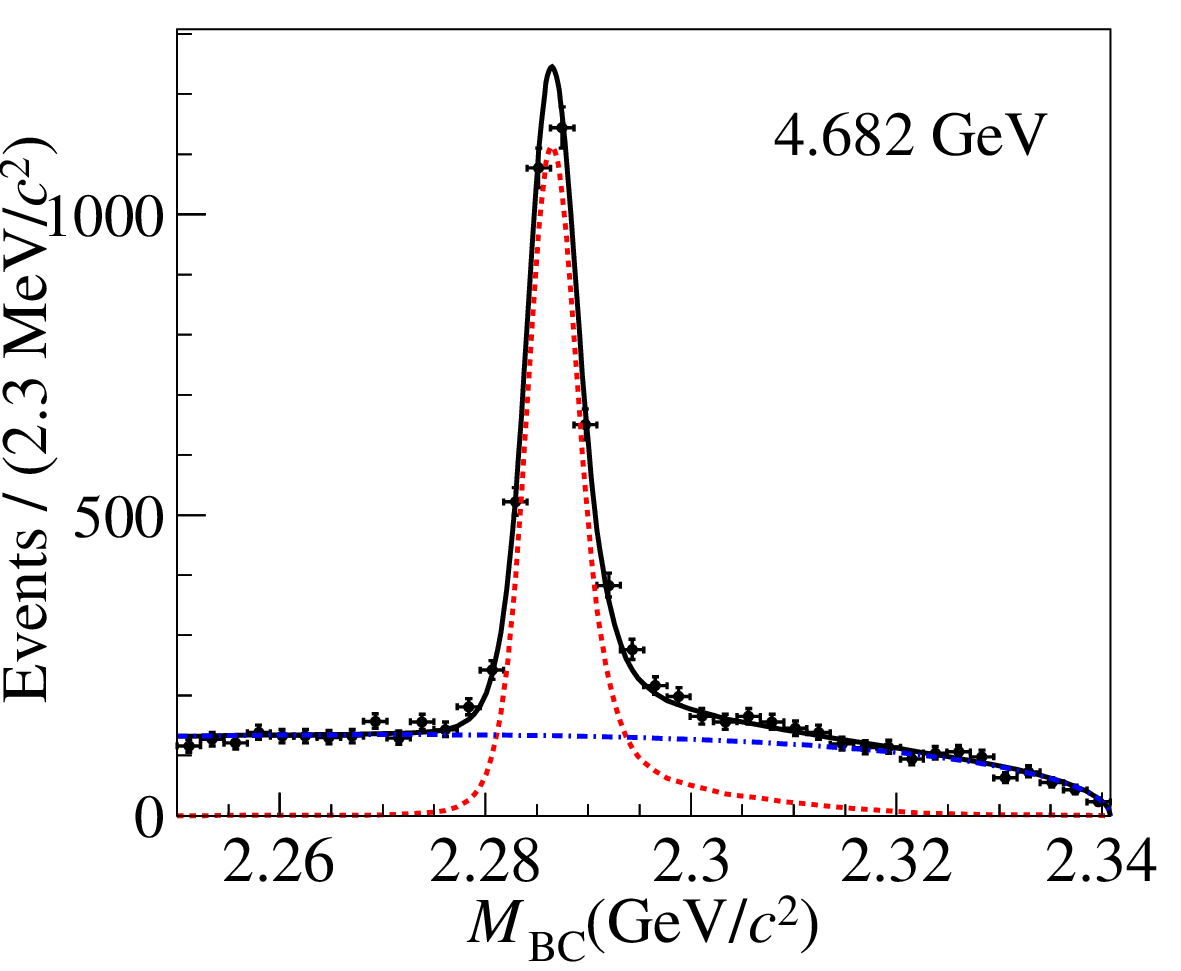}\\
\includegraphics[trim = 9mm 0mm 0mm 0mm, width=0.31\textwidth]{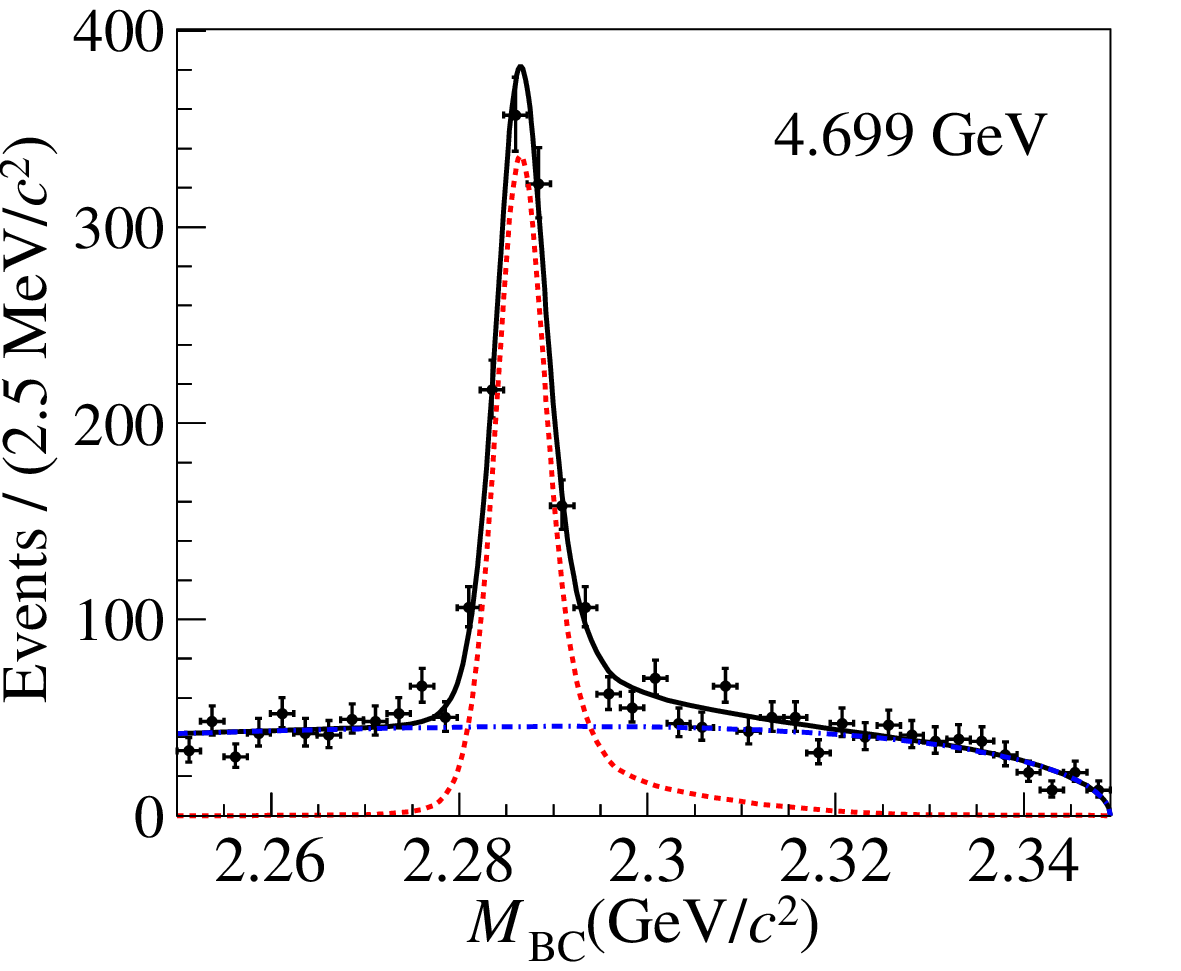}
\caption{Fits to the $\mbc$ distributions at each energy point for
  the $\lcp\to\Lambda\pip\piz$ decay. The dots with error bars are
  data, the (black) solid curve is the fit function, which is the sum
  of the signal shape (red dashed curve) and the background shape
  (blue dash-dotted curve). No obvious peaking background is observed.}
\label{fig:fitmbc}
\end{figure}

\begin{table*}[!htbp]
\caption{The results of signal and background (Bkg) yields in the $\mbc$ signal region at seven energy points, along with $\mbc$ requirement boundaries, signal purities and background fractions. The $\mbc$ requirement is applied to improve the signal purities.}
\setlength{\abovecaptionskip}{1.2cm}
\setlength{\belowcaptionskip}{0.2cm}
\label{tab:fitmbc}
\begin{center}
\vspace{-0.0cm}
\resizebox{\textwidth}{19mm}{
\begin{tabular}{c|ccccc}
	\hline \hline
	$\sqrt{s}$ (GeV)  & $\mbc$ requirement ($\gevcc$) & Signal yield & Bkg yield & Purity (\%) & Bkg fraction (\%)\\
	\hline
	4.600 & $(2.282, 2.291)$ & $1351\pm43$ & $217.4\pm8.2$ & $86.1\pm0.7$ & $13.9\pm0.7$\\
	4.612 & $(2.282, 2.291)$ & $233\pm17$  & $32.8\pm2.9$  & $87.7\pm1.4$ & $12.3\pm1.4$\\
	4.628 & $(2.282, 2.291)$ & $1040\pm37$ & $174.2\pm6.7$ & $85.7\pm0.8$ & $14.3\pm0.8$\\
	4.641 & $(2.282, 2.292)$ & $1200\pm39$ & $203.5\pm7.1$ & $85.5\pm0.7$ & $14.5\pm0.7$\\
	4.661 & $(2.282, 2.292)$ & $1047\pm35$ & $199.1\pm6.3$ & $84.0\pm0.7$ & $16.0\pm0.7$\\
	4.682 & $(2.282, 2.293)$ & $3120\pm63$ & $642\pm11$    & $82.9\pm0.4$ & $17.1\pm0.4$\\
	4.699 & $(2.282, 2.293)$ & $906\pm34$  & $201.4\pm5.9$ & $81.8\pm0.8$ & $18.2\pm0.8$\\
	\hline\hline
\end{tabular}
}
\end{center}
\end{table*}
\vspace{-0.0cm}

\section{Partial wave analysis}
\label{sec:PWA}
\hspace{1.5em} 

After applying all the selection criteria mentioned
above, around 10k signal events are selected in data and can be used
to fit the helicity amplitude of the $\lcp\to\Lambda\pip\piz$ decay.
To improve the momentum resolution, an
additional 3C kinematic fit under the hypothesis of $\ee \to \lcp (\to
\Lambda \pi^+\pi^0) \lcm$ is performed with the $p\pi^-$ system
constrained to the nominal $\Lambda$ mass, the $\Lambda \pi^+\pi^0$ 
constrained to the $\lcp$ mass and 
the recoil mass against $\Lambda \pi^+\pi^0$ constrained to the $\lcp$ mass. 

In this work, the decay amplitude is constructed using the helicity
amplitude formalism, and the full procedure is implemented based on the
open-source framework called TF-PWA~\cite{tfpwa}. The amplitude is
defined in the $\lcp$ rest frame to which all final state particles
are boosted.  Parameters describing the amplitude of the $\lcm$ decay
are related to those of $\lcp$ by performing a parity transformation
on the $\lcm$ candidates, under the assumption of CP conservation.

\subsection{Helicity angle definitions}
\label{sec:angle}
\hspace{1.5em}

The full decay amplitude of $\lcp\to\Lambda\pip\piz$ consists of three decay chains:
\begin{itemize}
\item $\lcp\to\Lambda\rho(770)^+(\theta^1_{\lcp}), \rho(770)^+\to \pip\piz (\theta_{\rho^+},\phi^{\rho^+}_{\piz}), \Lambda\to p\pim(\theta_{\Lambda_1},\phi_p^{\Lambda_1})$, 
\item $\lcp\to\Sigma^{*+}\pi^0(\theta^2_{\lcp}), \Sigma^{*+}\to\Lambda\pi^+(\theta_{\Sigma^{*+}},\phi^{\Sigma^{*+}}_{\Lambda}), \Lambda\to p\pim(\theta_{\Lambda_2},\phi_p^{\Lambda_2})$,
\item $\lcp\to\Sigma^{*0}\pi^+(\theta^3_{\lcp}), \Sigma^{*0}\to\Lambda\pi^0(\theta_{\Sigma^{*0}},\phi^{\Sigma^{*0}}_{\Lambda}), \Lambda\to p\pim(\theta_{\Lambda_3},\phi_p^{\Lambda_3})$.
\end{itemize}
The corresponding helicity angle definitions are shown in Figure~\ref{fig:angle}.
\vspace{-0.0cm}
\begin{figure}[htbp]
\setlength{\abovecaptionskip}{-1pt}
\setlength{\belowcaptionskip}{10pt}
\centering
\includegraphics[trim = 9mm 0mm 0mm 0mm, width=0.6\textwidth]{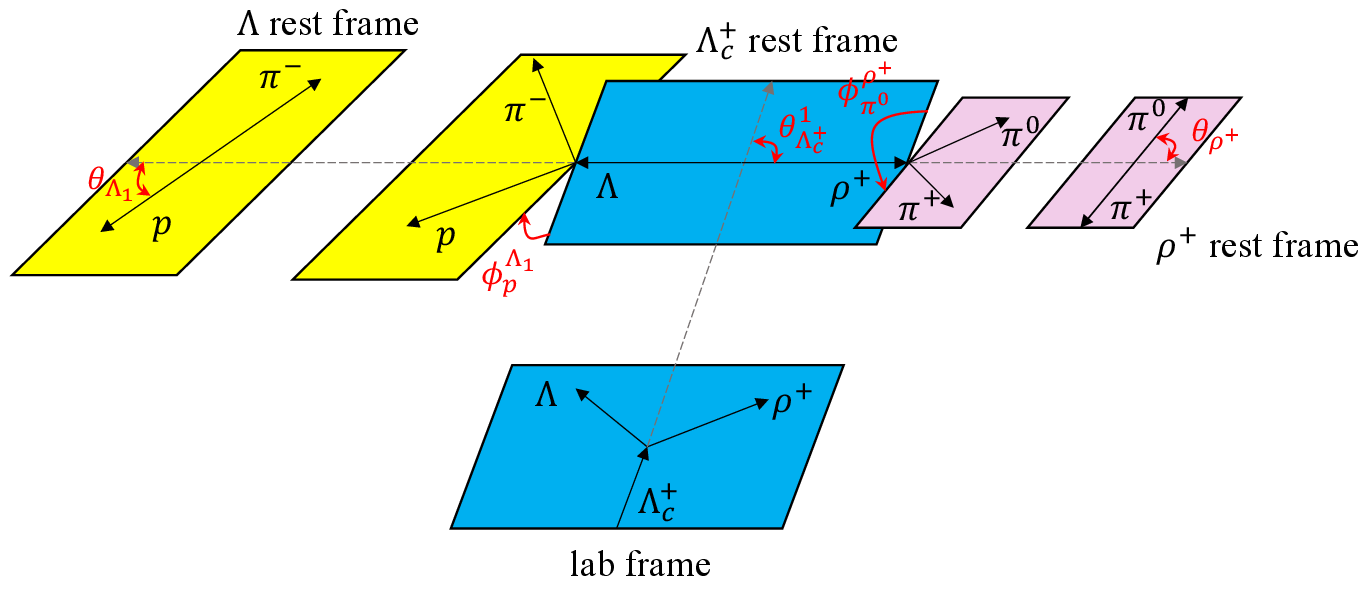}\\
\vspace{0.6cm}
\includegraphics[trim = 9mm 0mm 0mm 0mm, width=0.6\textwidth]{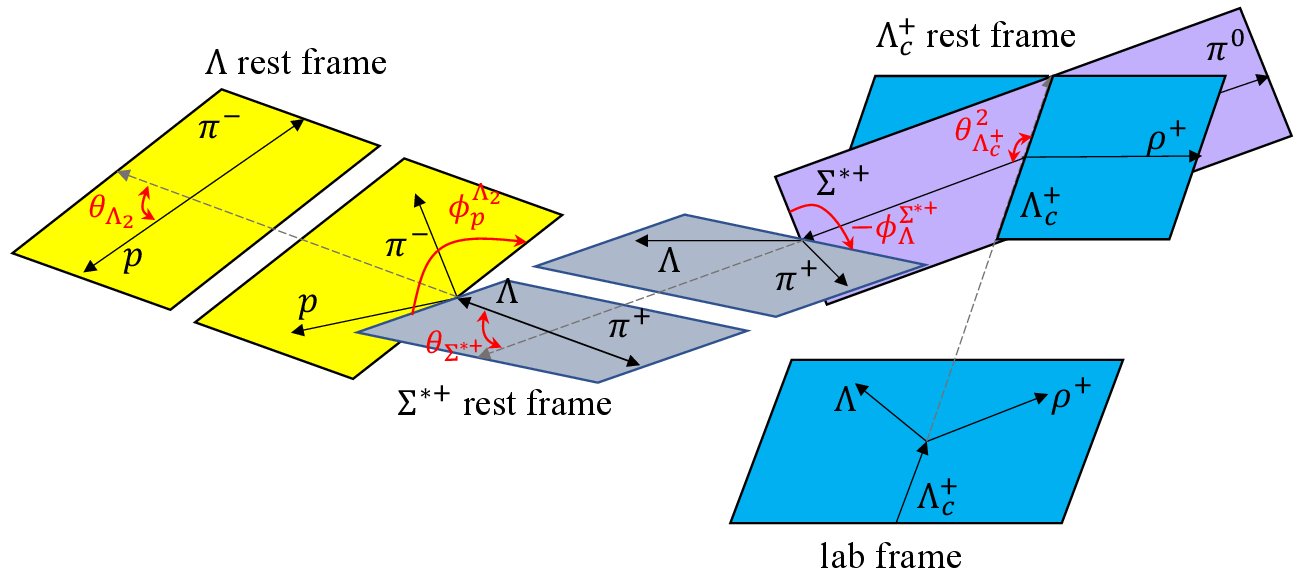}\\
\vspace{0.6cm}
\includegraphics[trim = 9mm 0mm 0mm 0mm, width=0.6\textwidth]{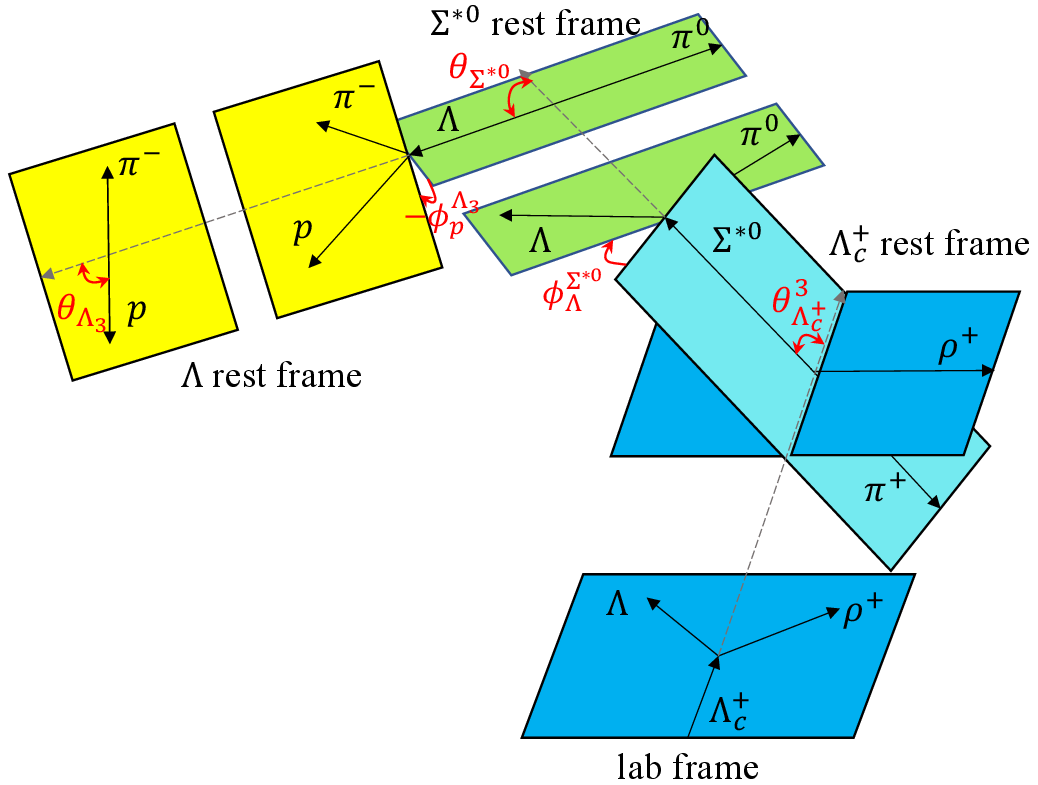}
\caption{Definitions for the helicity angles with different decay chains in the decay $\lcp\to\Lambda\pip\piz$. The notation $\theta_R$ denotes the $\theta$ helicity angle of the decay of resonance $R$. For the $\lcp$ helicity angle, the superscripts 1, 2, and 3 are used to differentiate three decay chains. The notation $\phi_B^A$ denotes the $\phi$ helicity angle between the plane of $B$ and the plane of its mother particle $A$. In convention, the right-hand frame is chosen as the normal direction for a $\phi$ rotation, and the anti-direction rotation is denoted as $-\phi$ in the figure. In addition, the subscripts 1, 2, and 3 for $\Lambda$ are also used to differentiate three decay chains. 
The exact calculations of different helicity angles are implemented according to the conventions in Ref.~\cite{helicity}.}
\label{fig:angle}
\end{figure}

\subsection{Helicity amplitude}
\label{sec:helicity}
\hspace{1.5em} To construct the full decay amplitude of the decay
$\lcp\to\Lambda\pip\piz$, the helicity formalism is used based on the
Isobar model describing the three-body decay as a two-step sequential
quasi-two-body decay. For each two-body decay $0\to1+2$, the helicity
amplitude can be written as
\begin{eqnarray}
    A^{0\rightarrow 1+2}_{\lambda_0, \lambda_1, \lambda_2} = H_{\lambda_1, \lambda_2}^{0\rightarrow 1+2} D^{J_0 *}_{\lambda_0,\lambda_1-\lambda_2}(\phi, \theta, 0),
\label{eq:helicty1}
\end{eqnarray}
where the amplitude $H_{\lambda_1, \lambda_2}^{0\rightarrow 1+2}$ is given
by the  LS coupling formula~\cite{LScoupling} along with
barrier factor terms
\begin{eqnarray}
    H_{\lambda_1, \lambda_2}^{0\rightarrow 1+2} = \sum_{l s} g_{ls} \sqrt{\frac{2l+1}{2J_0 + 1}} 
    \langle l 0,s \delta| J_0,\delta \rangle
    \langle J_1 J_2, \lambda_1\ -\lambda_2| s,\delta \rangle \left(\frac{q}{q_0}\right)^l B_l'(q,q_0,d),
\label{eq:helicty2}
\end{eqnarray}
where $g_{ls}$ is the partial wave amplitude, $J_{0,1,2}$ are the
spins of the particles 0, 1, and 2, $\lambda_{1,2}$ are the helicities
for the particles 1 and 2, and $\delta=\lambda_1-\lambda_2$ is the
helicity difference.  Here, $q$ is the three-momentum modulus of particle $1$
in the rest frame of particle $0$, which is calculated as
\begin{eqnarray}
    q = \frac{\sqrt{[m^2 - (m_1 +m_2)^2][m^2 - (m_1 - m_2)^2]}}{2 m},
\label{eq:helicty3}
\end{eqnarray}
where $m$, $m_1$ and $m_2$ are the masses of the particles 0, 1, and
2, respectively. The normalization factor $q_0$ is calculated at the
nominal resonance mass.  The factor $B_l'(q,q_0,d)$ is the reduced
Blatt-Weisskopf barrier factor~\cite{barrier_factor}, which is
explicitly expressed as
\begin{eqnarray}
\begin{aligned}
    B_0'(q,q_0,d) & =1,\\
    B_1'(q,q_0,d) & =\sqrt{\frac{1+(q_0d)^2}{1+(qd)^2}},\\
    B_2'(q,q_0,d) & =\sqrt{\frac{9+3(q_0d)^2+(q_0d)^4}{9+3(qd)^2+(qd)^4}},\\
    B_3'(q,q_0,d) & =\sqrt{\frac{225+45(q_0d)^2+6(q_0d)^4+(q_0d)^6}{225+45(qd)^2+6(qd)^4+(qd)^6}},\\
    B_4'(q,q_0,d) & =\sqrt{\frac{11025+1575(q_0d)^2+135(q_0d)^4+10(q_0d)^6+(q_0d)^8}{11025+1575(qd)^2+135(qd)^4+10(qd)^6+(qd)^8}}.
\end{aligned}
\label{eq:barrier}
\end{eqnarray}
In the Wigner $D$-function, $D^{J_0
  *}_{\lambda_0,\lambda_1-\lambda_2}(\phi, \theta, 0)$, $\phi$ and
$\theta$ are helicity angles, and are shown in
Figure~\ref{fig:angle}. The definitions can be found in
Ref.~\cite{helicity}. In Eq.~\eqref{eq:barrier}, the radius $d$ is
chosen as $d=0.73\;\mathrm{fm}$, which is the same as in
Ref.~\cite{barrierSet}.

The amplitude for a complete decay chain is constructed as the product
of each two body decay amplitude and the resonant propagator $R$. For
example, in the sequential decay $\Lambda_c^{+} \rightarrow \Lambda
\rho(770)^+$, $\rho(770)^+\to\pi^{+} \pi^{0}$, $\Lambda \rightarrow p
\pi^{-}$, the amplitude is written as
\begin{eqnarray}
    A^{\rho}_{\lambda_{\Lambda_c^{+}},\lambda_p} = \sum_{\lambda_\rho,\lambda_\Lambda} A^{\Lambda_c^{+} \rightarrow \rho \Lambda}_{\lambda_{\Lambda_c^{+}},\lambda_{\rho},\lambda_\Lambda}  R_{\rho} (M_{\pi^{+}\pi^{0}}) A^{\rho \rightarrow \pi^{+}\pi^{0}}_{\lambda_\rho,0,0} A^{\Lambda \rightarrow p \pi^{-}}_{\lambda_{\Lambda},\lambda_p,0}.
\label{eq:helicty4}
\end{eqnarray}
For the non-resonant (NR) decay, the amplitude $ A^{N\!R}_{\lambda_{\Lambda_c^{+}},\lambda_p}$ is replaced by setting $R_{N\!R} (M_{\pi^{+}\pi^{0}})$ as unity, expressed as
\begin{eqnarray}
    A^{N\!R}_{\lambda_{\Lambda_c^{+}},\lambda_p} = \sum_{\lambda_{N\!R},\lambda_\Lambda} A^{\Lambda_c^{+} \rightarrow {N\!R}+ \Lambda}_{\lambda_{\Lambda_c^{+}},\lambda_{N\!R},\lambda_\Lambda}  
    A^{{N\!R} \rightarrow \pi^{+}\pi^{0}}_{\lambda_{N\!R},0,0} A^{\Lambda \rightarrow p \pi^{-}}_{\lambda_{\Lambda},\lambda_p,0}.
\label{eq:helicty4b}
\end{eqnarray}

For the decay via the $\Sigma^*$ intermediate states, the amplitude reads
\begin{eqnarray}
    A^{\Sigma^*}_{\lambda_{\Lambda_c^{+}},\lambda_p} = \sum_{\lambda_{\Sigma^*},\lambda_{\Lambda}} A^{\Lambda_c^{+} \rightarrow \Sigma^* \pi}_{\lambda_{\Lambda_c^{+}},\lambda_\Sigma,0} R_{\Sigma^*} (M_{\Lambda \pi}) A^{\Sigma^* \rightarrow \Lambda \pi}_{\lambda_{\Sigma^*}, \lambda_\Lambda,0}A^{\Lambda \rightarrow p \pi^{-}}_{\lambda_\Lambda,\lambda_p,0}.
\label{eq:helicty5}
\end{eqnarray}

The propagator $R$ includes different models.
For $\Sigma^*$ resonances, the relativistic Breit-Wigner formula is taken as
\begin{eqnarray}
    R_{\Sigma^*}(m) = \frac{1}{m_0^2-m^2 -i m_0 \Gamma(m)},
\label{eq:helicty6}
\end{eqnarray}
where the mass dependent width is
\begin{eqnarray}
    \Gamma(m) = \Gamma_0 \left(\frac{q}{q_0}\right)^{2l+1}\frac{m_0}{m} B_l'^2(q,q_0,d).
\label{eq:helicty7}
\end{eqnarray}
For the $\rho(770)^{+}$ resonance, the Gounaris-Sakurai (GS) model~\cite{GSmodel} is used:
\begin{eqnarray}
    R_{\rho}(m) = \frac{1 + D \Gamma_0 / m_0}{(m_0^2 -m^2) + f(m) - i m_0 \Gamma(m)},
\label{eq:helicty8}
\end{eqnarray}
where the mass dependent width $\Gamma(m)$ is defined with the same parametrization as Eq.~\eqref{eq:helicty7}, and $f(m)$ and $D$ are defined as
\begin{eqnarray}
    f(m) = \Gamma_0 \frac{m_0 ^2 }{q_0^3} \left[q^2 [h(m)-h(m_0)] + (m_0^2 - m^2) q_0^2 \frac{\mathrm{d} h}{\mathrm{d} m}|_{m_0} \right],
\label{eq:helicty9}
\end{eqnarray}

\begin{eqnarray}
     h(m) = \frac{2}{\pi} \frac{q}{m} \ln \left(\frac{m+2q}{2m_{\pi}} \right),
\label{eq:helicty10}
\end{eqnarray}

\begin{eqnarray}
    \frac{\mathrm{d}h}{\mathrm{d}m}|_{m_0} = h(m_0) [(8q_0^2)^{-1} - (2m_0^2)^{-1}] + (2\pi m_0^2)^{-1},
\label{eq:helicty11}
\end{eqnarray}

\begin{eqnarray}
    D = \frac{f(0)}{\Gamma_0 m_0} = \frac{3}{\pi}\frac{m_\pi^2}{q_0^2} \ln \left(\frac{m_0 + 2q_0}{2 m_\pi }\right)
        + \frac{m_0}{2\pi q_0} - \frac{m_\pi^2 m_0}{\pi q_0^3}.
\label{eq:helicty12}
\end{eqnarray}
The full amplitude is the coherent sum of all possible resonances and NR amplitudes, given as
\begin{eqnarray}
\begin{aligned}
    \mathcal{A}_{\lambda_{\Lambda_c^{+}},\lambda_p} &=  \left (A^{\rho}_{\lambda_{\Lambda_c^{+}},\lambda_p}
    + A^{N\!R}_{\lambda_{\Lambda_c^{+}},\lambda_p} \right) \\ 
    & + \sum_{\lambda_p'} \left( \sum A^{\Sigma^{*+}}_{\lambda_{\Lambda_c^{+}},\lambda_p'}\right) D^{1/2}_{\lambda_p', \lambda_p}\left(\alpha_p, \beta_p, \gamma_p\right) \\
    & + \sum_{\lambda_p'} \left( \sum A^{\Sigma^{*0}}_{\lambda_{\Lambda_c^{+}},\lambda_p'}\right) D^{1/2}_{\lambda_p',\lambda_p}\left(\alpha_p', \beta_p', \gamma_p'\right),
\end{aligned}
\label{eq:helicty13}
\end{eqnarray}
where the extra alignment $D$-functions are added to align the helicities of the final state protons. The details of the alignment angle calculations can be found in Appendix~\ref{app:align_angle} and in Ref.~\cite{helicity}. 

For simplicity in the PWA fit, an overall resonance amplitude for each component term in Eq.~\eqref{eq:helicty13}, as those listed in Table~\ref{tab:nominal1}, can be derived to represent the overall magnitude and phase of the specific component, while in each cascade process, one of partial wave amplitudes $g_{ls}$ can be taken as reference amplitude with fixed amplitude of constant 1 (as the fixed amplitudes listed in Table~\ref{tab:nominal2}). Hence, the relative magnitudes and phases of the other partial wave amplitudes in each cascade process (as the free amplitudes listed in Table~\ref{tab:nominal2}) are left free in the fit.

\subsection{Likelihood function construction and fit fraction}
\label{sec:likelihood}
\hspace{1.5em}
The probability density function for a given event is constructed using the full amplitude as
\begin{eqnarray}
    P = \frac{|\mathcal{A}|^2}{\int |\mathcal{A}|^2 \mathrm{d} \Phi},\quad |\mathcal{A}|^2 = \frac{1}{2} \sum_{\lambda_{\Lambda_c^{+}}, \lambda_p} |\mathcal{A}_{\lambda_{\Lambda_c^{+}}, \lambda_p}|^2,
\label{eq:helicty14}
\end{eqnarray}
where the factor $1/2$ arises from the average of the initial $\lcp$
spin under the assumption of non polarization. Possible polarization
effects are considered with the systematic uncertainties. The
integration is calculated with a MC method via sufficiently large
PHSP samples passing the simulated detector reconstruction stage, 
and we have
\begin{eqnarray}
    \int |\mathcal{A}|^2 \mathrm{d}\Phi \propto \frac{1}{N_{\mathrm{PHSP}}} \sum_{i \in \mathrm{PHSP}} |\mathcal{A}(x_i)|^2.
\label{eq:helictyEx}
\end{eqnarray}

The negative log likelihood (NLL) is constructed by summing all signal
candidates and subtracting the $\mbc$ sideband backgrounds
\begin{eqnarray}
    -\ln L =  - \alpha \left[\sum_{i \in \text{data}} \ln P(x_i) - w'_{\text{bkg}} \sum_{i \in \text{sideband}} \ln P(x_i)\right],
\label{eq:helicty15}
\end{eqnarray}
where
$w'_{\text{bkg}}=w_{\text{bkg}}\cdot\frac{N_{\text{data}}}{N_{\text{sideband}}}$,
$w_{\text{bkg}}$ is the background fraction listed in
Table~\ref{tab:fitmbc}, $N_{\text{data}}$ and $N_{\text{sideband}}$
are the events in the $\mbc$ signal region and sideband region,
respectively. To achieve an unbiased uncertainty estimation, the
normalization factor derived from Ref.~\cite{sWeight} is taken into
account, expressed as
\begin{eqnarray}
    \alpha = \frac{N_{\text{data}} - N_{\text{sideband}}w'_{\text{bkg}}}{N_{\text{data}} + N_{\text{sideband}}{w'_{\text{bkg}}}^{2} }.
\label{eq:helicty16}
\end{eqnarray}
An individual NLL is first constructed separately for a given energy
point, and the joint NLL is obtained by summing over the NLL values of the
different energy points.  After minimizing the joint NLL, the
parameter error matrix is calculated by the inverse of the Hessian
matrix
\begin{eqnarray}
    V_{ij}^{-1} = - \frac{\partial^2 \ln L}{\partial X_i \partial X_j},
\label{eq:helicty18}
\end{eqnarray}
where $X_i$ is the $i$-th floating parameter in the fit. 

The fit fraction (FF) for each resonant component can be calculated as
\begin{eqnarray}
    \mathrm{FF}_i = \frac{\int |\mathcal{A}_i|^2 \mathrm{d} \Phi'}
    {\int |\sum_k \mathcal{A}_k|^2 \mathrm{d} \Phi'},
\label{eq:FF}
\end{eqnarray}
where $\mathcal{A}_i$ is the amplitude of the $i$-th component and the
integration is calculated by the sum of truth level PHSP MC samples
before requiring detector acceptance.  Hence, the FFs of the
interference part can be calculated as
\begin{eqnarray}
    \mathrm{FF}_{i,j} = \frac{\int |\mathcal{A}_i+\mathcal{A}_j|^2 \mathrm{d} \Phi'}
    {\int |\sum_k \mathcal{A}_k|^2 \mathrm{d} \Phi'}-\mathrm{FF}_i-\mathrm{FF}_j.
\label{eq:interference}
\end{eqnarray}
The statistical uncertainties for FFs are obtained with the standard form of error propagation. Let $Y$ be the variable whose error needs to be calculated, $\bm{X}$  the variables with a corresponding error matrix $V_{ij}$ in Eq.~\eqref{eq:helicty18} and $\bm{\mu}$  the nominal results of the floating variables $\bm{X}$, the squared uncertainty of $Y$ is estimated as
\begin{eqnarray}
    \sigma_Y^2 = \sum_{ij} 
    \left( \frac{\partial Y}{\partial X_i}\right)_{\bm{X}=\bm{\mu}}
    \cdot V_{ij} \cdot 
    \left( \frac{\partial Y}{\partial X_j}\right)_{\bm{X}=\bm{\mu}}.
\label{eq:errPropagation}
\end{eqnarray}

\subsection{Nominal fit results}
\label{sec:nominal}
\hspace{1.5em}

The starting point in the construction of the nominal fit hypothesis
of the PWA is the inclusion of the main resonances involved in the
decay, i.e. $\Sigma(1385)^+$, $\Sigma(1385)^0$ and $\rho(770)^+$.  The
$\rho(770)^+$ component is chosen as the reference channel due to its
dominant contribution, and the magnitude and phase of its total amplitude are fixed to one and
zero, respectively.

In addition, the statistical significance of the contribution from
excited $\Sigma$ states is evaluated, including $\Sigma(1660)$,
$\Sigma(1670)$, $\Sigma(1750)$, $\Sigma(1775)$, $\Sigma(1910)$,
$\Sigma(1915)$, and $\Sigma(2030)$ (considered as established by the
PDG~\cite{pdg} with a score of at least four stars), as well as from
the $\mathcal{S}$-wave ($N\!R_{0^+}$), $\mathcal{P}$-wave
($N\!R_{1^-}$), and $\mathcal{D}$-wave ($N\!R_{2^+}$) non-resonant
components in the $M_{\pippiz}$ spectrum, based on the change of the
NLL value when including singly each of these components.  The
statistical significance is calculated from the change of the NLL
values with and without including the component, by taking into
account the change of the number of degrees of freedom. The results
show that the resonances $\Sigma(1670)$ and $\Sigma(1750)$, along with
the non-resonant components $N\!R_{1^-}$, have statistical
significance larger than $5\sigma$, while none of the other tested
contributions exceeds this threshold.

Therefore, the nominal components are determined to be the $\rho(770)^+$, $\Sigma(1385)^+$, $\Sigma(1385)^0$, $\Sigma(1670)^+$, $\Sigma(1670)^0$, $\Sigma(1750)^+$, and $\Sigma(1750)^0$ states, as well as the $N\!R_{1^-}$ component, whose statistical significances are listed in Table~\ref{tab:nominal1}.
The mass and width parameters of the $\rho(770)^+$, $\Sigma(1385)^+$, and $\Sigma(1385)^0$ states are fixed to the corresponding world average values~\cite{pdg}, while the resonance parameters of the other $\Sigma^*$ resonances are taken from the most recent measurements~\cite{recent}. Moreover, the $\Lambda$ decay asymmetry parameter $\alpha_{\Lambda}=0.732\pm0.014$~\cite{pdg,BESIII:2018cnd,Ireland:2019uja,bes3LmdRecent} is used, as discussed in Sec.~\ref{sec:asymmetry}.

The constructed NLL of the nominal fit contains 38 floating parameters
in total, consisting of magnitudes and phases of the total amplitudes
(14) and partial wave amplitudes (24), whose fit results are listed in
Table~\ref{tab:nominal1} and Table~\ref{tab:nominal2}. The fractions
of the interference parts can be found in
Table~\ref{tab:interference}. The Dalitz plot distributions of data and fit results 
are shown in Figure~\ref{fig:dalitz}, and other one-dimensional projections are shown
in Figures~\ref{fig:nominal1} and \ref{fig:nominal2}. Using the FFs
listed in Table~\ref{tab:nominal1} multiplied by the total three-body
BF $\BR(\lcp\to\Lambda\pip\piz)=(7.1\pm 0.4)\%$~\cite{pdg,LcTagMode}
and considering the intermediate BF $\BR(\Sigma(1385)\to\Lambda\pi)$,
the absolute BFs of the involved resonances are obtained as listed in
Table~\ref{tab:BF_alpha}.  For the resonances $\rho(770)^+$ and
$\Sigma(1385)$, further studies on the decay asymmetries and the
relevant systematic uncertainties are performed, in order to better
confront them with the theoretical calculations.

\begin{table*}[!tp]
\caption{Numerical results of the total amplitudes for different
  components in the nominal fit, along with the FFs and the
  corresponding significance. The total FF is 141.8\%. Only
  statistical uncertainties are listed.}
\setlength{\abovecaptionskip}{1.2cm}
\setlength{\belowcaptionskip}{0.2cm}
\label{tab:nominal1}
\begin{center}
\vspace{-0.0cm}
\begin{tabular}{c  c c c c}
	\hline \hline
	Process & Magnitude & Phase $\phi$ (rad) & FF ($\%$) & Significance \\
	\hline
	$\Lambda\rho(770)^+$       & $ 1.0 $ (fixed)   & $ 0.0 $   (fixed)  & $ 57.2 \pm 4.2 $  & $36.9\,\sigma$\\
	$\Sigma(1385)^+\piz$    & $ 0.43 \pm 0.06 $ & $ -0.23 \pm 0.18 $ & $ 7.18 \pm 0.60 $   & $14.8\,\sigma$\\
	$\Sigma(1385)^0\pip$    & $ 0.37 \pm 0.07 $ & $ 2.84 \pm 0.23 $  & $ 7.92 \pm 0.72 $   & $16.0\,\sigma$\\
	$\Sigma(1670)^+\piz$    & $ 0.31 \pm 0.08 $ & $ -0.77 \pm 0.23 $ & $ 2.90 \pm 0.63 $   & $5.1\,\sigma$\\
	$\Sigma(1670)^0\pip$    & $ 0.41 \pm 0.07 $ & $ 2.77 \pm 0.20 $  & $ 2.65 \pm 0.58 $   & $5.2\,\sigma$\\
	$\Sigma(1750)^+\piz$    & $ 1.75 \pm 0.21 $ & $ -1.73 \pm 0.11 $ & $ 16.6 \pm 2.2 $  & $10.1\,\sigma$\\
	$\Sigma(1750)^0\pip$    & $ 1.83 \pm 0.21 $ & $ 1.34 \pm 0.11 $  & $ 17.5 \pm 2.3 $  & $10.2\,\sigma$\\
	$\Lambda+N\!R_{1^-}$          & $ 4.05 \pm 0.47 $ & $ 2.16 \pm 0.13 $  & $ 29.7 \pm 4.5 $  & $10.5\,\sigma$\\
	\hline\hline
\end{tabular}
\end{center}
\end{table*}
\vspace{-0.0cm}

\vspace{-0.0cm}
\begin{figure}[htbp]
\setlength{\abovecaptionskip}{-1pt}
\setlength{\belowcaptionskip}{10pt}
\centering
\includegraphics[trim = 9mm 0mm 0mm 0mm, width=0.45\textwidth]{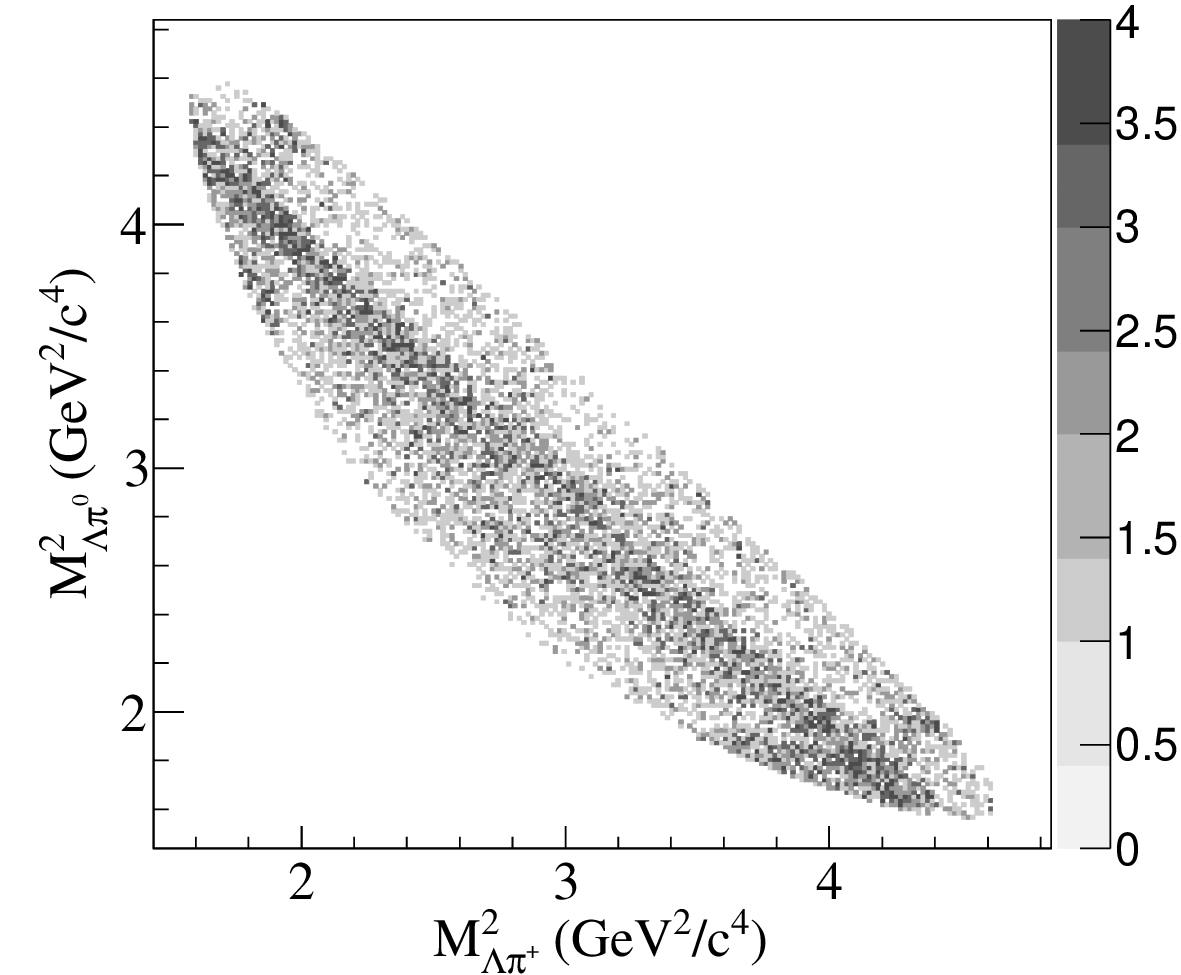}\;\;
\includegraphics[trim = 9mm 0mm 0mm 0mm, width=0.45\textwidth]{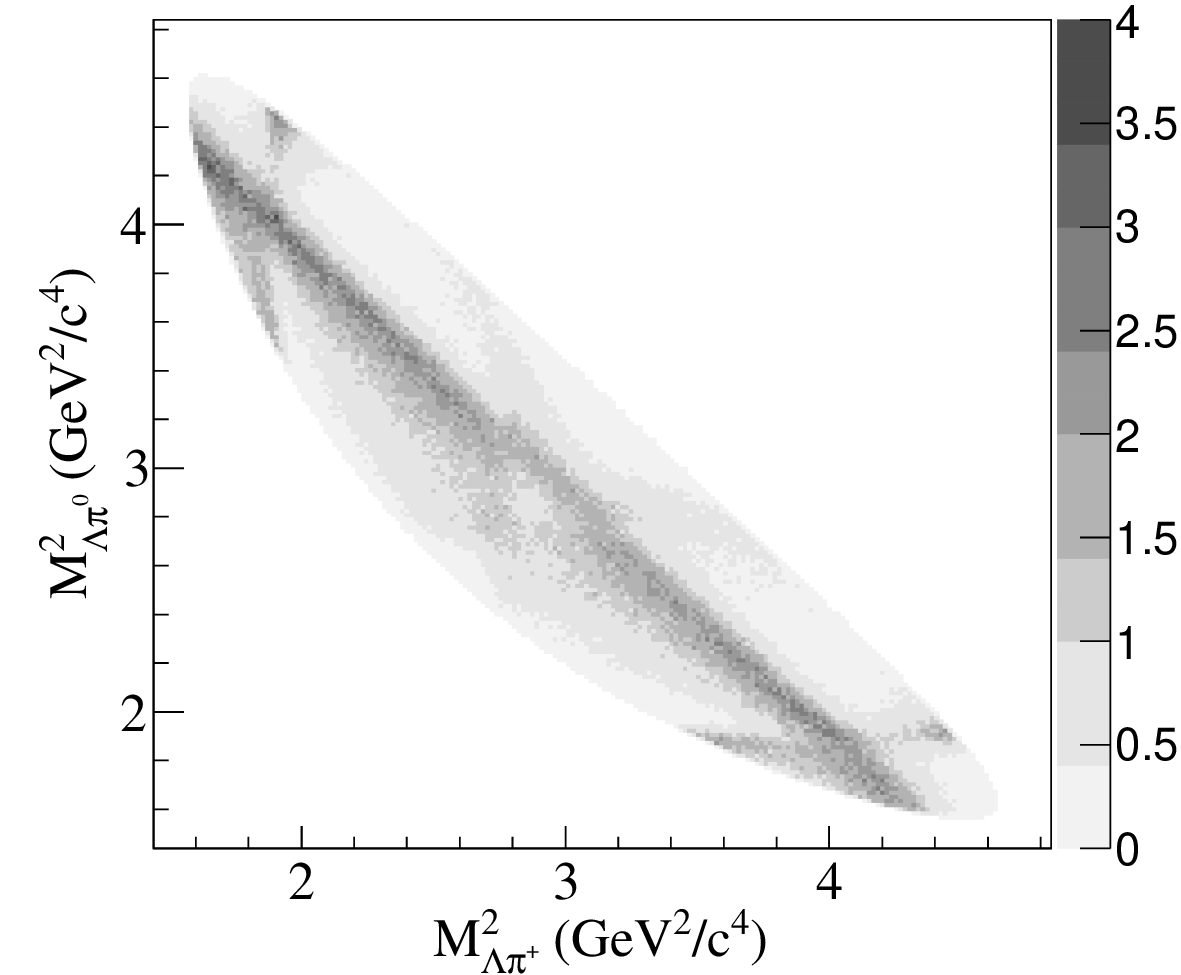}
\caption{The Dalitz plot distributions of data (left) and fit results (right).}
\label{fig:dalitz}
\end{figure}

\vspace{-0.0cm}
\begin{figure}[htbp]
\setlength{\abovecaptionskip}{-1pt}
\setlength{\belowcaptionskip}{10pt}
\centering
\includegraphics[trim = 9mm 0mm 0mm 0mm, width=0.45\textwidth]{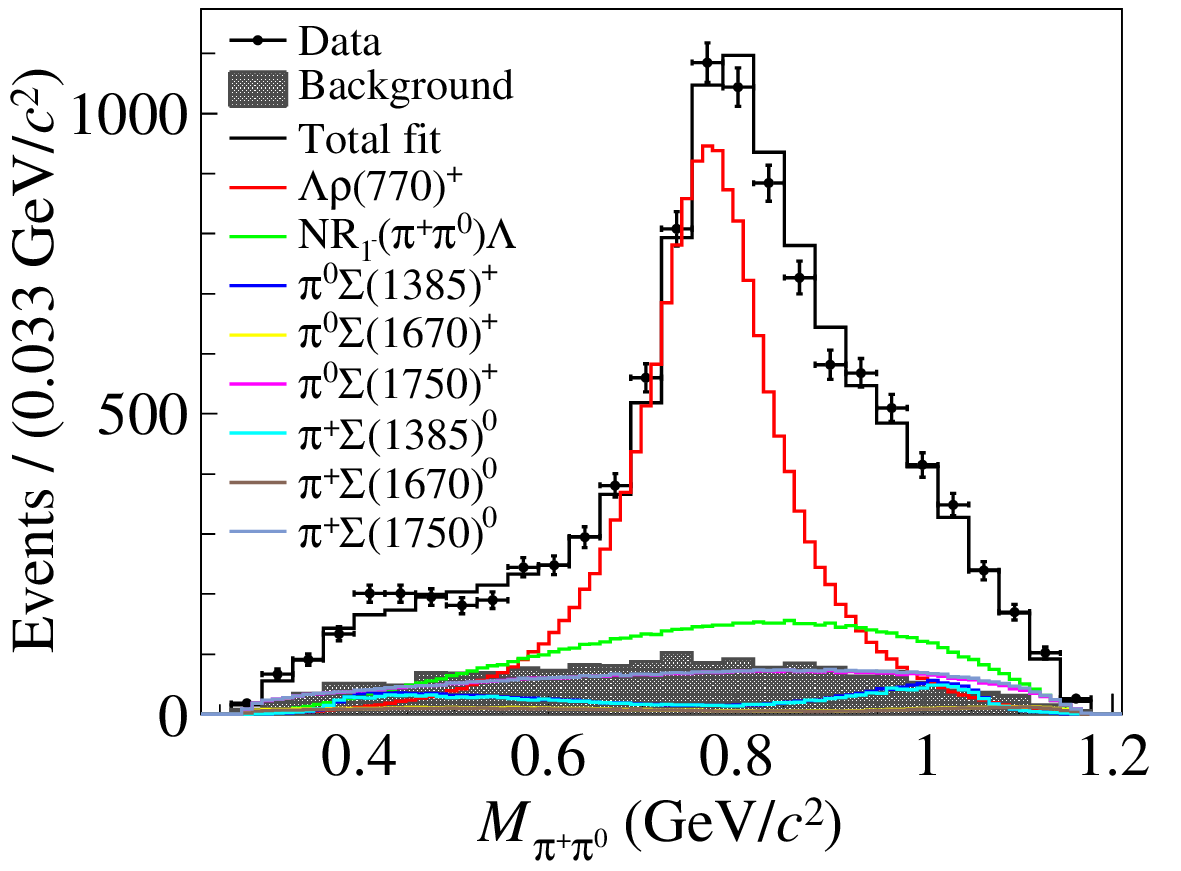}
\includegraphics[trim = 9mm 0mm 0mm 0mm, width=0.45\textwidth]{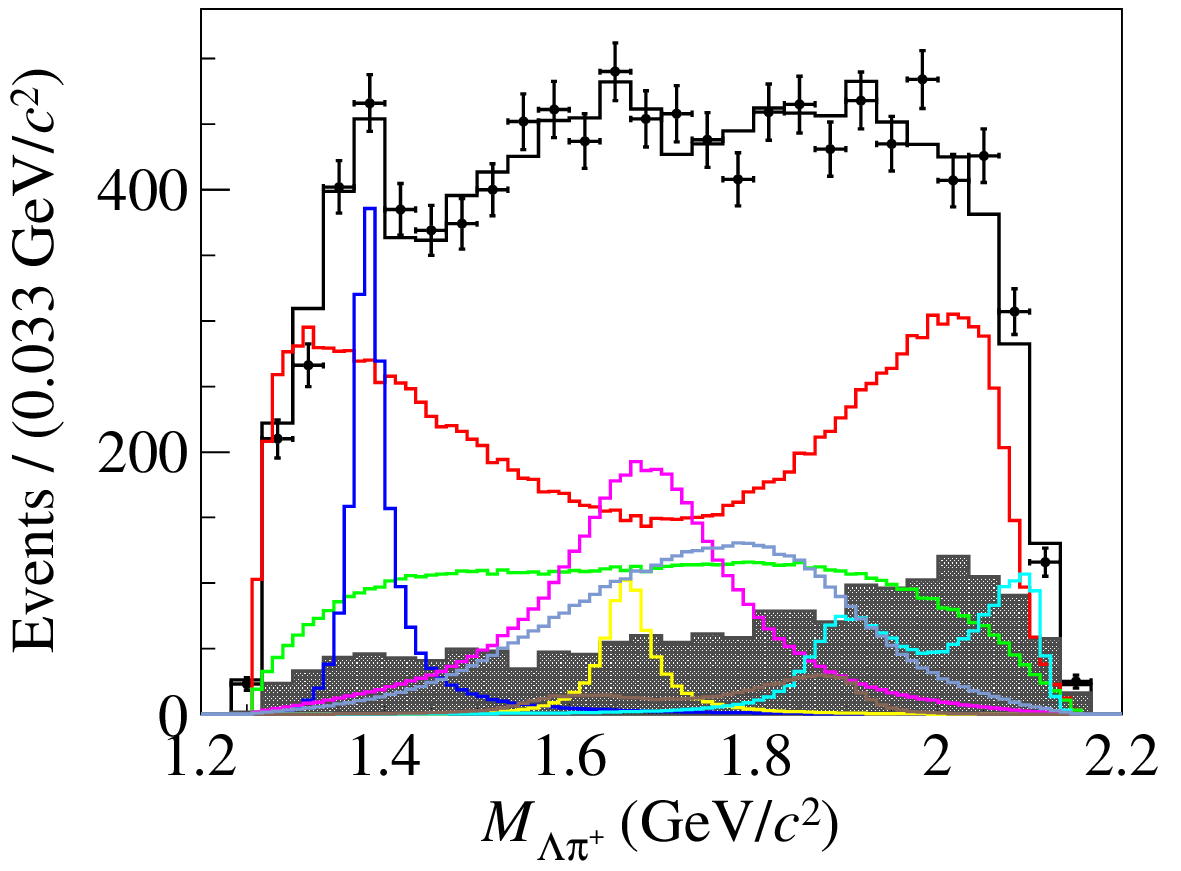}\\
\includegraphics[trim = 9mm 0mm 0mm 0mm, width=0.45\textwidth]{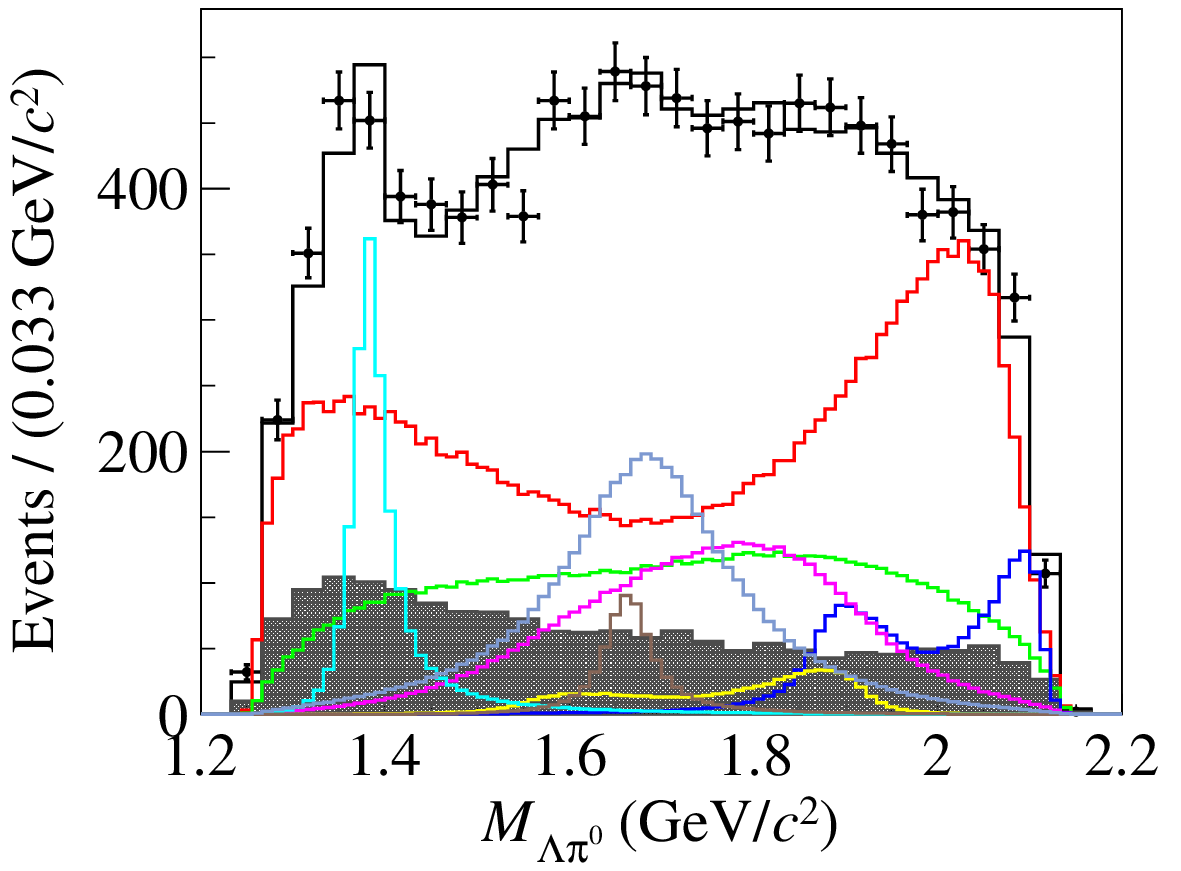}
\caption{Projections of the fit results in the invariant mass spectra
  $M_{\pip\piz}$, $M_{\Lambda\pip}$ and $M_{\Lambda\piz}$. Points with
  error bars denote data. Different styles of the curves denote
  different components.}
\label{fig:nominal1}
\end{figure}

\begin{table*}[!tp]
\caption{Numerical results of the partial wave amplitudes $g_{ls}$ for
  different resonances in the nominal fit. Only statistical
  uncertainties are listed.}  \setlength{\abovecaptionskip}{1.2cm}
\setlength{\belowcaptionskip}{0.2cm}
\label{tab:nominal2}
\begin{center}
\vspace{-0.0cm}
\begin{tabular}{c c c | c c c}
	\hline \hline
	\multicolumn{3}{c|}{$\frac{1}{2}^{+}(\lcp)\to\frac{3}{2}^{+}(\Sigma(1385)^+)+0^-(\piz)$} 
	& \multicolumn{3}{c}{$\frac{1}{2}^{+}(\lcp)\to\frac{3}{2}^{+}(\Sigma(1385)^0)+0^-(\pip)$}\\
	Amplitude & Magnitude & Phase $\phi$ (rad)
	& Amplitude & Magnitude & Phase $\phi$ (rad)\\
	\hline
	$g^{\Sigma(1385)^+}_{1,\frac{3}{2}}$   & $1.0$ (fixed) & $0.0$ (fixed)
	& $g^{\Sigma(1385)^0}_{1,\frac{3}{2}}$ & $1.0$ (fixed) & $0.0$ (fixed) \\
	$g^{\Sigma(1385)^+}_{2,\frac{3}{2}}$   & $ 1.29 \pm 0.25 $ & $ 2.82 \pm 0.18 $
	& $g^{\Sigma(1385)^0}_{2,\frac{3}{2}}$ & $ 1.70 \pm 0.38 $ & $ 2.70 \pm 0.22 $\\
	\hline\hline
	
	\multicolumn{3}{c|}{$\frac{1}{2}^{+}(\lcp)\to\frac{3}{2}^{-}(\Sigma(1670)^+)+0^-(\piz)$} 
	& \multicolumn{3}{c}{$\frac{1}{2}^{+}(\lcp)\to\frac{3}{2}^{-}(\Sigma(1670)^0)+0^-(\pip)$}\\
	Amplitude & Magnitude & Phase $\phi$ (rad)
	& Amplitude & Magnitude & Phase $\phi$ (rad)\\
	\hline
	$g^{\Sigma(1670)^+}_{1,\frac{3}{2}}$   & $1.0$ (fixed)  & $0.0$ (fixed)
	& $g^{\Sigma(1670)^0}_{1,\frac{3}{2}}$ & $1.0$ (fixed)  & $0.0$ (fixed) \\
	$g^{\Sigma(1670)^+}_{2,\frac{3}{2}}$   & $ 1.39 \pm 0.42 $ & $ 0.85 \pm 0.26 $
	& $g^{\Sigma(1670)^0}_{2,\frac{3}{2}}$ & $ 0.74 \pm 0.18 $ & $ 0.29 \pm 0.24 $\\
	\hline\hline

	\multicolumn{3}{c|}{$\frac{1}{2}^{+}(\lcp)\to\frac{1}{2}^{-}(\Sigma(1750)^+)+0^-(\piz)$} 
	& \multicolumn{3}{c}{$\frac{1}{2}^{+}(\lcp)\to\frac{1}{2}^{-}(\Sigma(1750)^0)+0^-(\pip)$}\\
	Amplitude & Magnitude & Phase $\phi$ (rad)
	& Amplitude & Magnitude & Phase $\phi$ (rad)\\
	\hline
	$g^{\Sigma(1750)^+}_{0,\frac{1}{2}}$   & $1.0$ (fixed)  & $0.0$ (fixed)
	& $g^{\Sigma(1750)^0}_{0,\frac{1}{2}}$ & $1.0$ (fixed)  & $0.0$ (fixed) \\
	$g^{\Sigma(1750)^+}_{1,\frac{1}{2}}$   & $ 0.45 \pm 0.10 $ & $ -2.28 \pm 0.22 $
	& $g^{\Sigma(1750)^0}_{1,\frac{1}{2}}$ & $ 0.38 \pm 0.10 $ & $ -2.03 \pm 0.20 $\\
	\hline\hline
	
	\multicolumn{3}{c|}{$\frac{1}{2}^{+}(\lcp)\to\frac{1}{2}^{+}(\Lambda)+1^-(\rho(770)^+)$} 
	& \multicolumn{3}{c}{$\frac{1}{2}^{+}(\lcp)\to\frac{1}{2}^{+}(\Lambda)+1^-(N\!R_{1^-})$}\\
	Amplitude & Magnitude & Phase $\phi$ (rad) 
	& Amplitude & Magnitude & Phase $\phi$ (rad)\\
	\hline
	$g^{\rho}_{0,\frac{1}{2}}$ & $1.0$ (fixed) & $0.0$ (fixed)
	& $g^{N\!R}_{0,\frac{1}{2}}$ & $1.0$ (fixed) & $0.0$ (fixed) \\
	$g^{\rho}_{1,\frac{1}{2}}$ & $ 0.48 \pm 0.12 $ & $ -1.69 \pm 0.12 $
	& $g^{N\!R}_{1,\frac{1}{2}}$ & $ 0.94 \pm 0.12 $ & $ -0.49 \pm 0.16 $\\
	$g^{\rho}_{1,\frac{3}{2}}$ & $ 0.90 \pm 0.10 $ & $ 0.48 \pm 0.13 $
	& $g^{N\!R}_{1,\frac{3}{2}}$ & $ 0.21 \pm 0.09 $ & $ -2.84 \pm 0.53 $ \\
	$g^{\rho}_{2,\frac{3}{2}}$ & $ 0.55 \pm 0.08 $ & $ -0.04 \pm 0.18 $
	& $g^{N\!R}_{2,\frac{3}{2}}$ & $ 0.33 \pm 0.14 $ & $ -1.92 \pm 0.30 $\\
	\hline\hline
	
	\multicolumn{3}{c|}{$\frac{1}{2}^{+}(\Lambda)\to\frac{1}{2}^{+}(p)+0^-(\pim)$}\\
	Amplitude & Magnitude & Phase $\phi$ (rad)\\
	\cline{1-3}
	$g^{\Lambda}_{0,\frac{1}{2}}$ & $1.0$ (fixed) & $0.0$ (fixed)\\
	$g^{\Lambda}_{1,\frac{1}{2}}$ & $0.435376$ (fixed) & $0.0$ (fixed)\\
	\hline\hline
\end{tabular}
\end{center}
\end{table*}
\vspace{-0.0cm}

\begin{table*}[!htbp]
\caption{Interference fractions (I.F.) between $\lcp$ amplitudes in
  units of percentage. The uncertainties are statistical only.
}
\setlength{\abovecaptionskip}{1.2cm}
\setlength{\belowcaptionskip}{0.2cm}
\label{tab:interference}
\begin{center}
\vspace{-0.0cm}
\resizebox{\textwidth}{16mm}{
\begin{tabular}{c | c c c c c c c}
	\hline \hline
	I.F. & $\Lambda +N\!R_{1^-}$ & $\Sigma(1385)^0\pip$ & $\Sigma(1385)^+\piz$ & $\Sigma(1670)^0\pip$ & $\Sigma(1670)^+\piz$ & $\Sigma(1750)^0\pip$ & $\Sigma(1750)^+\piz$\\
	\hline
    $\Sigma(1385)^0\pip$ & $ -0.50 \pm 0.38 $ \\
    $\Sigma(1385)^+\piz$ & $ -0.76 \pm 0.36 $ & $ -0.05 \pm 0.04 $ \\
    $\Sigma(1670)^0\pip$ & $ -0.36 \pm 0.17 $ & $ -0.00 \pm 0.00 $ & $ -0.66 \pm 0.09 $  \\
    $\Sigma(1670)^+\piz$ & $ -0.34 \pm 0.15 $ & $ -0.58 \pm 0.12 $ & $ 0.00 \pm 0.00 $ & $ 0.04 \pm 0.02 $ \\
    $\Sigma(1750)^0\pip$ & $ -8.1 \pm 3.1 $ & $ -0.03 \pm 0.00 $ & $ 0.43 \pm 0.07 $ & $ -0.01 \pm 0.00 $ & $ 0.08 \pm 0.05 $ \\
   $\Sigma(1750)^+\piz$ & $ -7.2 \pm 3.1 $ & $ 0.35 \pm 0.08 $ & $ -0.02 \pm 0.00 $ & $ 0.23 \pm 0.05 $ & $ -0.00 \pm 0.00 $ & $ -6.23 \pm 0.92 $ \\
    $\Lambda\rho(770)^+$ & $ -2.7 \pm 4.4 $ & $ -5.94 \pm 0.56 $ & $ -6.01 \pm 0.46 $ & $ 0.72 \pm 0.29 $ & $ 1.29 \pm 0.26 $ & $ -2.1 \pm 1.3 $ & $ -3.1 \pm 1.3 $ \\
	\hline\hline
\end{tabular}
}
\end{center}
\end{table*}
\vspace{-0.0cm}

\vspace{-0.0cm}
\begin{figure}[htbp]
\setlength{\abovecaptionskip}{-1pt}
\setlength{\belowcaptionskip}{10pt}
\centering
\includegraphics[trim = 9mm 0mm 0mm 0mm, width=0.62\textwidth]{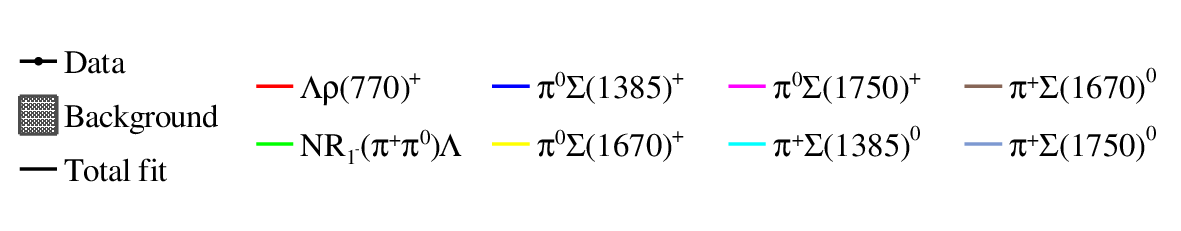}\\
\includegraphics[trim = 9mm 0mm 0mm 0mm, width=0.32\textwidth]{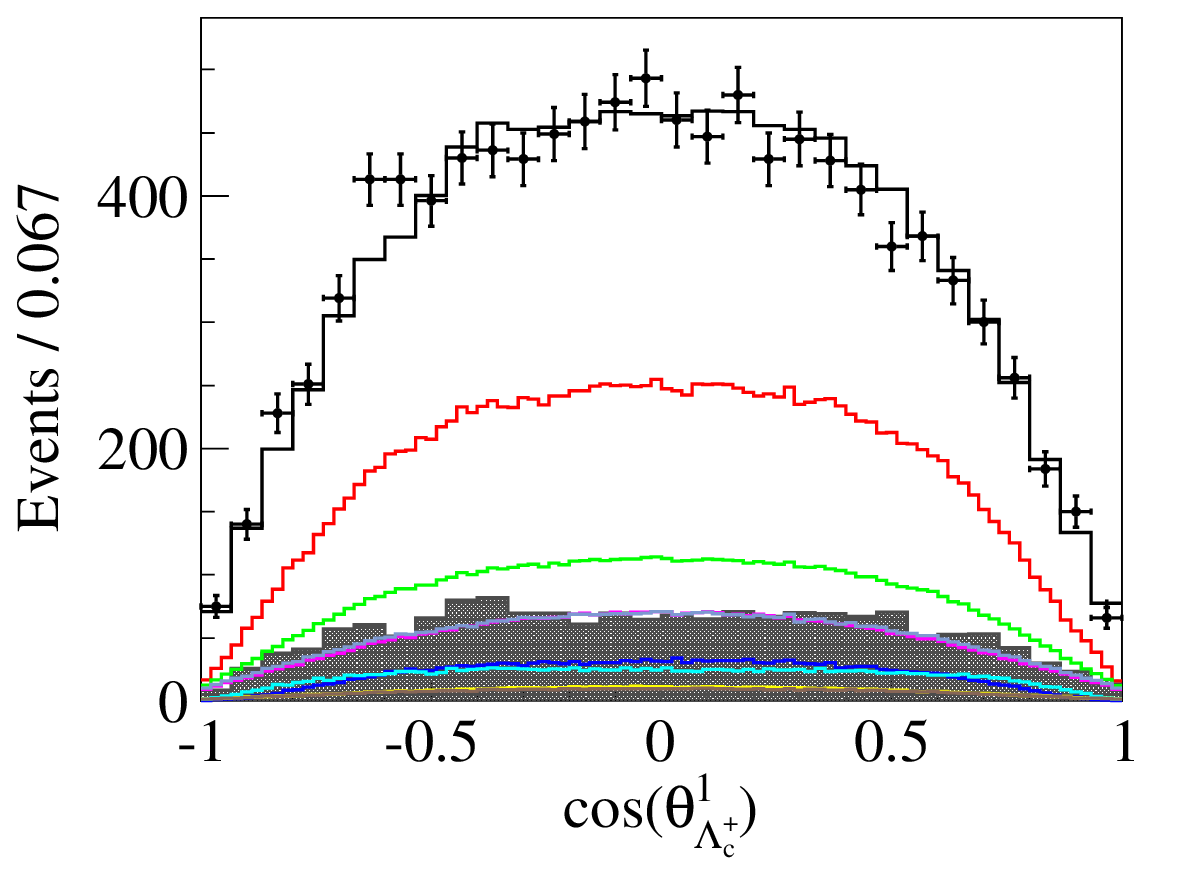}
\includegraphics[trim = 9mm 0mm 0mm 0mm, width=0.32\textwidth]{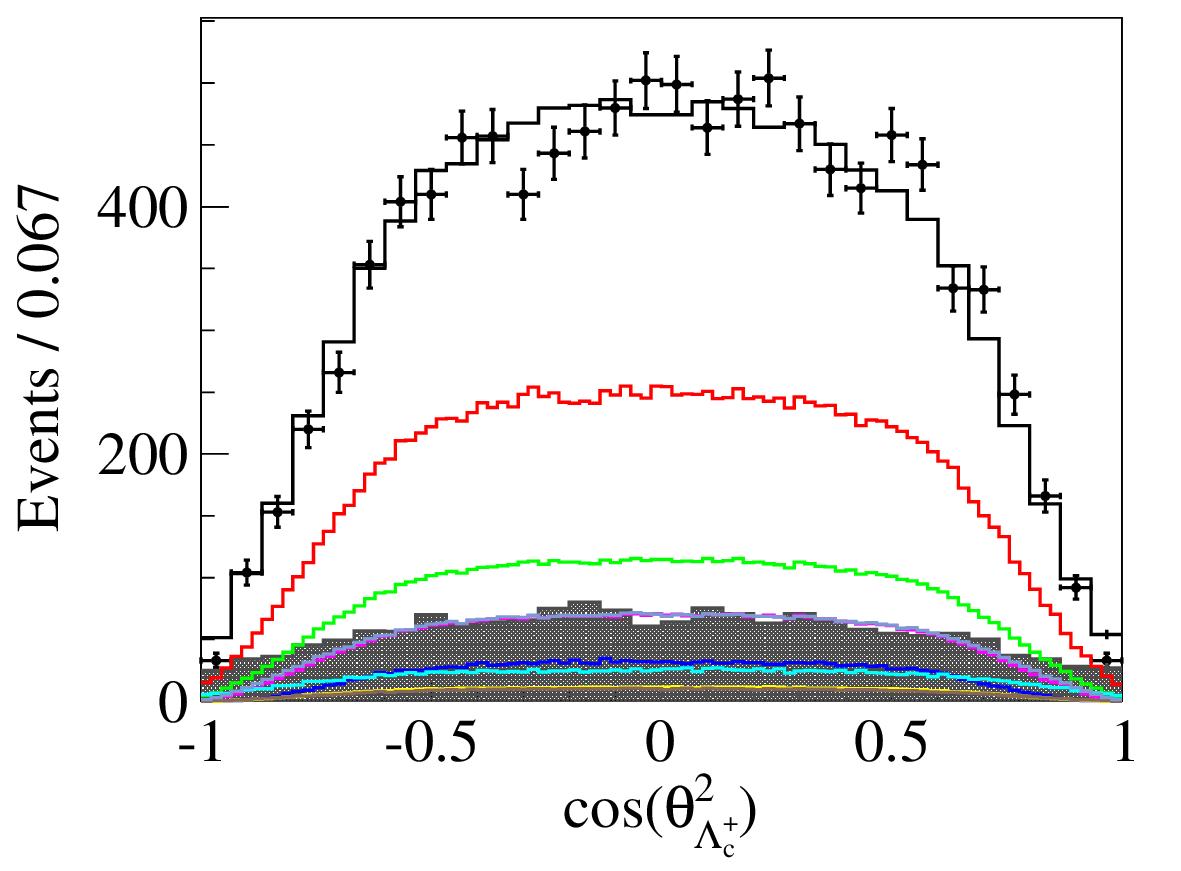}
\includegraphics[trim = 9mm 0mm 0mm 0mm, width=0.32\textwidth]{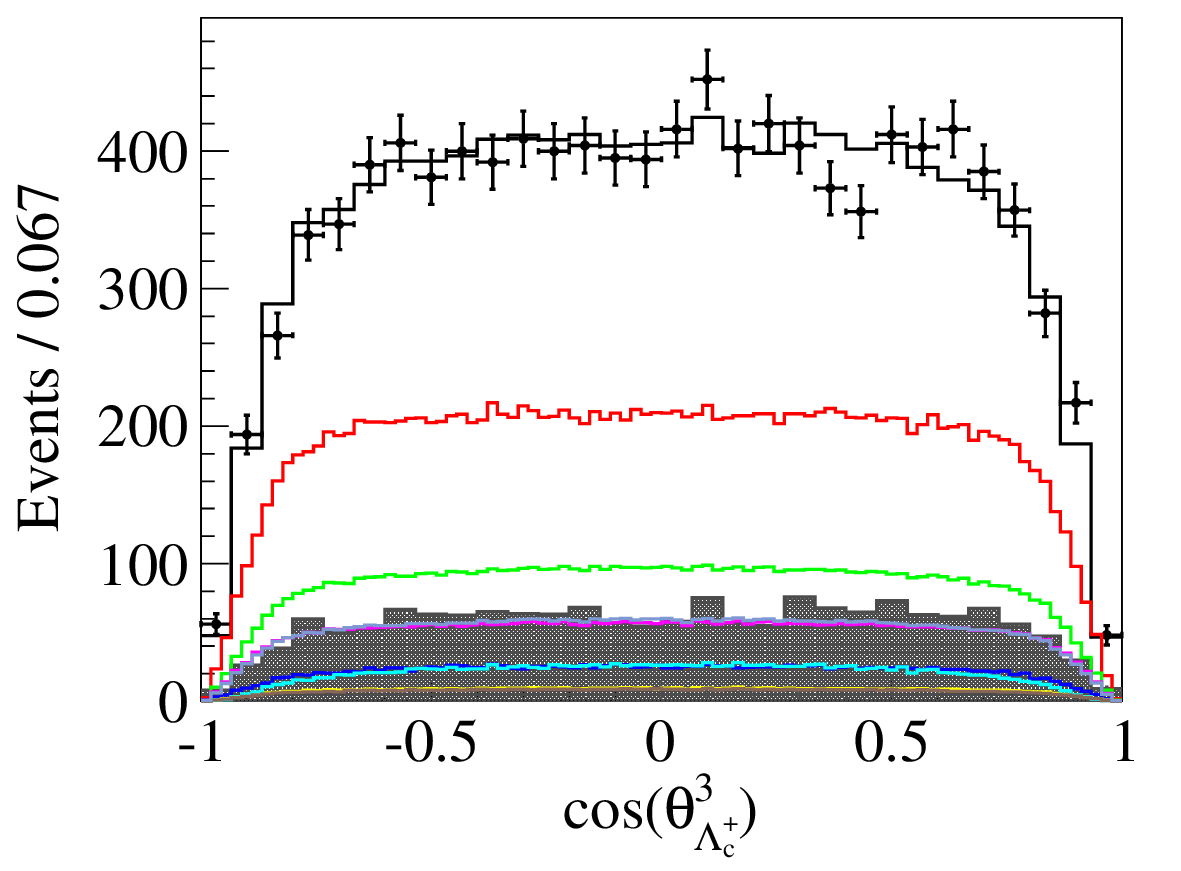}\\
\includegraphics[trim = 9mm 0mm 0mm 0mm, width=0.32\textwidth]{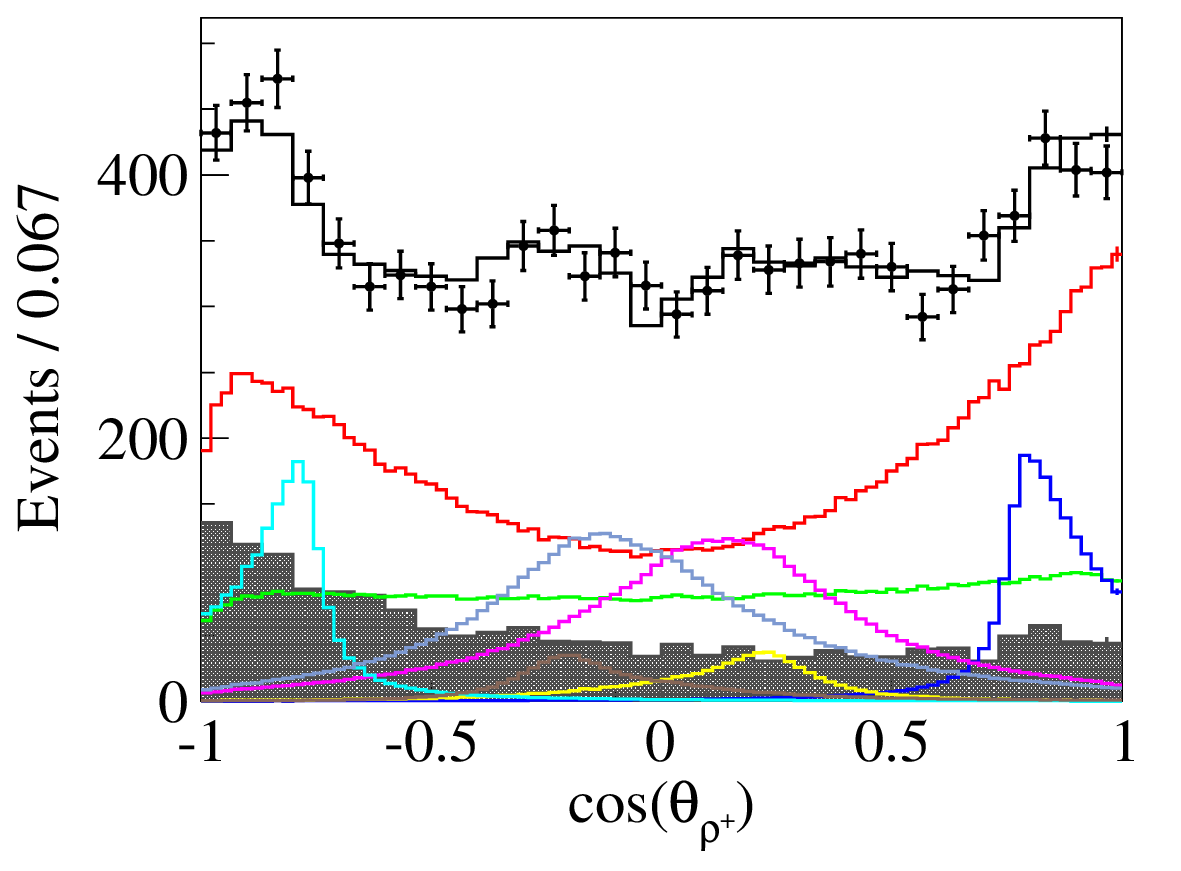}
\includegraphics[trim = 9mm 0mm 0mm 0mm, width=0.32\textwidth]{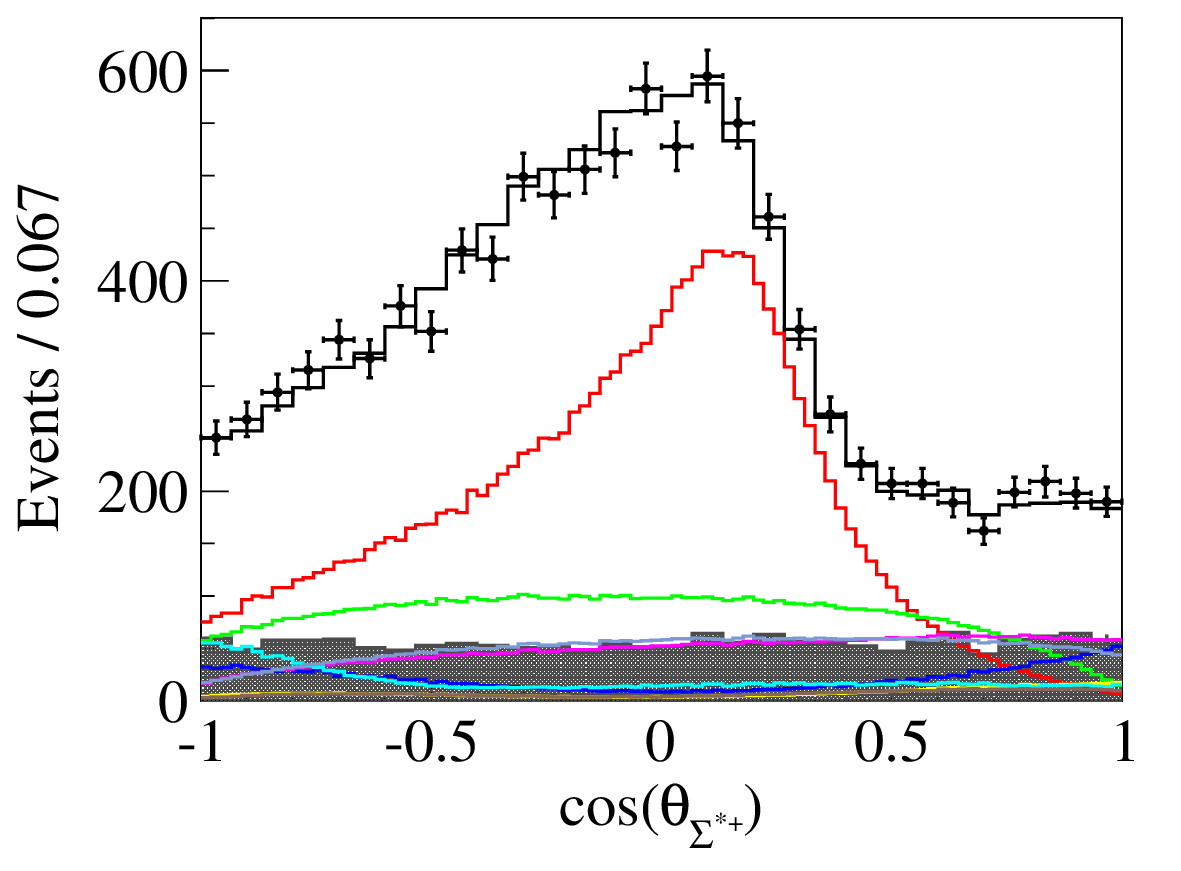}
\includegraphics[trim = 9mm 0mm 0mm 0mm, width=0.32\textwidth]{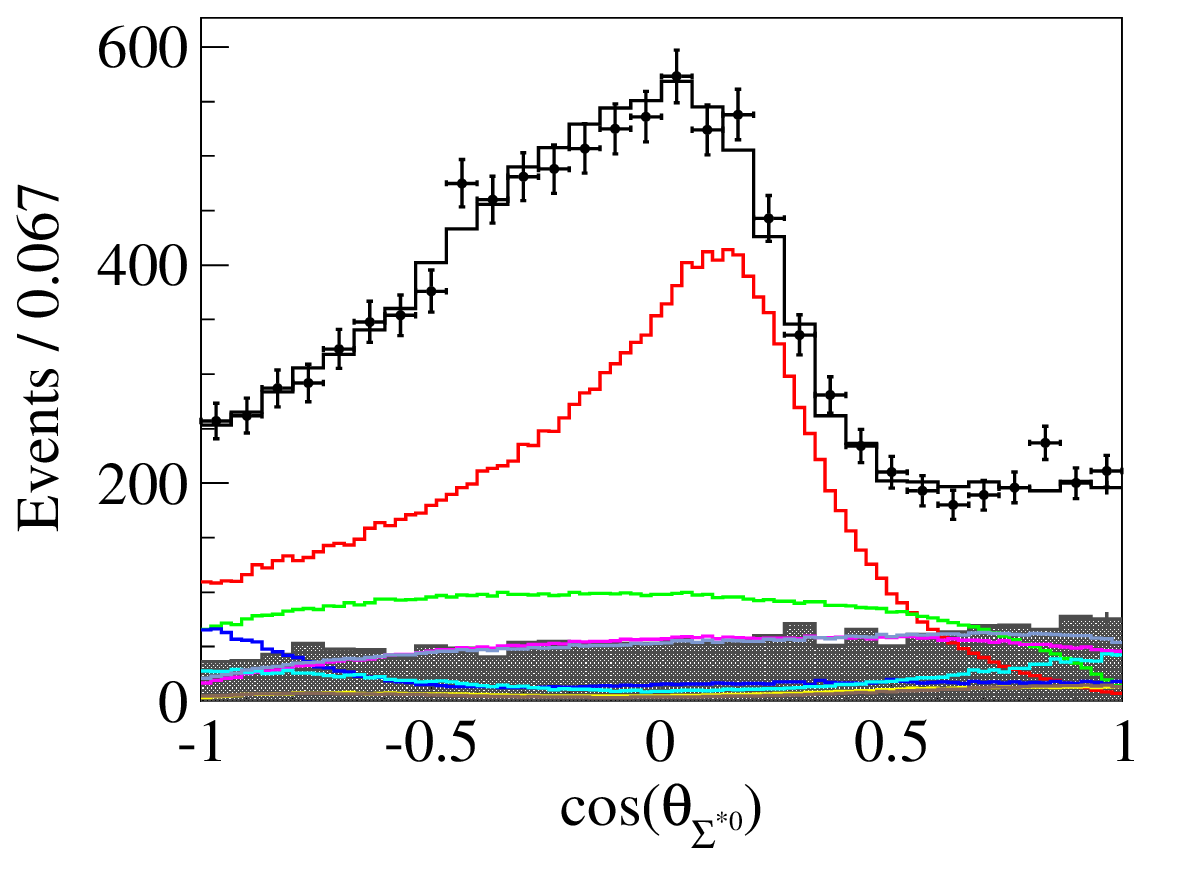}\\
\includegraphics[trim = 9mm 0mm 0mm 0mm, width=0.32\textwidth]{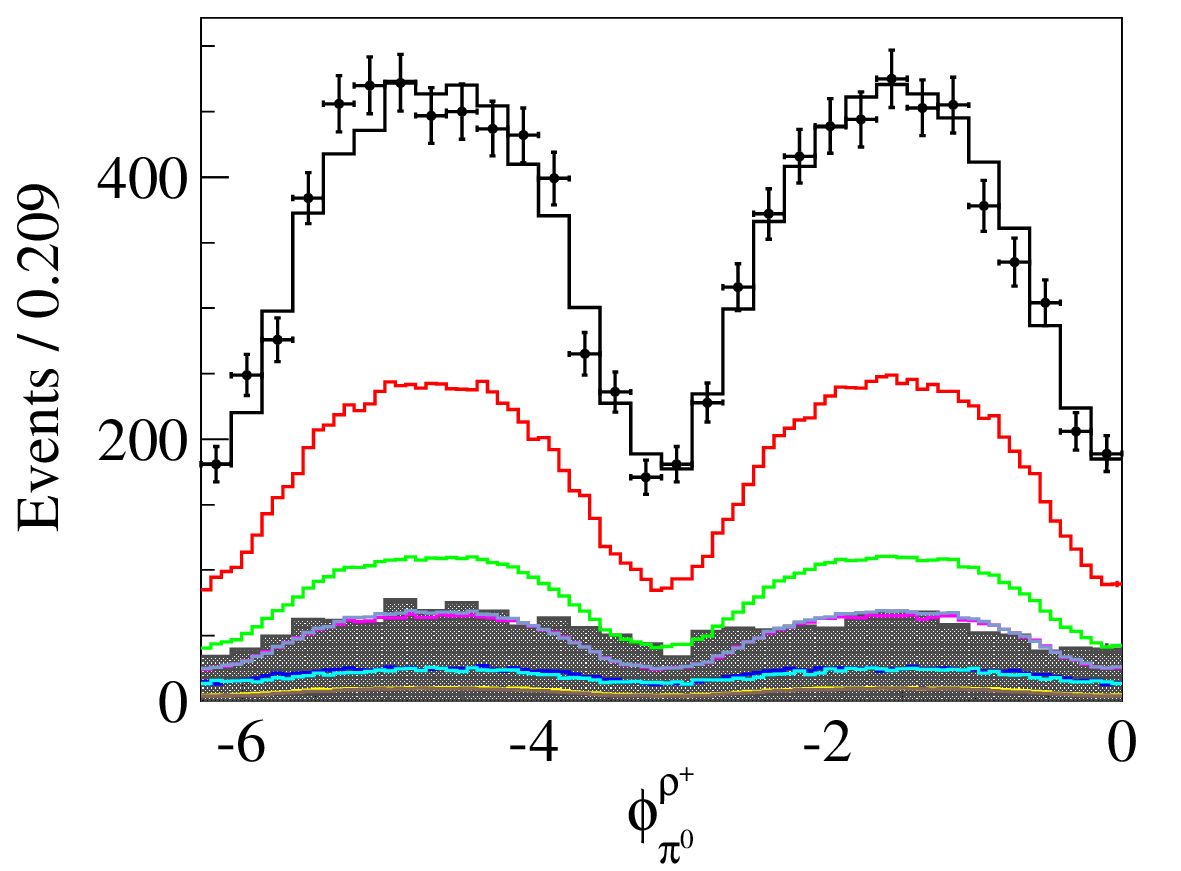}
\includegraphics[trim = 9mm 0mm 0mm 0mm, width=0.32\textwidth]{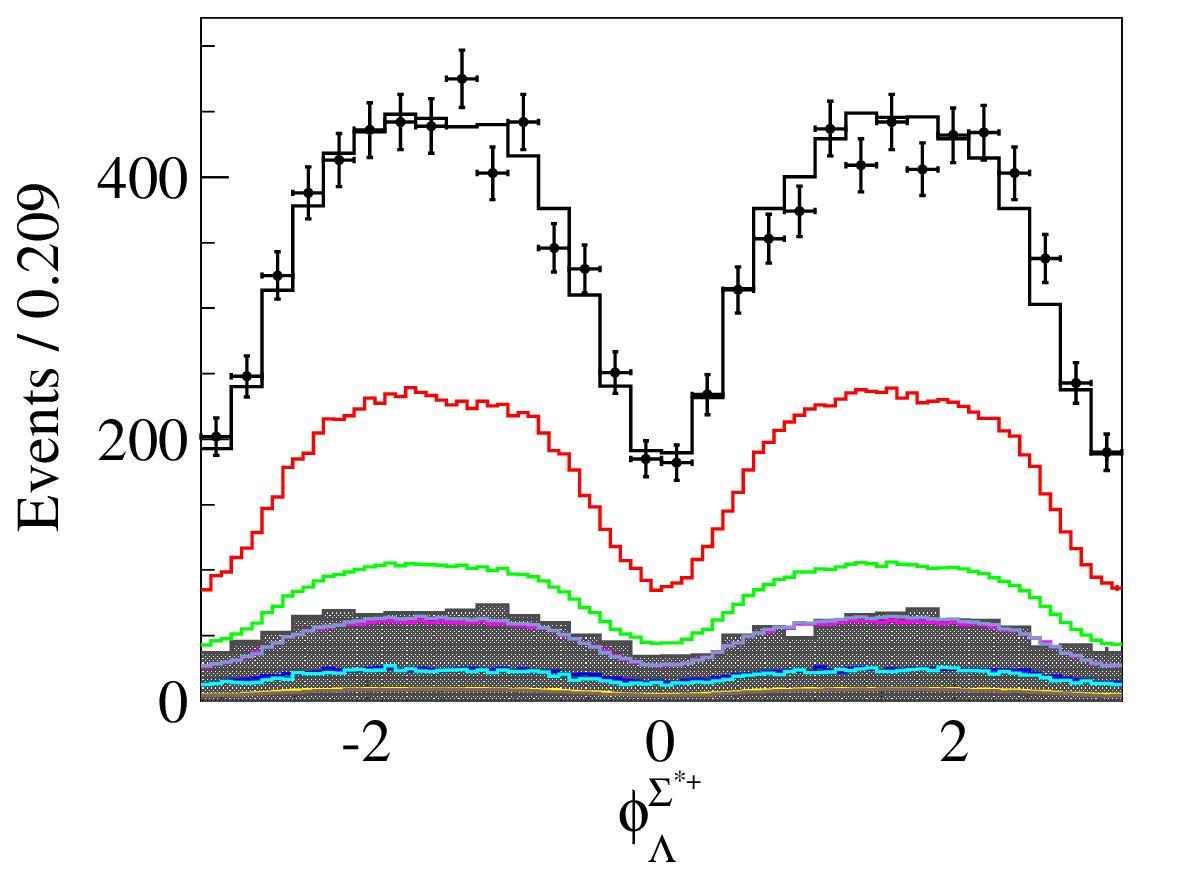}
\includegraphics[trim = 9mm 0mm 0mm 0mm, width=0.32\textwidth]{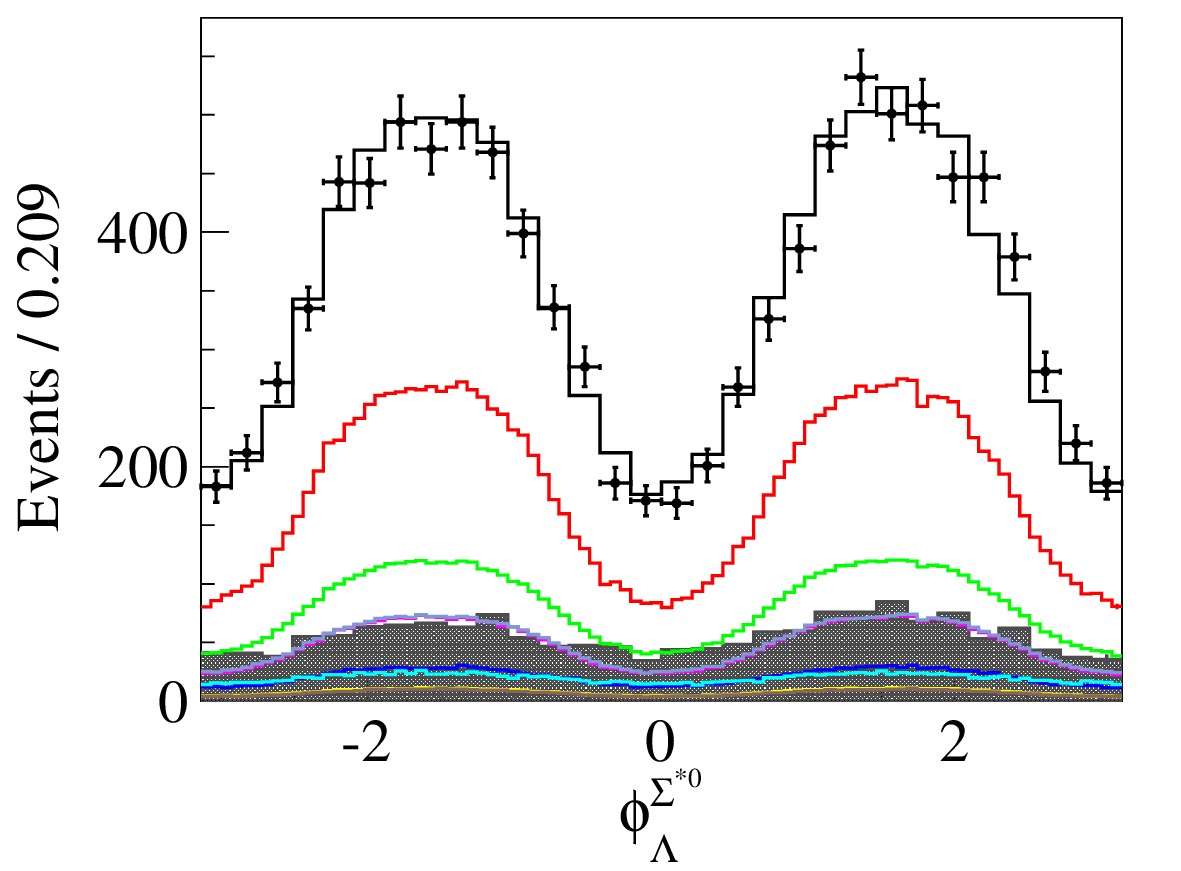}\\
\includegraphics[trim = 9mm 0mm 0mm 0mm, width=0.32\textwidth]{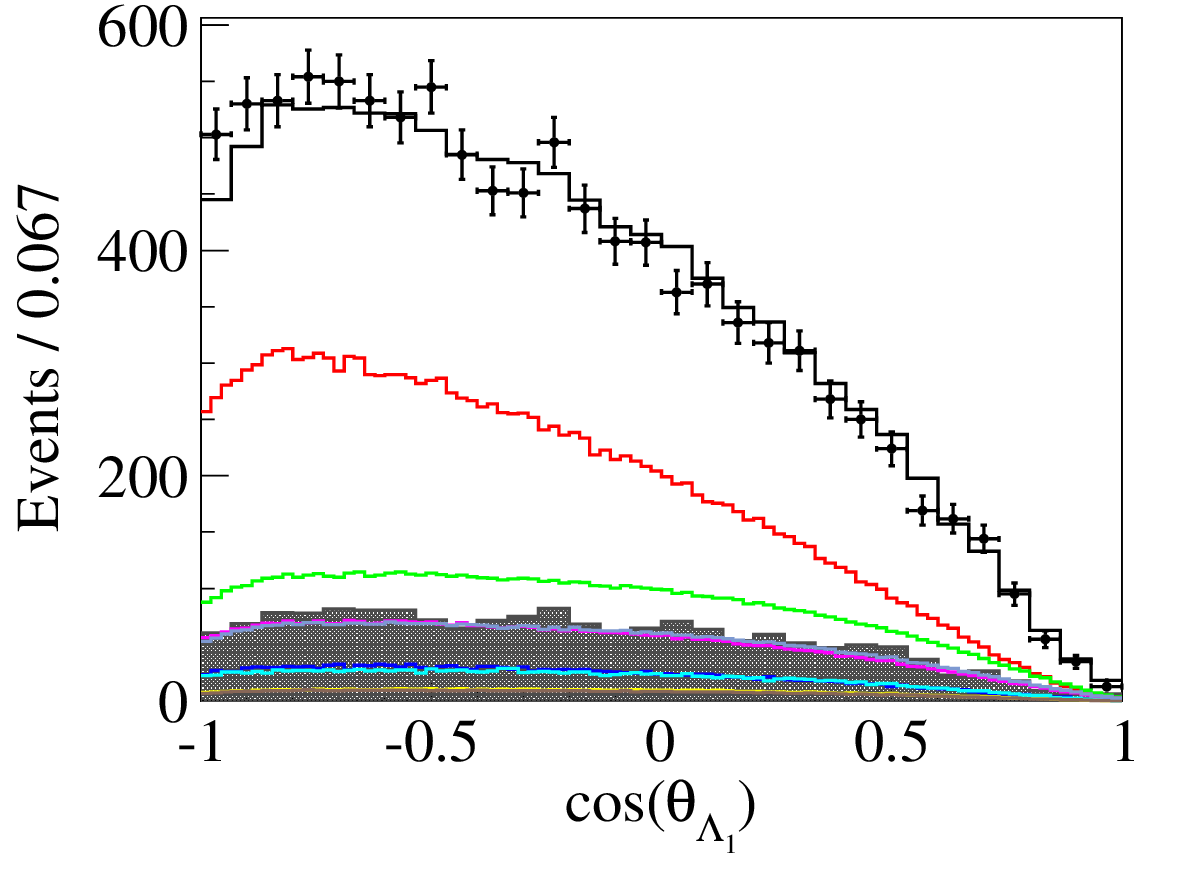}
\includegraphics[trim = 9mm 0mm 0mm 0mm, width=0.32\textwidth]{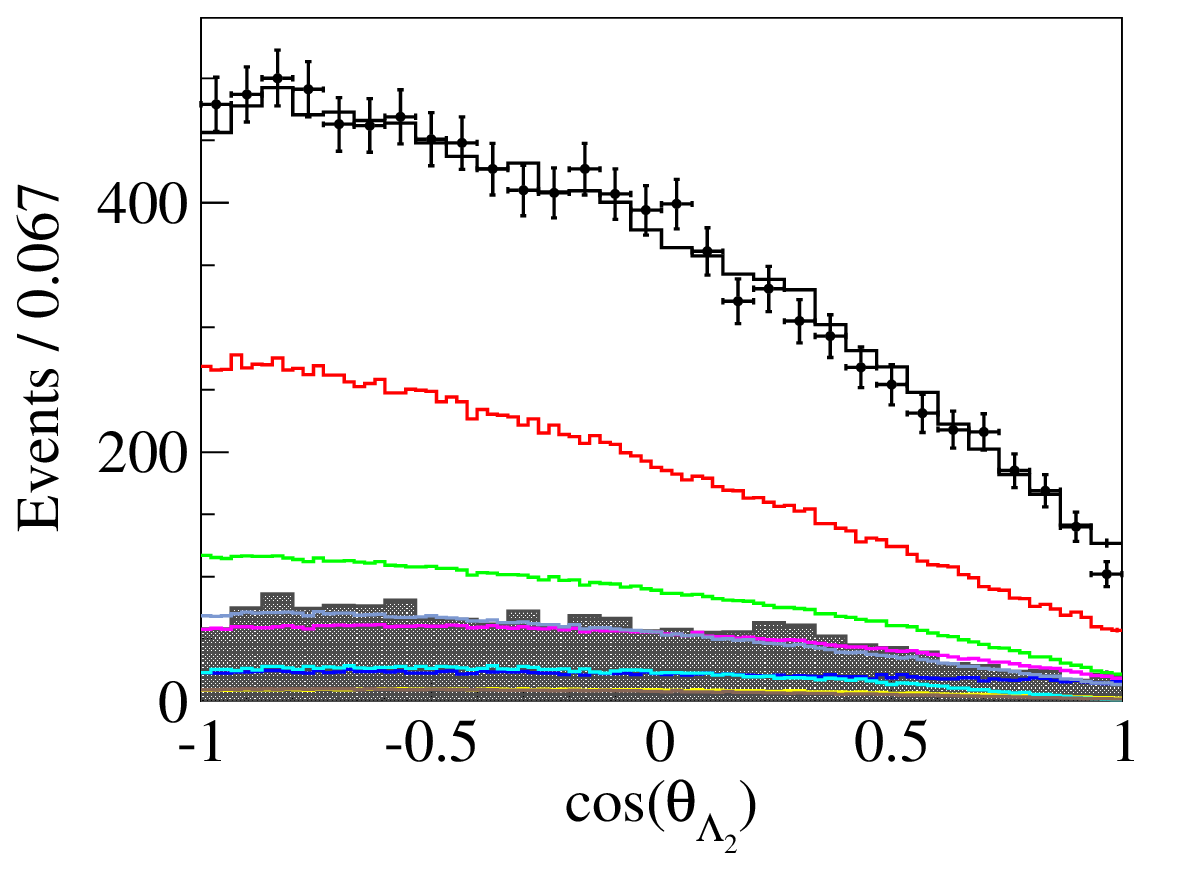}
\includegraphics[trim = 9mm 0mm 0mm 0mm, width=0.32\textwidth]{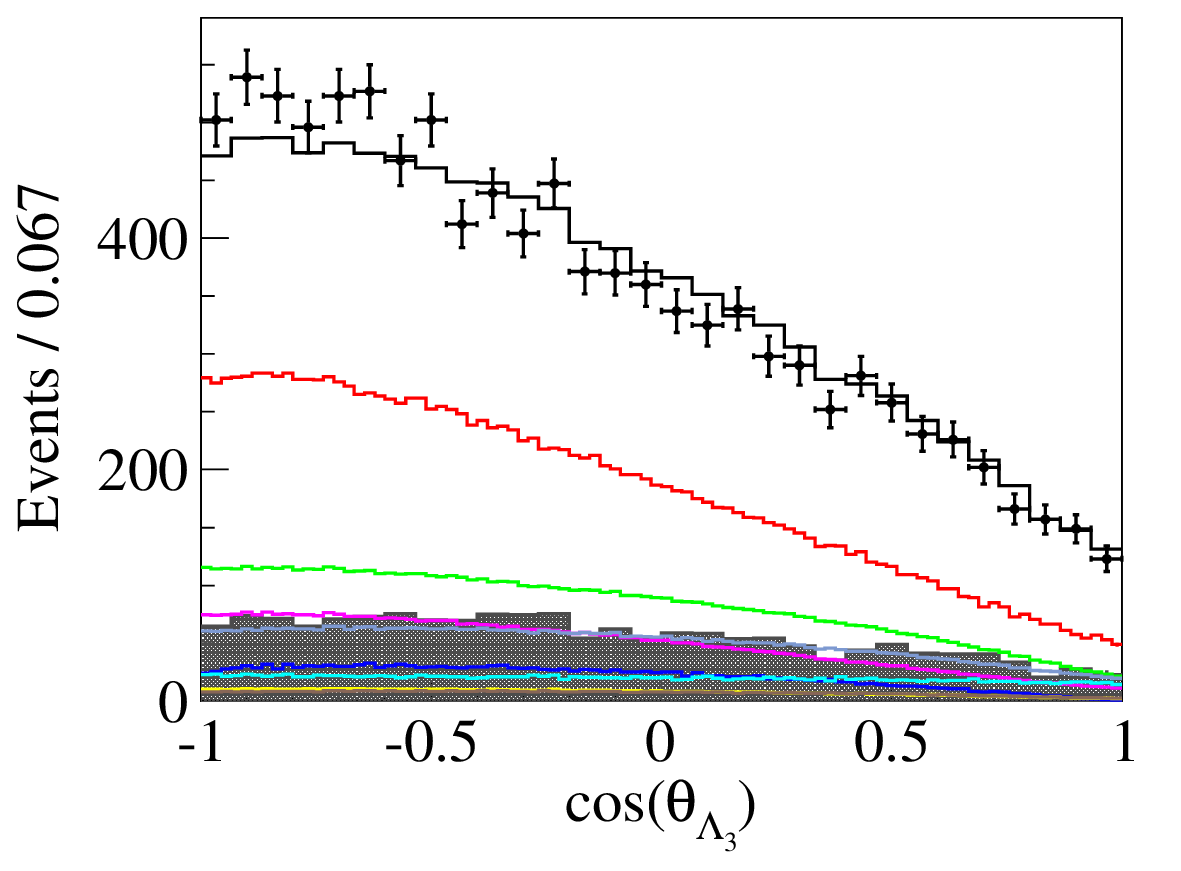}\\
\includegraphics[trim = 9mm 0mm 0mm 0mm, width=0.32\textwidth]{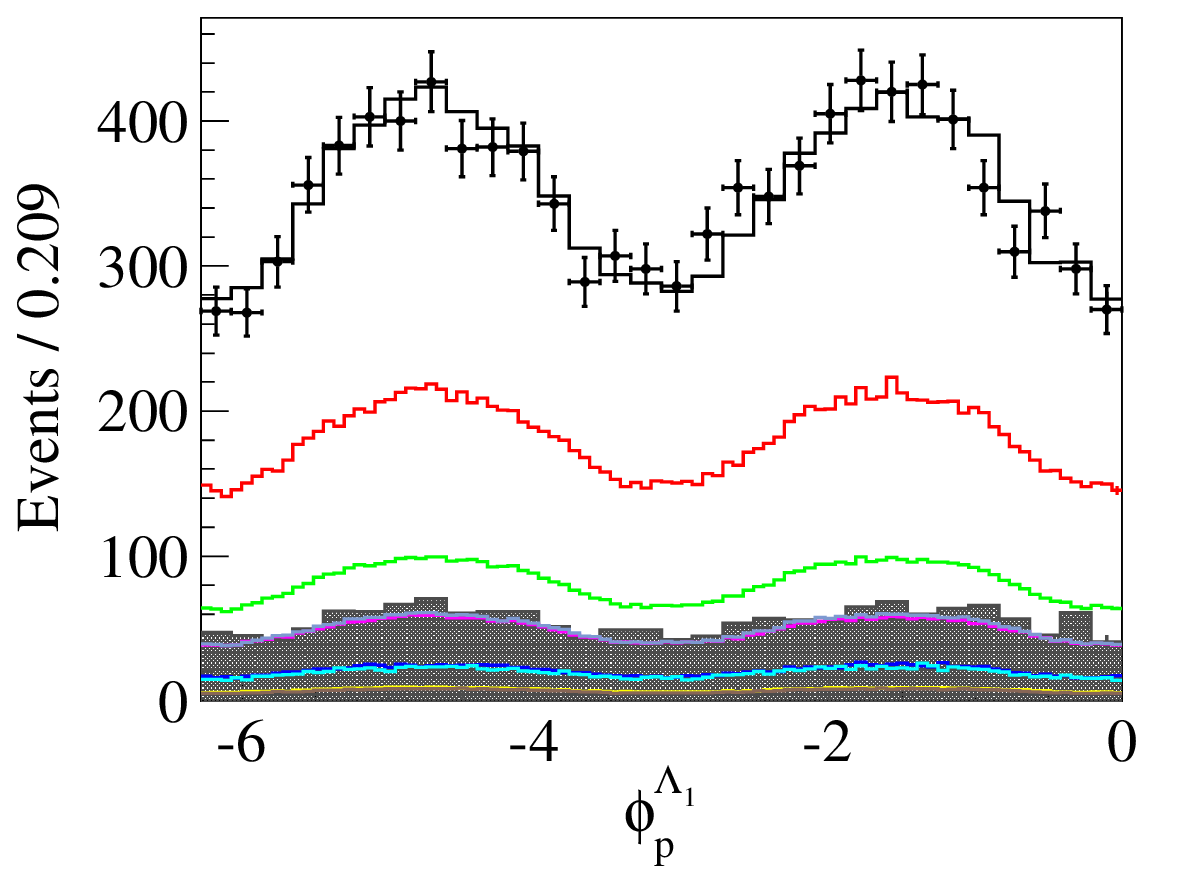}
\includegraphics[trim = 9mm 0mm 0mm 0mm, width=0.32\textwidth]{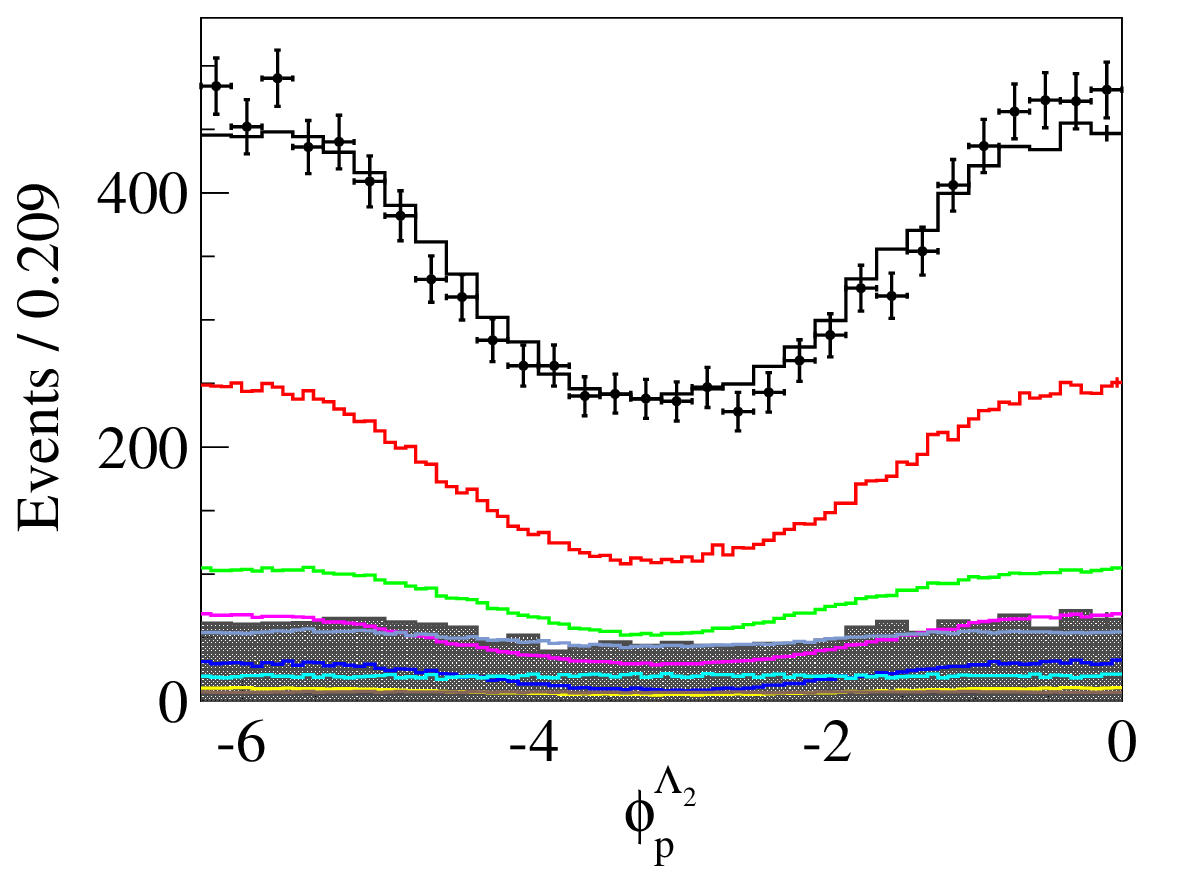}
\includegraphics[trim = 9mm 0mm 0mm 0mm, width=0.32\textwidth]{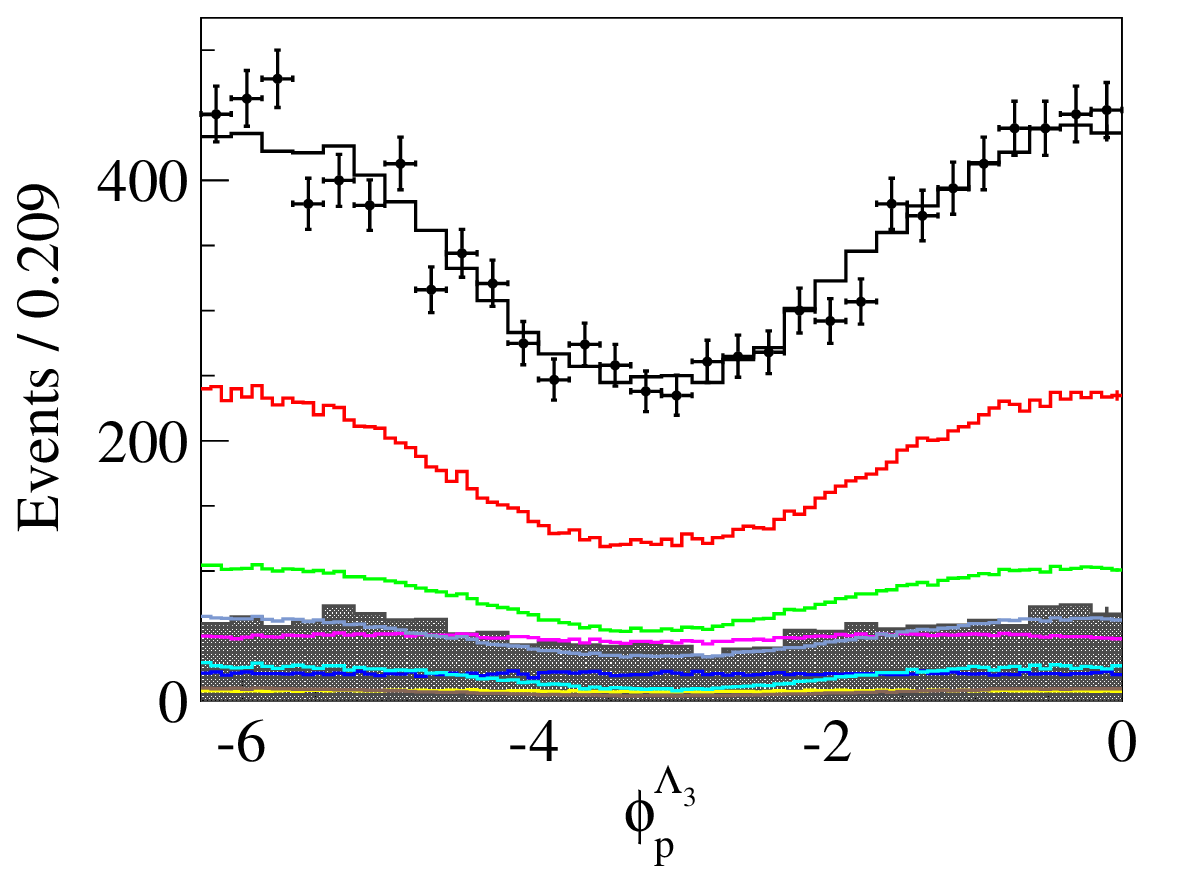}\\
\caption{Projections of the fit results in the distributions of helicity angles.}
\label{fig:nominal2}
\end{figure}

\begin{table*}[!htbp]
\caption{Numerical results of the BFs and decay asymmetry parameters,
  where the first uncertainties are statistical and the second
  systematic. For the absolute BFs, the third uncertainties are due to
  the quoted external BFs $\BR(\lcp\to\Lambda\pip\piz)$ and
  $\BR(\Sigma(1385)\to\Lambda\pi)$. The relative BFs are equivalent to
  the FFs listed in Table~\ref{tab:nominal1}.}
\setlength{\abovecaptionskip}{1.2cm}
\setlength{\belowcaptionskip}{0.2cm}
\label{tab:BF_alpha}
\begin{center}
\vspace{-0.0cm}
\begin{tabular}{c | c}
	\hline \hline
	 & Result \\
	\hline
	$\frac{\BR(\lcp\to\Lambda\rho(770)^+)}{\BR(\lcp\to\Lambda\pip\piz)}$ & $( 57.2 \pm 4.2 \pm 4.9)\%$ \\
	$\frac{\BR(\lcp\to\Sigma(1385)^+\piz)\cdot\BR(\Sigma(1385)^+ \to \Lambda\pip)}{\BR(\lcp\to\Lambda\pip\piz)}$ & 
	$(7.18 \pm 0.60 \pm 0.64)\%$ \\
	$\frac{\BR(\lcp\to\Sigma(1385)^0\pip)\cdot\BR(\Sigma(1385)^0 \to \Lambda\piz)}{\BR(\lcp\to\Lambda\pip\piz)}$ & 
	$(7.92 \pm 0.72 \pm 0.80)\%$ \\
	$\BR(\lcp\to\Lambda\rho(770)^+)$ & $(4.06\pm0.30\pm0.35\pm0.23)\times10^{-2}$ \\
	$\BR(\lcp\to\Sigma(1385)^+\piz)$ & $(5.86\pm0.49\pm0.52\pm0.35)\times10^{-3}$ \\
	$\BR(\lcp\to\Sigma(1385)^0\pip)$ & $(6.47\pm0.59\pm0.66\pm0.38)\times10^{-3}$ \\
	$\alphaA$ & $-0.763\pm0.053\pm0.045$ \\
	$\alphaB$ & $-0.917\pm0.069\pm0.056$ \\
	$\alphaC$ & $-0.789\pm0.098\pm0.056$ \\
	\hline\hline
\end{tabular}
\end{center}
\end{table*}
\vspace{-0.0cm}

\subsection{Decay asymmetry parameters}
\label{sec:asymmetry}
\hspace{1.5em} The decay asymmetry parameters $\alpha$ are related to
the interference effects among the different partial waves, and the
corresponding expressions are considered in the formula of the partial
wave amplitudes. When the intermediate state is at its nominal mass, the helicity
amplitude of a two-body decay can be written under LS coupling
expansion~\cite{LScoupling} as
\begin{eqnarray}
    H_{\lambda_1, \lambda_2} = \sum_{l s} g_{ls} \sqrt{\frac{2l+1}{2J_0 + 1}} 
    \langle l 0,s  \delta| J_0,\delta \rangle
    \langle J_1 J_2, \lambda_1\ -\lambda_2| s,\delta \rangle,
\label{eq:ls2}
\end{eqnarray}
where $g_{ls}$ is the partial wave amplitude.

When considering the decay $\Lambda\to p \pim$, the two helicity
amplitudes are obtained as
\begin{eqnarray}
    H^{\Lambda}_{ 0, \pm\frac{1}{2} } = \frac{\left(g^{\Lambda}_{0,\frac{1}{2}} \pm g^{\Lambda}_{1,\frac{1}{2}}\right)}{\sqrt{2}},
\label{eq:lmdasy}
\end{eqnarray}
and the $\Lambda$ decay asymmetry parameter $\alpha_{\Lambda}$ is expressed as
\begin{eqnarray}
    \alpha_{\Lambda} =  
    \frac{|H^{\Lambda}_{0,\frac{1}{2}}|^2-|H^{\Lambda}_{0,-\frac{1}{2} }|^2}
    {|H^{\Lambda}_{ 0,\frac{1}{2}}|^2+|H^{\Lambda}_{ 0,-\frac{1}{2}}|^2}
    =  \frac{2\Re\left(g^{\Lambda}_{0,\frac{1}{2}}\cdot \bar{g}^{\Lambda}_{1,\frac{1}{2}}\right)}
    {|g^{\Lambda}_{0,\frac{1}{2}}|^2+|g^{\Lambda}_{1,\frac{1}{2}}|^2},
\label{eq:lmdasy2}
\end{eqnarray}
which represents the interference between the $\mathcal{S}$ and $\mathcal{P}$ partial wave amplitudes. In Eq.~\eqref{eq:helicty14}, the helicity of the proton $\lambda_p=-1/2$ and $\lambda_p=1/2$ is summed directly outside the module square. This reduces the relative partial wave amplitude $g^{\Lambda}_{1,\frac{1}{2}}/g^{\Lambda}_{0,\frac{1}{2}}$ by one degree, by fixing the phase of $g^{\Lambda}_{1,\frac{1}{2}}$ to zero and the magnitude to the value listed in Table~\ref{tab:nominal2} according to Eq.~\eqref{eq:lmdasy2} with the input value $\alpha_{\Lambda}=0.732\pm0.014$.

The decay $\lcp\to\Lambda\rho(770)^+$ is described by four helicity
amplitudes and the differential decay width depends on the decay
asymmetry $\alphaA$~\cite{asymmetry1}
\begin{eqnarray}
    \frac{\mathrm{d}\Gamma}{\mathrm{d}\cos\Theta_\Lambda}\propto
    1+\alphaA\cdot\alpha_\Lambda\cdot\cos\Theta_\Lambda,
\label{eq:alphaA1}
\end{eqnarray}
where $\Theta_\Lambda$ is the $\Lambda$ helicity angle, denoted as $\theta_{\Lambda_1}$ in Figure~\ref{fig:angle}. Using Eq.~\eqref{eq:ls2}, the four helicity amplitudes are denoted as
\begin{eqnarray}
\begin{aligned}
   H^{\rho}_{ -\frac{1}{2}, -1 } =& - \frac{g^{\rho}_{0,\frac{1}{2}}}{\sqrt{3}} + \frac{g^{\rho}_{1,\frac{1}{2}}}{\sqrt{3}} - \frac{g^{\rho}_{1,\frac{3}{2}}}{\sqrt{6}} + \frac{g^{\rho}_{2,\frac{3}{2}}}{\sqrt{6}}, \\
   H^{\rho}_{ -\frac{1}{2}, 0 } =& - \frac{g^{\rho}_{0,\frac{1}{2}}}{\sqrt{6}} - \frac{g^{\rho}_{1,\frac{1}{2}}}{\sqrt{6}} - \frac{g^{\rho}_{1,\frac{3}{2}}}{\sqrt{3}} - \frac{g^{\rho}_{2,\frac{3}{2}}}{\sqrt{3}}, \\
   H^{\rho}_{ \frac{1}{2}, 0 } =& \frac{g^{\rho}_{0,\frac{1}{2}}}{\sqrt{6}} - \frac{g^{\rho}_{1,\frac{1}{2}}}{\sqrt{6}} - \frac{g^{\rho}_{1,\frac{3}{2}}}{\sqrt{3}} + \frac{g^{\rho}_{2,\frac{3}{2}}}{\sqrt{3}}, \\
   H^{\rho}_{ \frac{1}{2}, 1 } =& \frac{g^{\rho}_{0,\frac{1}{2}}}{\sqrt{3}} + \frac{g^{\rho}_{1,\frac{1}{2}}}{\sqrt{3}} - \frac{g^{\rho}_{1,\frac{3}{2}}}{\sqrt{6}} - \frac{g^{\rho}_{2,\frac{3}{2}}}{\sqrt{6}}. \\
\end{aligned}
\label{eq:alphaA2}
\end{eqnarray}
The decay asymmetry $\alphaA$ can be expressed with the partial wave amplitudes as
\begin{eqnarray}
\begin{aligned}
    \alphaA & =
    \frac{
     |H^{\rho}_{ \frac{1}{2}, 1 }|^2 -
     |H^{\rho}_{-\frac{1}{2}, -1 }|^2 + 
     |H^{\rho}_{ \frac{1}{2}, 0 }|^2 -
     |H^{\rho}_{-\frac{1}{2}, 0 }|^2}
    {|H^{\rho}_{ \frac{1}{2}, 1 }|^2 +
     |H^{\rho}_{ -\frac{1}{2}, -1 }|^2 + 
     |H^{\rho}_{ \frac{1}{2}, 0 }|^2 +
     |H^{\rho}_{ -\frac{1}{2}, 0 }|^2}
    \\
    = & \frac{\sqrt{\frac{1}{9}}\cdot
    2\cdot\Re\left(g^{\rho}_{0,\frac{1}{2}}\cdot \bar{g}^{\rho}_{1,\frac{1}{2}}
    -g^{\rho}_{1,\frac{3}{2}}\cdot \bar{g}^{\rho}_{2,\frac{3}{2}}\right)
    -\sqrt{\frac{8}{9}}\cdot
    2\cdot\Re\left(g^{\rho}_{0,\frac{1}{2}}\cdot \bar{g}^{\rho}_{1,\frac{3}{2}}
    +g^{\rho}_{1,\frac{1}{2}}\cdot \bar{g}^{\rho}_{2,\frac{3}{2}}\right)}
    {|g^{\rho}_{0,\frac{1}{2}}|^2+|g^{\rho}_{1,\frac{1}{2}}|^2
    +|g^{\rho}_{1,\frac{3}{2}}|^2+|g^{\rho}_{2,\frac{3}{2}}|^2}.
\end{aligned}
\label{eq:alphaA3}
\end{eqnarray}

For the process $\lcp\to\Sigma(1385)\pi$, the corresponding decay asymmetry appears in the differential width expression~\cite{asy2}
\begin{eqnarray}
\begin{aligned}
&\frac{\mathrm{d}\Gamma }{\mathrm{d}\cos\theta_{ee}\mathrm{d}\cos\theta_{\Lambda_c^+}\mathrm{d}\cos\theta_{\Sigma^*}\mathrm{d}\phi^{ee}_{\Lambda_c^+}}  \propto  \left(7+9\cos(2\theta_{\Sigma^*})\right) \cdot \\
& \left(1 + \alpha_0\cos^2(\theta_{ee}) + \alpha_{\Sigma^*\pi}\sqrt{1-\alpha_0^2} \sin\Delta_0 \cos\theta_{ee} \sin\theta_{ee}\sin\theta_{\Lambda_c^+}\sin\phi^{ee}_{\Lambda_c^+}\right),
\end{aligned}
\label{eq:asym1385}
\end{eqnarray}
where $\alpha_0$ and $\Delta_0$ are constants related to the
$\lcp\lcm$ production, and the definitions of the
individual helicity angles can be found in
Figure~\ref{fig:asym1385}. There are two different helicity amplitudes
in this process, which can be written using Eq.~\eqref{eq:ls2} as
\begin{eqnarray}
    H^{\Sigma(1385)}_{ 0, \pm\frac{1}{2} } = \mp \frac{ \left(g^{\Sigma(1385)}_{2,\frac{3}{2}}\pm g^{\Sigma(1385)}_{1,\frac{3}{2}}\right)}{\sqrt{2}}.
\label{eq:alphaBC2}
\end{eqnarray}
Thus, the decay asymmetry $\alpha_{\Sigma(1385)\pi}$ can be expressed with the partial wave amplitudes as
\begin{eqnarray}
    \alpha_{\Sigma(1385)\pi} = 
    \frac{|H^{\Sigma(1385)}_{0,\frac{1}{2}}|^2-|H^{\Sigma(1385)}_{0,-\frac{1}{2} }|^2}
    {|H^{\Sigma(1385)}_{ 0,\frac{1}{2}}|^2+|H^{\Sigma(1385)}_{ 0,-\frac{1}{2}}|^2}
    =  \frac{2\Re\left(g^{\Sigma(1385)}_{1,\frac{3}{2}}\cdot \bar{g}^{\Sigma(1385)}_{2,\frac{3}{2}}\right)}
    {|g^{\Sigma(1385)}_{1,\frac{3}{2}}|^2+|g^{\Sigma(1385)}_{2,\frac{3}{2}}|^2}. 
\label{eq:alphaBC3}
\end{eqnarray}

\vspace{-0.0cm}
\begin{figure}[htbp]
\setlength{\abovecaptionskip}{-1pt}
\setlength{\belowcaptionskip}{10pt}
\centering
\quad\quad
\includegraphics[trim = 9mm 0mm 0mm 0mm, width=0.6\textwidth]{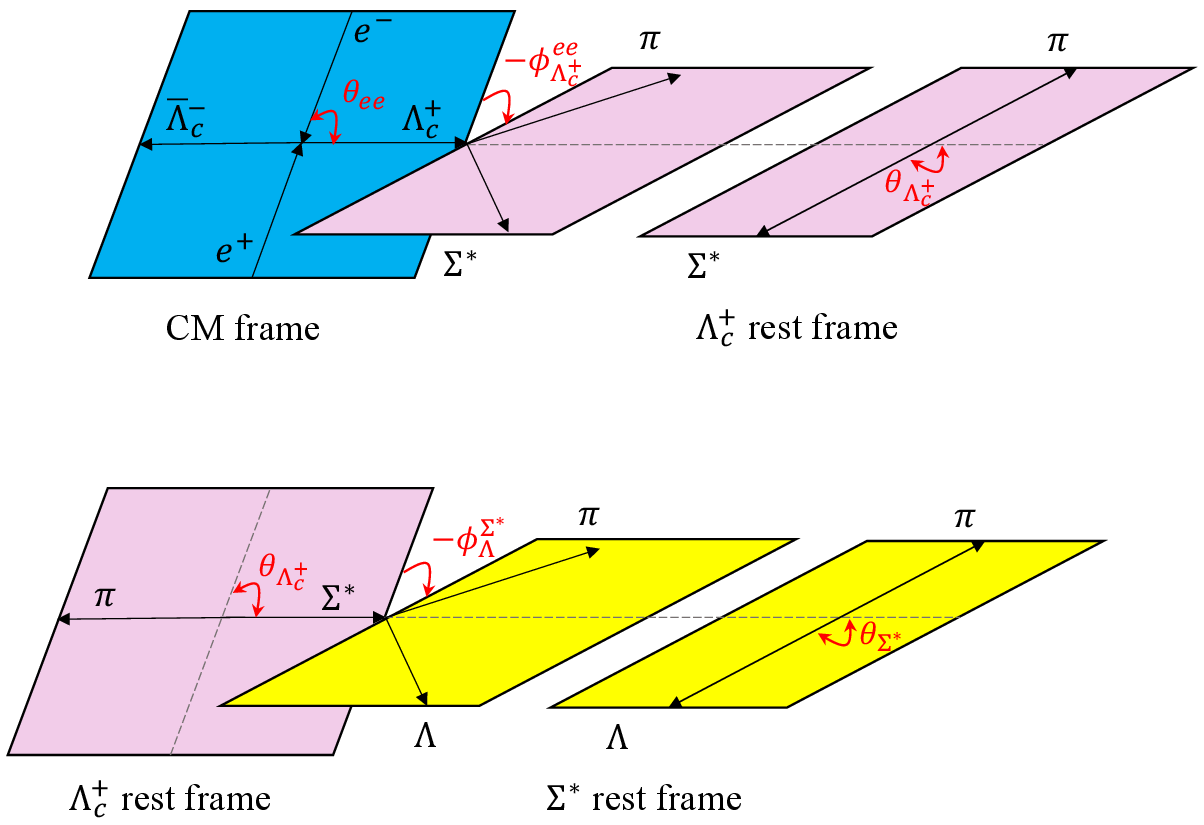}
\caption{Definitions of helicity angles for the cascade process $e^+e^-\to\lcp\lcm$, $\lcp\to\Sigma^*\pi$, $\Sigma^*\to\Lambda\pi$, where $\Sigma^*$ denotes $\Sigma(1385)^+$ or $\Sigma(1385)^0$. The convention of the notation is the same with Figure~\ref{fig:angle}.}
\label{fig:asym1385}
\end{figure}

Through the nominal fit results of the partial wave amplitudes listed in Table~\ref{tab:nominal2}, the decay asymmetry parameters can be obtained, as listed in Table~\ref{tab:BF_alpha}. The statistical uncertainties of the asymmetry parameters are calculated via error propagation in Eq.~\eqref{eq:errPropagation}.

\section{Systematic uncertainties}
\label{sec:sys}
\hspace{1.5em} The systematic uncertainties of the observables,
including the FFs of the components $\rho(770)^+$, $\Sigma(1385)^+$,
and $\Sigma(1385)^0$, as well as their decay asymmetry parameters
$\alphaA$, $\alphaB$, and $\alphaC$, are estimated. In general, the
following sources are considered: (\romanOne) fixed parameters,
(\romanTwo) mass dependent width, (\romanThree) radius parameter,
(\romanFour) background description, (\romanFive) data-MC differences,
(\romanSix) resonance components, (\romanSeven) polarization of the
initial $\lcp$, and (\romanEight) fit bias. All the systematic
uncertainties are listed in Table~\ref{tab:sys_err}.

\begin{table*}[!tp]
\caption{Systematic uncertainties (in units of corresponding statistical uncertainties) on the FFs of $\Sigma(1385)^+$, $\Sigma(1385)^0$ and $\rho(770)^+$, and the corresponding decay asymmetry parameters $\alphaA$, $\alphaB$ and $\alphaC$. The total systematic uncertainties are obtained by summing up all contributions in quadrature.}
\setlength{\abovecaptionskip}{1.2cm}
\setlength{\belowcaptionskip}{0.2cm}
\label{tab:sys_err}
\begin{center}
\vspace{-0.0cm}
\begin{tabular}{c c c c c c c}
	\hline \hline 
	& $\mathrm{FF}_{\rho(770)^+}$ & $\mathrm{FF}_{\Sigma(1385)^+}$ & $\mathrm{FF}_{\Sigma(1385)^0}$ & $\alphaA$ & $\alphaB$ & $\alphaC$\\
	\hline
	\romanOne   & 0.47 & 0.32 & 0.62 & 0.15 & 0.18 & 0.32 \\
	\romanTwo   & 0.15 & 0.14 & 0.15 & 0.16 & 0.01 & 0.01 \\
	\romanThree & 0.88 & 0.39 & 0.36 & 0.50 & 0.07 & 0.29 \\
	\romanFour  & 0.11 & 0.41 & 0.26 & 0.21 & 0.22 & 0.21 \\
	\romanFive  & 0.06 & 0.25 & 0.43 & 0.03 & 0.07 & 0.03 \\
	\romanSix   & 0.49 & 0.79 & 0.65 & 0.57 & 0.74 & 0.20 \\
	\romanSeven & 0.08 & 0.13 & 0.01 & 0.20 & 0.16 & 0.23 \\
	\romanEight & 0.20 & 0.10 & 0.13 & 0.08 & 0.06 & 0.04 \\
	\hline
	Total       & 1.15 & 1.08 & 1.11 & 0.84 & 0.81 & 0.57 \\
	\hline\hline
\end{tabular}
\end{center}
\end{table*}
\vspace{-0.0cm}

\begin{itemize}
\item [\romanOne.] \emph{Fixed parameters.} In the nominal fit, the mass and width parameters of resonances and the $\Lambda$ decay asymmetry parameter are fixed according to Refs.~\cite{pdg,recent,BESIII:2018cnd,Ireland:2019uja,bes3LmdRecent}. To estimate the relevant systematic uncertainties, the fixed parameters are varied within $\pm1\,\sigma$ and the fit procedure is repeated. The quadratic sums of the largest variations from each parameter are assigned as systematic uncertainties.

\item [\romanTwo.] \emph{Mass dependent width.} In the nominal fit,
  the running width is used as in Eq.~\eqref{eq:helicty7} to describe
  the width of a resonance. To estimate the potential systematic
  uncertainties, an alternative method using a pseudo coupled channel
  approach is considered, where the full width consists of partial
  widths from the main decay channels $N\bar{K}$, $\Sigma\pi$, and
  $\Lambda\pi$, as described in Refs.~\cite{recent,coupled}. Using the
  coupled channel widths, the fit procedure is repeated, and the
  resulting variations are assigned as systematic uncertainties.

\item [\romanThree.] \emph{Radius parameter.} In the nominal fit, the
  radius parameter $d$ is chosen as
  $d=0.73\;\mathrm{fm}$~\cite{barrierSet}.  To estimate the potential
  bias, the fits are performed by setting $d$ at the alternative
  values derived from Ref.~\cite{barrierSet}, $d=0.53\;\mathrm{fm}$
  and $d=1.16\;\mathrm{fm}$, and the largest variations are assigned
  as systematic uncertainties.

\item [\romanFour.] \emph{Background description.} The systematic
  uncertainties arising from the background description
  originate from two sources: background fractions and background
  shape.

In the nominal fit, the background fractions are fixed according to
the fit results given in Table~\ref{tab:fitmbc}. To estimate
systematic uncertainties, the background fraction for each energy
point is varied within $\pm1\,\sigma$, and the fit procedure is
repeated. The quadratic sums of the largest variations from each
fraction are assigned as the associated systematic uncertainties.

For the background shape, the $\mbc$ sideband is used to describe the background components in the $\mbc$ signal region. To understand the possible bias coming from the background modelling based on sideband data, this approach is tested on a large toy MC sample generated according to the fitted amplitudes in the nominal fit plus the background processes in the inclusive MC samples. The PWA fit is applied to the toy sample and the variations of the output fit results from the input amplitudes are assigned as systematic uncertainties.

Finally, the systematic uncertainties arising from background description are assigned as quadratic sums of the above two sources.

\item [\romanFive.] \emph{Data-MC differences.} To estimate possible systematic uncertainties due to the difference between the MC-determined efficiency and the experimental one, the effects from tracking and PID of $\pip$ candidates, and the reconstruction of $\Lambda$ and $\piz$ candidates are considered. The reconstruction efficiency differences between data and MC simulations have been investigated in previous studies, such as $\Lambda$ reconstruction in Ref.~\cite{sys_Lmd}, $\piz$ reconstruction in Ref.~\cite{sys_piz}, and $\pip$ tracking and PID in Ref.~\cite{sys_pip}. The correction factors $w=\varepsilon_\mathrm{Data}/\varepsilon_\mathrm{MC}$ are assigned as the weighting factors of the PHSP MC sample and the fit procedure is repeated. Including the weighting factors, Eq.~\eqref{eq:helictyEx} becomes
\begin{eqnarray}
    \int |\mathcal{A}|^2 \mathrm{d}\Phi \propto \frac{1}{\sum_{i \in \mathrm{PHSP}} w(x_i)} \sum_{i \in \mathrm{PHSP}} w(x_i)\cdot|\mathcal{A}(x_i)|^2,
\label{eq:helictyEx2}
\end{eqnarray}
where $w(x_i)$ is the weighting factor for the $i$-th event.
The resulting variations on the final results are considered as systematic uncertainties.

\item [\romanSix.] \emph{Resonance components.} To estimate the
  systematic uncertainties from resonance components, 
  the amplitude fit is repeated by including each possible 
  additional resonance among those listed in Sec.~\ref{sec:nominal}.
  The corresponding largest changes on the fit results among all variations 
  of resonance components are taken into account as systematic uncertainties.

\item [\romanSeven.] \emph{Polarization of initial $\lcp$.} In the nominal fit, the total amplitude is constructed using Eq.~\eqref{eq:helicty14}, where the initial state $\lcp$ is assumed to be unpolarized. To estimate the systematic uncertainties due to polarization of initial $\lcp$, the polarization parameters $P_{x,y,z}$ are introduced as free parameters, and the total amplitude is rewritten as
\begin{eqnarray}
    |\mathcal{A}'|^2 = \sum_{\lambda_p} \left[
    \begin{pmatrix}
        \mathcal{A}_{-\frac{1}{2},\lambda_p} & \mathcal{A}_{\frac{1}{2},\lambda_p}
    \end{pmatrix}\cdot
    \frac{1}{2}\cdot
    \begin{pmatrix}
        1 + P_z & P_x - i P_y \\ P_x + i P_y & 1 - P_z
    \end{pmatrix}\cdot
    \begin{pmatrix}
        \mathcal{A}^*_{-\frac{1}{2},\lambda_p} \\ \mathcal{A}^*_{\frac{1}{2},\lambda_p}
    \end{pmatrix}
    \right].
\label{eq:polarization}
\end{eqnarray}
Here, the polarization parameters indicate the average polarization of a single $\lcp$. Then, using the alternative total amplitude, the fit procedure is repeated, and the corresponding variations are assigned as systematic uncertainties.

\item [\romanEight.] \emph{Fit bias.} To estimate the potential fit
  bias effect, an input-output check is performed. Several toy MC
  samples are generated, where the signal events are generated
  according to the results of the nominal fit, and the background
  events are generated using a bootstrap method~\cite{bootstrap}.  The
  PWA fit is performed to each toy sample and the pull distributions
  for every fitted variable are plotted. The mean values of the pull
  distributions, which indicate a possible fit bias, are considered as
  systematic uncertainties.
\end{itemize}

The total systematic uncertainty is obtained by summing up all contributions in quadrature, as listed in Table~\ref{tab:BF_alpha}. The systematic uncertainties of the BFs are of a similar size as the corresponding statistical uncertainties, while the systematic uncertainties for the decay asymmetry parameters are rather smaller.

\section{Summary}
\label{sec:summary}
\hspace{1.5em} To summarize, based on the $\ee$ collision samples
corresponding to an integrated luminosity of 4.4 $\ifb$ collected with
the BESIII detector at c.m. energies between $4.6\;\gev$ and
$4.7\;\gev$, the first PWA of the charmed baryon hadronic decay
$\lcp\to\Lambda\pip\piz$ is performed.  Based on the analysis results,
the relative BFs for the resonant components are measured to be
$$\frac{\BR(\lcp\to\Lambda\rho(770)^+)}{\BR(\lcp\to\Lambda\pip\piz)}=(57.2\pm4.2\pm4.9)\%,$$ 
$$\frac{\BR(\lcp\to\Sigma(1385)^+\piz)\cdot\BR(\Sigma(1385)^+ \to \Lambda\pip)}{\BR(\lcp\to\Lambda\pip\piz)}=(7.18\pm0.60\pm0.64)\%,$$ 
$$\frac{\BR(\lcp\to\Sigma(1385)^0\pip)\cdot \BR(\Sigma(1385)^0 \to \Lambda\piz)}{\BR(\lcp\to\Lambda\pip\piz)}=(7.92\pm0.72\pm0.80)\%.$$
Here, the first uncertainty is statistical and the second is systematic.
After taking into account $\BR(\lcp\to\Lambda\pip\piz)=(7.1\pm 0.4)\%$ and $\BR(\Sigma(1385)\to\Lambda\pi)=(87.5\pm1.5)\%$~\cite{pdg}, the absolute BFs are obtained for the first time:
$$\BR(\lcp\to\Lambda\rho(770)^+)=(4.06\pm0.30\pm0.35\pm0.23)\%,$$
$$\BR(\lcp\to\Sigma(1385)^+\piz)=(5.86\pm0.49\pm0.52\pm0.35)\times10^{-3},$$
$$\BR(\lcp\to\Sigma(1385)^0\pip)=(6.47\pm0.59\pm0.66\pm0.38)\times10^{-3},$$
where the third uncertainty is due to the input $\BR(\lcp\to\Lambda\pip\piz)$ and $\BR(\Sigma(1385)\to\Lambda\pi)$.
Based on the results of the PWA, the decay asymmetry parameters for the resonant components are determined for the first time:
$$\alphaA=-0.763\pm0.053\pm0.045,$$
$$\alphaB=-0.917\pm0.069\pm0.056,$$
$$\alphaC=-0.789\pm0.098\pm0.056.$$ 

\begin{table*}[!htbp]
\caption{The comparison among this work, various theoretical calculations and PDG results. Here, the uncertainties of this work are the combined uncertainties. ``---'' means unavailable.}
\setlength{\abovecaptionskip}{1.2cm}
\setlength{\belowcaptionskip}{0.2cm}
\label{tab:compa}
\begin{center}
\vspace{-0.0cm}
\resizebox{\textwidth}{20mm}{
\begin{tabular}{c|cc|c|c}
\hline\hline
 & \multicolumn{2}{c|}{Theoretical calculation} & This work & PDG\\
\hline
$10^2\times\BR(\lcp\to\Lambda\rho(770)^+)$  & $4.81\pm0.58$~\cite{theo2} & $4.0$~\cite{theo31,theo32} & $4.06\pm0.52$ & $<6$\\
$10^3\times\BR(\lcp\to\Sigma(1385)^+\piz)$  & $2.8\pm0.4$~\cite{theo4} & $2.2\pm0.4$~\cite{theo5} & $5.86\pm0.80$ & ---\\
$10^3\times\BR(\lcp\to\Sigma(1385)^0\pip)$  & $2.8\pm0.4$~\cite{theo4} & $2.2\pm0.4$~\cite{theo5} & $6.47\pm0.96$ & ---\\
$\alphaA$ & $-0.27\pm0.04$~\cite{theo2} & $-0.32$~\cite{theo31,theo32} & $-0.763\pm0.070$ & ---\\
$\alphaB$ & \multicolumn{2}{c|}{$-0.91^{+0.45}_{-0.10}$~\cite{theo5}}  & $-0.917\pm0.089$ & ---\\
$\alphaC$ & \multicolumn{2}{c|}{$-0.91^{+0.45}_{-0.10}$~\cite{theo5}}  & $-0.79\pm0.11$ & ---\\
\hline\hline
\end{tabular}
}
\end{center}
\end{table*}
\vspace{-0.0cm}

The comparison between this work and various theoretical calculations
can be found in Table~\ref{tab:compa}.  The
$\lcp\to\Lambda\rho(770)^+$ BF presented in this paper has good
agreement with the theoretical predictions evaluated in
Refs.~\cite{theo2,theo31,theo32}. However, the corresponding asymmetry
parameter differs significantly with their calculations.  The
comparison of the measurements presented here with the prediction of
Refs.~\cite{theo4,theo5} for the decays
$\lcp\to\Sigma(1385)^{+(0)}\pi^{0(+)}$ presents an opposite situation;
the asymmetry parameters are in good agreement, but a more than
$3\,\sigma$ difference is observed between the BF values.  As none of the
theoretical models is able to explain both the BFs and the decay
asymmetries, the results reported in this paper provide a crucial
input to improve and extend the current understanding of the dynamics
of the charmed baryon hadronic decays.

\appendix
\section*{APPENDIX}
\section{Alignment Angle Calculation}
\label{app:align_angle}
\indent\indent In the helicity formalism, several alignment angles are used in constructing the full amplitude as expressed in Eq.~\eqref{eq:helicty13}. Here, we detail the calculations of the alignment angles. The alignment angle is the correction of different definition of $z$-axis in helicity amplitude. 
Suppose an arbitrary decay $0\to 1+2$, of which the helicity formula can be expressed as
\begin{equation}
    \langle p_1, \lambda_1; p_2,\lambda_2 | U | p_0,\lambda_0 \rangle =H_{\lambda_1,\lambda_2} D^{J_0*}_{\lambda_0, \lambda_1 - \lambda_2}(\phi, \theta, 0),
\end{equation}
where $p_{0,1,2}$ are momenta, and $\lambda_{0,1,2}$ are helicities for the particles. This means the evolution operator $U$ can be divided into two parts, the rotation part $R$ 
\begin{equation}
\langle p_0,\lambda_1-\lambda_2|R(\phi, \theta, 0)|p_0,\lambda_0 \rangle = D^{J_0*}_{\lambda_0, \lambda_1 - \lambda_2}(\phi, \theta, 0)
\end{equation}
and the energy depended part $H_{\lambda_1,\lambda_2}$. The rotation operator gives a clear definition of $\lambda_1$ and $\lambda_2$ that the $z$-axis is the direction of $\vec{p_1}$ in the rest frame of $p_0$, which can be denoted as Euler rotation $R(\phi, \theta, 0)$. In sequence decay, all rotation from initial state to the final states should be considered.

In our $\Lambda_{c}^{+} \rightarrow \Lambda \pi^{+}\pi^{0},\Lambda \rightarrow p \pi^{-}$ process, there are three decay chains, which have been shown in Figure~\ref{fig:angle}. Following the rotation sequence, we can get the expressions below.
\begin{itemize}

\item For the decay chain $\Lambda_{c}^{+} \rightarrow \Lambda \rho(770)^+$, the rotation is expressed as:
\begin{equation}
    R_{\rho} =  
    R(\phi_{p}^{\Lambda_1}, \theta_{\Lambda_1}, 0) 
    B_z(p_{\Lambda}^{\Lambda_{c}^{+}}) 
    R_x(\pi)
    R(\phi_{\Lambda_{c}^{+}}^{1}, \theta_{\Lambda_{c}^{+}}^{1}, 0),
\end{equation}
where the Euler rotation $R(\phi,\theta, 0)$ can be expanded as $R_y(\theta) R_z(\phi)$, $B_z$ denotes boost operation and $R_x(\pi)$ part means rotation from $\rho(770)^+$ direction to the $\Lambda$ direction, where the rotation $R_x(\pi)$ will satisfy the relation $R_x(\pi)R(\phi,\theta, 0)=R(\phi - \pi, \pi - \theta, 0)$.

\item For $\Lambda_{c}^{+} \rightarrow \Sigma^{*+}\pi^{0}$, it is expressed as:
\begin{equation}
    R_{\Sigma^{*+}} =  
    R(\phi_{p}^{\Lambda_2}, \theta_{\Lambda_2}, 0) 
    B_z(p_{\Lambda}^{\Sigma^{*+}}) 
    R(\phi_{\Lambda}^{\Sigma^{*+}}, \theta_{\Sigma^{*+}}, 0)  
    B_z(p_{\Sigma^{*+}}) 
    R(\phi_{\Lambda_{c}^{+}}^{2}, \theta_{\Lambda_{c}^{+}}^{2}, 0).
\end{equation}

\item For $\Lambda_{c}^{+} \rightarrow \Sigma^{*0}\pi^{+}$, it is expressed as:
\begin{equation}
    R_{\Sigma^{*0}} =  
    R(\phi_{p}^{\Lambda_3}, \theta_{\Lambda_3}, 0) 
    B_z(p_{\Lambda}^{\Sigma^{*0}})  
    R(\phi_{\Lambda}^{\Sigma^{*0}}, \theta_{\Sigma^{*0}}, 0) 
    B_z(p_{\Sigma^{*0}})
    R(\phi_{\Lambda_{c}^{+}}^{3}, \theta_{\Lambda_{c}^{+}}^{3}, 0).
\end{equation}
\end{itemize}

When we take the decay chain $\Lambda_{c}^{+} \rightarrow \Lambda \rho(770)^+$ as a reference chain, the rotation of alignment is defined as:
\begin{equation}
    R_{\Sigma^{*+}}^{\mathrm{align}} = R_{\rho} R_{\Sigma^{*+}}^{-1},  \ 
    R_{\Sigma^{*0}}^{\mathrm{align}} = R_{\rho} R_{\Sigma^{*0}}^{-1}, 
\end{equation}
Using the 2-dimensional representation of $\mathrm{SU}(2)$ group, the rotation and boost operations can be expressed with:
\begin{equation}
R_z(\phi) = \begin{pmatrix}
e^{-i\frac{\phi}{2}} & 0 \\
0 & e^{i\frac{\phi}{2}}
\end{pmatrix}
,\,
R_y(\theta) = \begin{pmatrix}
\cos\frac{\theta}{2} & -\sin\frac{\theta}{2} \\
\sin\frac{\theta}{2} & \cos\frac{\theta}{2}
\end{pmatrix}
,\,
B_z(\omega) = \begin{pmatrix}
e^{-\omega/2} & 0 \\
0 & e^{\omega/2}
\end{pmatrix},
\end{equation}
where $\omega = \mathrm{arccosh}(\frac{1}{\sqrt{1-(\frac{p}{E})^2}})$ ,
we can convert $R^{\mathrm{align}}_{i}$ into several rotations with Euler angles which are the expected alignment angles and can be solved with the equations:
\begin{equation}
    R_{\Sigma^{*+}}^{\mathrm{align}} = B_z(\omega) R_z(\gamma_{p})R_y(\beta_{p})R_z(\alpha_{p}), \ 
    R_{\Sigma^{*0}}^{\mathrm{align}} = B_z(\omega') R_z(\gamma_{p}')R_y(\beta_{p}')R_z(\alpha_{p}').
\end{equation}

\acknowledgments
\hspace{1.5em}
The BESIII collaboration thanks the staff of BEPCII and the IHEP computing center for their strong support. This work is supported in part by National Key R\&D Program of China under Contracts Nos. 2020YFA0406400, 2020YFA0406300; National Natural Science Foundation of China (NSFC) under Contracts Nos. 11635010, 11735014, 11835012, 11935015, 11935016, 11935018, 11961141012, 12022510, 12025502, 12035009, 12035013, 12192260, 12192261, 12192262, 12192263, 12192264, 12192265, 12221005; the Chinese Academy of Sciences (CAS) Large-Scale Scientific Facility Program; Joint Large-Scale Scientific Facility Funds of the NSFC and CAS under Contract No. U1832207; the CAS Center for Excellence in Particle Physics (CCEPP); 100 Talents Program of CAS; Fundamental Research Funds for the Central Universities, Lanzhou University, University of Chinese Academy of Sciences; The Institute of Nuclear and Particle Physics (INPAC) and Shanghai Key Laboratory for Particle Physics and Cosmology; ERC under Contract No. 758462; European Union's Horizon 2020 research and innovation programme under Marie Sklodowska-Curie grant agreement under Contract No. 894790; German Research Foundation DFG under Contracts Nos. 443159800, Collaborative Research Center CRC 1044, GRK 2149; Istituto Nazionale di Fisica Nucleare, Italy; Ministry of Development of Turkey under Contract No. DPT2006K-120470; National Science and Technology fund; National Science Research and Innovation Fund (NSRF) via the Program Management Unit for Human Resources \& Institutional Development, Research and Innovation under Contract No. B16F640076; STFC (United Kingdom); Suranaree University of Technology (SUT), Thailand Science Research and Innovation (TSRI), and National Science Research and Innovation Fund (NSRF) under Contract No. 160355; The Royal Society, UK under Contracts Nos. DH140054, DH160214; The Swedish Research Council; U. S. Department of Energy under Contract No. DE-FG02-05ER41374.

\bibliographystyle{JHEP}
\bibliography{BAM550}

\newpage
\section*{The BESIII collaboration}
\addcontentsline{toc}{section}{The BESIII collaboration}

\begin{small}
\begin{center}
M.~Ablikim$^{1}$, M.~N.~Achasov$^{11,b}$, P.~Adlarson$^{70}$, M.~Albrecht$^{4}$, R.~Aliberti$^{31}$, A.~Amoroso$^{69A,69C}$, M.~R.~An$^{35}$, Q.~An$^{66,53}$, X.~H.~Bai$^{61}$, Y.~Bai$^{52}$
% ...
\\
\vspace{0.2cm}
(BESIII Collaboration)\\
\vspace{0.2cm} {\it
$^{1}$ Institute of High Energy Physics, Beijing 100049, People's Republic of China\\
% ...
$^{4}$ Bochum Ruhr-University, D-44780 Bochum, Germany\\
% ...
$^{11}$ G.I. Budker Institute of Nuclear Physics SB RAS (BINP), Novosibirsk 630090, Russia\\
% ...
$^{31}$ Johannes Gutenberg University of Mainz, Johann-Joachim-Becher-Weg 45, D-55099 Mainz, Germany\\
% ...
$^{35}$ Liaoning Normal University, Dalian 116029, People's Republic of China\\
% ...
$^{52}$ Southeast University, Nanjing 211100, People's Republic of China\\
$^{53}$ State Key Laboratory of Particle Detection and Electronics, Beijing 100049, Hefei 230026, People's Republic of China\\
% ...
$^{61}$ University of Jinan, Jinan 250022, People's Republic of China\\
% ...
$^{66}$ University of Science and Technology of China, Hefei 230026, People's Republic of China\\
% ...
$^{69}$ University of Turin and INFN, (A)University of Turin, I-10125, Turin, Italy; (B)University of Eastern Piedmont, I-15121, Alessandria, Italy; (C)INFN, I-10125, Turin, Italy\\
$^{70}$ Uppsala University, Box 516, SE-75120 Uppsala, Sweden\\
% ...
$^{b}$ Also at the Novosibirsk State University, Novosibirsk, 630090, Russia\\
% ...
}\end{center}
\vspace{0.4cm}
\end{small}
\newpage

\end{document}